\documentclass[10pt]{article}

\usepackage[dvips]{graphicx}
\usepackage{eepic,epic}
\usepackage{amsfonts}
\newcommand{\bela}[1]{\begin{equation}\label{#1}}
\newcommand{\ela}{\end{equation}}
\newcommand{\bear}[1]{\begin{array}{#1}}
\newcommand{\ear}{\end{array}}

\newcommand{\Chi}{\mbox{\boldmath $\chi$}}

\renewcommand{\Psi}{\mbox{\boldmath $\psi$}}

\renewcommand{\v}{\mbox{\boldmath $v$}}

\newcommand{\n}{\mbox{\boldmath $n$}}

\newcommand{\as}{\\[.6em]}

\newcommand{\AS}{\\[1.2em]}

\newcommand{\dis}{\displaystyle}
\renewcommand{\i}{\mbox{\rm i}}
\newcommand{\ii}{\mbox{\rm \scriptsize i}}

\newcommand{\text}{\textstyle}
\newcommand{\del}{\partial}

\newtheorem{theorem}{Theorem}
\newtheorem{corollary}{Corollary}
\newtheorem{definition}{Definition}

\begin{document}
\begin{center}
  \Large\bf
  Reciprocal figures, graphical statics and inversive geometry of
  the Schwarzian BKP hierarchy\\[8mm]
 \large\sc B.G.\ Konopelchenko\footnote{Permanent address: Dipartimento di
 Fisica, Universit\`a di Lecce and  Sezione INFN, 73100 Lecce, Italy}
  {\rm and} W.K.\ Schief\\[2mm]
  \small\sl School of Mathematics, The University of New South Wales,\\
  Sydney, NSW 2052, Australia\\[9mm]
\end{center}
\begin{abstract}
A remarkable connection between soliton theory
and an important and beautiful branch of the theory of graphical statics 
developed by Maxwell and his contemporaries is revealed. 
Thus, it is demonstrated that reciprocal triangles
which constitute the simplest pair of reciprocal figures representing
both a framework and a self-stress encapsulate 
the integrable discrete BKP equation and its Schwarzian version. The 
inherent M\"obius invariant nature
of the Schwarzian BKP equation is then exploited to define reciprocity in an
inversive geometric setting. Integrable pairs of lattices of non-trivial
combinatorics consisting of reciprocal
triangles and their natural generalizations are discussed. Particular
reductions of these BKP lattices are related to
the integrable discrete versions of Darboux's \mbox{2+1-dimensional} 
sine-Gordon 
equation and the classical Tzitz\'eica 
equation of affine geometry. Furthermore,
it is shown that octahedral figures and their hexahedral reciprocals
as considered by Maxwell likewise give rise to discrete integrable systems
and associated integrable lattices.

\end{abstract}

\section{Introduction}
\setcounter{equation}{0}

The soliton with its novel interaction property represents one of the most
intriguing of nonlinear phenomena in modern physics. Solitons occur in such
diverse areas as hydrodynamics, plasma and solid state physics, as well as
in general relativity \cite{SolAppl1,SolAppl2,SolAppl3}.
They have important current technological
applications in optical fibre communication systems and Josephson junction
superconducting devices~\cite{Has89,KocLub91}. 
Nonlinear equations which describe solitonic
phenomena (integrable systems) are ubiquitous and of great mathematical
interest. Thus, in particular, they are generically amenable to the 
Inverse Scattering Transform (IST) method~\cite{ZakManNovPit84,AblCla91} and 
admit invariance under B\"acklund transformations \cite{RogSha82,MatSal91}
with associated nonlinear superposition
principles (permutability theorems) whereby analytic expressions descriptive 
of multi-soliton interaction may be constructed.

Here, it demonstrated that there exists a deep connection between discrete
integrable systems and reciprocal figures of graphical statics 
(structural \mbox{geometry)}. 
The theory of reciprocal figures was developed by the physicist and 
geometer James Clerk Maxwell in the XIX$^{\rm \scriptstyle th}$ century in 
connection with diagrams of forces representing stresses in frameworks 
\cite{Max64}. Maxwell's contributions \cite{Niv65} 
culminated in the paper {\em On reciprocal
figures, frames and diagrams of forces} \cite{Max70} for which he received the
Keith Prize of the Royal Society of Edinburgh. However, Maxwell points out
that the construction of diagrams of forces in which each force is represented
by one line had been discovered independently and earlier by the practical
draughtsman W.P.\ Taylor and he also refers to William Rankine's 
{\em Applied Mechanics} \cite{Ran58}. 
Subsequently, in \cite{Jen69}, Fleeming Jenkin gave a 
variety of practical applications of reciprocal figures to the calculation
of strains on frameworks. He received the Keith Gold Medal for his paper
{\em On the application of graphic methods to the determination of the
efficiency of machinery} \cite{Jen77}. His contribution {\em Bridges} to the
Encyclopaedia Britannica \cite{EB} contains reciprocal figures associated with
a variety of bridges including suspension bridges. References to the 
relevant literature of this period may be found in Luigi Cremona's two 
treatises {\em Le Figure Reciproche Nella Statica Grafica} \cite{Cre72}.

It is interesting to note that Karl Culmann in his {\em Graphische Statik} 
\cite{Cul66}
makes great use of diagrams of forces, some of which are reciprocal. Culmann's
student Maurice Koechling, in turn, was one of two chief engineers in Gustave
Eiffel's company and carried out in a graphical manner many calculations 
for the Eiffel Tower built in 1889.
Remarkably, a century later, Maxwell's observation that 
there exists a close relationship between plane rectilinear reciprocal figures 
and 
perspective representations of closed polyhedra was rediscovered in the context
of `artificial intelligence', namely the recognition and `realizability' of 
plane line drawings as three-dimensional polyhedral scenes 
\cite{Mac73,ElcMic77}.

The connection with integrable systems is made by identifying the 
{\em geometric} compatibility  which guarantees the existence of reciprocal 
figures with the {\em algebraic} compatibility of {\em linear} 
difference equations 
which give rise to {\em nonlinear}
integrable difference equations. We begin with
the simplest pair of reciprocal figures, namely reciprocal triangles. It is
demonstrated that the dilation coefficients which encode the reciprocal 
relation between the two triangles satisfy the integrable discrete BKP 
equation. The 
discrete BKP (dBKP) equation is known to encapsulate the entire $B$-type 
Kadomtsev-Petviashvili hierarchy of integrable equations \cite{MIWA} 
and may also
be regarded as a nonlinear superposition principle for eight solutions of this
hierarchy generated by B\"acklund transformations \cite{NimSch97}. 
The vertices of 
the reciprocal triangles and their
`interior points' are shown to obey a simple 8-point relation which is likewise
integrable and may be regarded as a `Schwarzian' version of the dBKP equation.
Since this 8-point relation is formulated in terms of two cross-ratios, it is
invariant under the group of inversive transformations, that is M\"obius
transformations and complex conjugation. This observation is used to
formulate reciprocity in a purely inversive geometric manner. Thus, the
Schwarzian BKP equation constitutes a natural object of inversive geometry.

Integrable lattices on the complex plane 
consisting of reciprocal triangles and their
`circle geometric' extensions 
are defined in terms of pairs of face-centred cubic (fcc) lattices
and it is shown that there exist canonical geometric 
reductions which are associated with the discrete analogue of Darboux's
\mbox{2+1-dimensional} sine-Gordon equation set down in \cite{NimSch98} and the
discrete Tzitz\'eica equation which has been obtained earlier 
in both an algebraic and purely geometric 
manner~\mbox{\cite{Sch96}-\cite{BobSch99b}}. We 
discuss another class of reciprocal figures which is
obtained via a `truncation' of Maxwell's example of an `octahedral' figure 
\cite{Max64}. Once again, it is shown that the pair of reciprocal
figures so obtained may be considered `integrable'. We conclude the paper
with a discussion of Maxwell's complete \mbox{octahedral}
figure and determine a
canonical class of hexahedral reciprocals which may be defined in terms of 
multi-ratio relations. This class is then shown to be associated with 
integrable lattices
of octahedral-hexahedral combinatorics.

In an earlier paper \cite{KonSch01}, we established a connection between
the Schwarzian Kadomtsev-Petviashvili hierarchy and Menelaus' fundamental 
theorem of plane geometry. Therein, we first set down the discrete Schwarzian 
KP equation in the context of soliton theory and then interpreted it 
geometrically. Here, we adopt an inverse procedure. We discuss the geometry
of reciprocal figures and then retrieve discrete integrable systems and
their associated hierarchies. This is
to underline the intimate relation between canonical hierarchies of 
soliton theory and plane configurations in inversive geometry.

\section{Reciprocal figures, frames and diagrams of $\mbox{ }$ forces}
\setcounter{equation}{0}

Some 140 years ago, James Clerk Maxwell \cite{Max64} discovered a fascinating
connection between planar reciprocal figures, diagrams of forces and orthogonal
projections of polyhedra. Subsequently, Jenkin \cite{Jen69,Jen77} applied 
diagrams of
forces and reciprocal figures to the most important cases occurring in practice
at the time. Other names associated with the theory of graphical statics and,
in particular, reciprocal diagrams of forces are Bow, Culmann, Cremona, Rankine
and W.P.\ Taylor \cite{Ran58,Cre72,Cul66,Bow73}. 
Maxwell's geometric definition of a {\em frame} is
a system of straight lines connecting a number of points. These lines may 
be thought of as material pieces such as beams, rods or wires and the forces 
which act on each piece joining two points may be interpreted as two 
forces of the same magnitude but opposite direction acting 
between the two points. 

\begin{figure}[h]
\begin{center}
\setlength{\unitlength}{0.0004in}
\begin{picture}(11724,4989)(0,-10)
\path(4062,3162)(6312,912)
\path(1812,912)(4062,2037)
\path(4062,2037)(6312,912)
\path(4062,2037)(4062,3162)
\path(1812,912)(4062,3162)
\path(1812,912)(6312,912)
\path(8787,237)(8787,2037)
\path(8787,237)(10587,1137)
\path(9687,1137)(10587,1137)
\path(9687,1137)(8787,237)
\path(8787,2037)(9687,1137)
\path(10587,1137)(8787,2037)
\put(4062,3162){\shade\ellipse{100}{100}}
\put(1812,912){\shade\ellipse{100}{100}}
\put(6312,912){\shade\ellipse{100}{100}}
\put(4062,2037){\shade\ellipse{100}{100}}
\put(8787,2037){\shade\ellipse{100}{100}}
\put(8787,237){\shade\ellipse{100}{100}}
\put(9687,1137){\shade\ellipse{100}{100}}
\put(10587,1137){\shade\ellipse{100}{100}}
\dottedline{45}(6312,912)(8112,12)
\blacken\path(7991.252,38.833)(8112.000,12.000)(8018.085,92.498)(7991.252,38.833)
\dottedline{45}(1812,912)(12,12)
\blacken\path(105.915,92.498)(12.000,12.000)(132.748,38.833)(105.915,92.498)
\dottedline{45}(4062,2037)(4062,237)
\blacken\path(4032.000,357.000)(4062.000,237.000)(4092.000,357.000)(4032.000,357.000)
\dottedline{45}(4062,3162)(4062,4962)
\blacken\path(4092.000,4842.000)(4062.000,4962.000)(4032.000,4842.000)(4092.000,4842.000)
\dottedline{45}(4062,2037)(5862,2937)
\blacken\path(5768.085,2856.502)(5862.000,2937.000)(5741.252,2910.167)(5768.085,2856.502)
\dottedline{45}(4062,2037)(2262,2937)
\blacken\path(2382.748,2910.167)(2262.000,2937.000)(2355.915,2856.502)(2382.748,2910.167)
\dottedline{45}(1812,912)(2712,1812)
\blacken\path(2648.360,1705.934)(2712.000,1812.000)(2605.934,1748.360)(2648.360,1705.934)
\dottedline{45}(1812,912)(2712,912)
\blacken\path(2592.000,882.000)(2712.000,912.000)(2592.000,942.000)(2592.000,882.000)
\dottedline{45}(6312,912)(5412,912)
\blacken\path(5532.000,942.000)(5412.000,912.000)(5532.000,882.000)(5532.000,942.000)
\dottedline{45}(6312,912)(5412,1812)
\blacken\path(5518.066,1748.360)(5412.000,1812.000)(5475.640,1705.934)(5518.066,1748.360)
\dottedline{45}(4062,3162)(3162,2262)
\blacken\path(3225.640,2368.066)(3162.000,2262.000)(3268.066,2325.640)(3225.640,2368.066)
\dottedline{45}(4062,3162)(4962,2262)
\blacken\path(4855.934,2325.640)(4962.000,2262.000)(4898.360,2368.066)(4855.934,2325.640)
\path(8337,3837)(7437,4737)
\blacken\path(7543.066,4673.360)(7437.000,4737.000)(7500.640,4630.934)(7543.066,4673.360)
\path(9237,3837)(8337,3837)
\blacken\path(8457.000,3867.000)(8337.000,3837.000)(8457.000,3807.000)(8457.000,3867.000)
\path(7437,4737)(9237,3837)
\blacken\path(9116.252,3863.833)(9237.000,3837.000)(9143.085,3917.498)(9116.252,3863.833)
\path(9237,3387)(7437,2487)
\blacken\path(7530.915,2567.498)(7437.000,2487.000)(7557.748,2513.833)(7530.915,2567.498)
\path(8337,3387)(9237,3387)
\blacken\path(9117.000,3357.000)(9237.000,3387.000)(9117.000,3417.000)(9117.000,3357.000)
\path(7437,2487)(8337,3387)
\blacken\path(8273.360,3280.934)(8337.000,3387.000)(8230.934,3323.360)(8273.360,3280.934)
\path(7887,3612)(6987,2712)
\blacken\path(7050.640,2818.066)(6987.000,2712.000)(7093.066,2775.640)(7050.640,2818.066)
\path(6987,4512)(7887,3612)
\blacken\path(7780.934,3675.640)(7887.000,3612.000)(7823.360,3718.066)(7780.934,3675.640)
\path(6987,2712)(6987,4512)
\blacken\path(7017.000,4392.000)(6987.000,4512.000)(6957.000,4392.000)(7017.000,4392.000)
\path(9912,4512)(9912,2712)
\blacken\path(9882.000,2832.000)(9912.000,2712.000)(9942.000,2832.000)(9882.000,2832.000)
\path(11712,3612)(9912,4512)
\blacken\path(10032.748,4485.167)(9912.000,4512.000)(10005.915,4431.502)(10032.748,4485.167)
\path(9912,2712)(11712,3612)
\blacken\path(11618.085,3531.502)(11712.000,3612.000)(11591.252,3585.167)(11618.085,3531.502)
\put(1582,907){$\scriptstyle1$}
\put(2632,3657){\scriptsize (a)}
\put(9132,4657){\scriptsize (b)}
\put(11132,1057){\scriptsize (c)}
\put(10732,1057){$\scriptstyle6$}
\put(9382,1057){$\scriptstyle8$}
\put(8742,2157){$\scriptstyle7$}
\put(8742,-43){$\scriptstyle5$}
\put(6432,907){$\scriptstyle2$}
\put(4152,3207){$\scriptstyle3$}
\put(3882,1727){$\scriptstyle4$}
\put(8472,3157){$\scriptstyle1$}
\put(8472,3927){$\scriptstyle2$}
\put(7302,3567){$\scriptstyle3$}
\put(10542,3527){$\scriptstyle4$}
\end{picture}
\end{center}
\caption{A frame and its reciprocal diagram of forces}
\label{rec}
\end{figure}
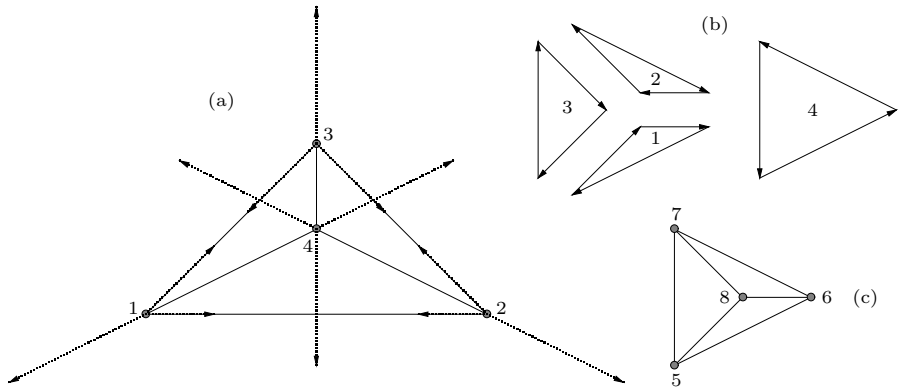

The simplest frame consists of four points joined by six lines 
as displayed in Figure \ref{rec}(a). 
One may now inquire as to whether there exist forces acting on the points
along the lines such that the frame is in {\em equilibrium}. 
Figure~\ref{rec}(a) depicts six forces and their counterparts 
which are parallel to the lines. In 
order to verify that the frame is in equilibrium, it must be possible to 
rearrange any three force vectors which act on a point in such a way that they
form a closed polygon. Figure~\ref{rec}(b) indicates that 
the four polygons of forces associated
with the points $1,2,3$ and 4 are indeed closed. Thus, the frame is in 
equilibrium and any other forces which guarantee equilibrium must
be multiples of those drawn in Figure~\ref{rec}(a). Hence, the ratios and the
directions of the forces are determined. 

\begin{figure}[h]
\begin{center}
\setlength{\unitlength}{0.00040489in}
\begin{picture}(7140,6745)(0,-10)
\thicklines
\path(2976,5362)(5001,2212)
\path(1176,3787)(2751,4462)
\path(2751,4462)(2976,5362)
\path(2751,4462)(5001,2212)
\path(1176,3787)(2976,5362)
\path(1176,3787)(5001,2212)
\dottedline{45}(2758,2250)(4978,4432)
\dottedline{45}(2758,2250)(1491,5257)
\dottedline{45}(1491,5250)(4978,4425)
\dottedline{45}(3193,3262)(2766,2242)
\dottedline{45}(4963,4432)(3193,3255)
\dottedline{45}(1498,5242)(3201,3255)
\thinlines
\put(2976,5362){\blacken\ellipse{100}{100}}
\put(1176,3787){\blacken\ellipse{100}{100}}
\put(2751,4462){\blacken\ellipse{100}{100}}
\put(5001,2212){\blacken\ellipse{100}{100}}
\put(1498,5242){\shade\ellipse{100}{100}}
\put(4971,4432){\shade\ellipse{100}{100}}
\put(3193,3262){\shade\ellipse{100}{100}}
\put(2766,2242){\shade\ellipse{100}{100}}
\put(1491,5239){\ellipse{2966}{2966}}
\put(4941,4406){\ellipse{4382}{4382}}
\put(2766,2239){\ellipse{4464}{4464}}
\put(3208,3285){\ellipse{4182}{4182}}
\end{picture}
\end{center}
\caption{Geometric construction of reciprocal triangles}
\label{max}
\end{figure}
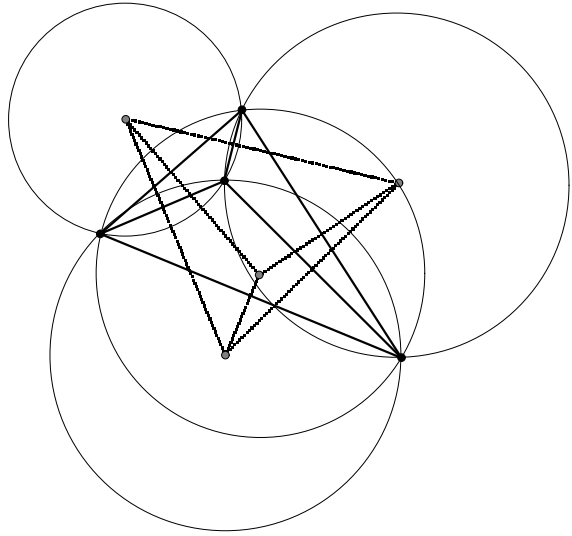

Now, the crucial observation is that the four polygons of forces may be fitted
together to form another frame. This is depicted in Figure \ref{rec}(c).
Consequently, it is also possible to interpret the directed lines connecting
the points $1,2,3$ and~4 as force vectors which act on the points of the second
frame in such a way that equilibrium is maintained. Hence, the two figures
\ref{rec}(a) and \ref{rec}(c) constitute {\em reciprocal} figures. Maxwell
gave a beautiful geometric 
construction of the second frame (cf.\ Figure \ref{max}). 
Thus, if we draw 
four circles passing through any three points of the original figure then
the line segments which connect the centres of these four circumcircles are
orthogonal to the six lines and form another figure consisting of four points
and six lines. If we rotate this figure by 90 degrees then a reciprocal figure
is obtained. This figure is definite in size and position but any figure
similar to it is still reciprocal to the original figure. 

It has been seen that the reciprocal of a figure of the type \ref{rec}(a) is
uniquely determined up to a scaling and its position on the plane. 
Based on the following definition, Maxwell \cite{Max64}
investigated in detail the conditions of indeterminateness and impossibility
in drawing reciprocal figures:
\vspace{-1.3mm}

\begin{definition} {\bf (Maxwell (1864)).} Two plane figures are reciprocal
when they consist of an equal number of lines, so that 
corresponding lines in the two figures are parallel, and corresponding lines
which converge to a point in one figure form a closed polygon in the other.
\end{definition}

\begin{figure}[h]
\begin{center}
\setlength{\unitlength}{0.00040489in}
\begin{picture}(8141,4146)(-500,-10)
\put(0403,3703){\blacken\ellipse{100}{100}}
\put(0178,2128){\blacken\ellipse{100}{100}}
\put(1078,1903){\blacken\ellipse{100}{100}}
\put(-943,328){\blacken\ellipse{100}{100}}
\put(0628,1228){\blacken\ellipse{100}{100}}
\put(2653,328){\blacken\ellipse{100}{100}}
\put(3428,1898){\blacken\ellipse{100}{100}}
\put(6243,1898){\blacken\ellipse{100}{100}}
\put(6653,1868){\blacken\ellipse{100}{100}}
\put(7133,558){\blacken\ellipse{100}{100}}
\put(6878,3478){\blacken\ellipse{100}{100}}
\put(7253,4078){\blacken\ellipse{100}{100}}
\put(8588,1403){\blacken\ellipse{100}{100}}
\put(7643,53){\blacken\ellipse{100}{100}}
\path(6873,3483)(6238,1903)
\path(6243,1898)(7138,558)
\path(6878,3478)(6648,1873)
\path(6648,1873)(7133,563)
\path(6870,3471)(7253,4078)
\path(3428,1903)(7253,4078)
\path(7133,561)(7643,51)
\path(3428,1903)(7635,51)
\path(3428,1903)(6240,1903)
\path(7643,43)(8588,1408)
\path(7253,4071)(8588,1401)
\path(6653,1866)(8588,1401)
\path(-953,328)(2653,328)
\path(0403,3703)(2653,328)
\path(-953,328)(0403,3703)
\path(-953,328)(0628,1228)
\path(0628,1228)(2653,328)
\path(0403,3703)(1078,1903)
\path(0403,3703)(0178,2128)
\path(0178,2128)(0628,1228)
\path(0628,1228)(1078,1903)
\path(0178,2128)(1078,1903)
\path(0178,2128)(-953,328)
\path(1078,1903)(2653,328)
\put(0808,148){$\scriptscriptstyle1$}
\put(1663,1903){$\scriptscriptstyle2$}
\put(-528,1903){$\scriptscriptstyle3$}
\put(-1000,3000){\scriptsize (a)}
\put(3500,3000){\scriptsize (b)}
\put(-98,688){$\scriptscriptstyle4$}
\put(1348,688){$\scriptscriptstyle5$}
\put(0673,2488){$\scriptscriptstyle6$}
\put(0313,2488){$\scriptscriptstyle7$}
\put(0223,1588){$\scriptscriptstyle8$}
\put(0878,1408){$\scriptscriptstyle9$}
\put(0493,2083){$\scriptscriptstyle10$}
\put(-228,1228){$\scriptscriptstyle11$}
\put(1503,1093){$\scriptscriptstyle12$}
\put(6488,1138){$\scriptscriptstyle2$}
\put(6353,2623){$\scriptscriptstyle3$}
\put(5228,3028){$\scriptscriptstyle4$}
\put(5543,733){$\scriptscriptstyle5$}
\put(5138,1993){$\scriptscriptstyle1$}
\put(6918,1273){$\scriptscriptstyle6$}
\put(6828,2488){$\scriptscriptstyle7$}
\put(7973,2713){$\scriptscriptstyle8$}
\put(8198,643){$\scriptscriptstyle9$}
\put(7568,1723){$\scriptscriptstyle10$}
\put(7073,3568){$\scriptscriptstyle11$}
\put(7393,373){$\scriptscriptstyle12$}
\end{picture}
\end{center}
\caption{`Indeterminate' or `conditionally' reciprocal figures}
\label{indeterminate}
\end{figure}
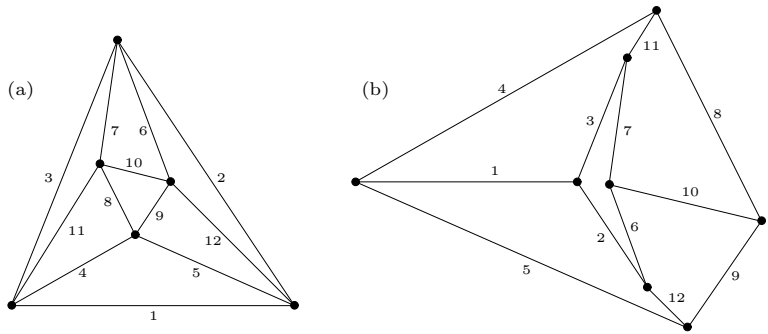

\vspace{-1.3mm}
\noindent
He first analysed this problem in an algebraic manner. Thus, the 
construction of the reciprocal of Figure \ref{rec}(a) is possible and unique
because the number of points equals the number of polygons in this figure.
Figure \ref{indeterminate}(a) displays a figure of `octahedral' combinatorics.
The number of polygons exceeds the number of points by two. Accordingly, the
construction of a reciprocal figure is possible but there exist two degrees
of freedom. This is reflected in the following construction of the reciprocal
figure. We first draw three concurrent lines which are parallel to the edges
$1,4$ and 5 and arbitrarily choose a point on each line. The
remainder of the reciprocal figure is then uniquely determined by drawing 
successively lines parallel to the edges $2, 12, 3, 11, 6, 7, 9$ and 8. The 
latter two lines and the line which is parallel to the edge 10 and passes 
through the intersection of the lines parallel to the edges 6 and 7 may be
shown to be concurrent. Hence, in addition to a scaling, the reciprocal figure
is indeed capable of two degrees of variability. It is noted that the 
reciprocal figures \ref{indeterminate} have the combinatorics of 
an octahedron and a cube respectively.

The construction of the reciprocal figure \ref{indeterminate}(b) shows that
if we begin with a figure of the combinatorics of Figure 
\ref{indeterminate}(b) then 
it is not always possible to construct a reciprocal. 
Indeed, there exist two 
conditions on the directions of the lines which are required to hold. From a
physical point of view, this implies that if these conditions are satisfied
then the frame is able to support forces. However, any small variation of
the frame may instantly lead to instability. Maxwell 
realized that these conditions are elegantly expressed
by the fact that Figure \ref{indeterminate}(b) constitutes an orthogonal 
projection of a closed polyhedron with planar faces. Remarkably, he proved 
in a constructive manner that any plane figure which may be regarded as an
orthogonal projection of a three-dimensional polyhedron with planar faces 
admits a reciprocal and may therefore be identified with a frame which is in 
equilibrium. Indeed, let a plane figure be the orthogonal projection of a
polyhedron and draw concurrent lines which are parallel to the normals to the 
faces of the polyhedron. The points of intersection
with the plane on which the figure lies define
another plane figure if we join by a line any two points which are 
associated with two adjacent faces. By construction, the lines of the
second figure will be orthogonal to the corresponding lines of the original
figure. Rotation by 90 degrees then produces a reciprocal figure.

The converse of Maxwell's theorem is not strictly true. It is not difficult to
find plane figures which admit reciprocals but are not related to spatial
polyhedra in the above-mentioned manner and frames which are in equilibrium 
but do not
have (proper) reciprocal diagrams of forces. Nevertheless, if one makes the
assumption that the combinatorics of the frame is that of a spherical
polyhedron then the existence of a reciprocal, a spatial polyhedron and 
a self-stress (that is, a frame in equilibrium) are equivalent 
\cite{CraWhi93}.

In view of the following, it is now useful to return to the construction
of the `reciprocal triangles' displayed in Figure \ref{rec}. Thus, given a
figure of the type~\ref{rec}(a), we first draw three lines which are parallel
to the edges $(1,4)$, $(2,4)$ and $(3,4)$ and denote the points of intersection
by $5, 6$ and 7 as shown in Figure~\ref{rec}(c). The two lines which pass
through the points 5 and 7 and are parallel to the edges $(1,3)$ and $(2,3)$
respectively determine the point 8. The existence of the reciprocal figure
then guarantees that the line which passes through 5 and 8 is indeed parallel
to the edge $(1,2)$. The phenomenom that during the construction of reciprocal 
figures certain lines turn out to be concurrent (cf.\ the construction of 
Figure \ref{indeterminate}(b)) provides the key to the link with integrable 
systems. Thus, in the following, we set in correspondence this {\em geometric
compatibility} of reciprocal figures with the {\em algebraic compatibility} of
linear systems defining discrete integrable systems.

\section{Reciprocal triangles, dilations and an 8-point relation}
\setcounter{equation}{0}

\begin{figure}
\begin{center}
\setlength{\unitlength}{0.00057489in}
\begin{picture}(7714,3705)(0,-10)
\put(7205,1420){$\beta'$}
\dottedline{45}(532.000,1658.000)(555.631,1728.609)(567.471,1802.120)
        (567.206,1876.578)(554.843,1950.003)(530.710,2020.442)
        (495.450,2086.022)(450.000,2145.000)
\dottedline{45}(1560.000,3128.000)(1607.028,3073.701)(1662.530,3028.100)
        (1724.920,2992.500)(1792.416,2967.918)(1863.089,2955.058)
        (1934.917,2954.286)(2005.850,2965.624)(2073.858,2988.750)
        (2137.000,3023.001)
\dottedline{45}(3525.000,878.000)(3460.271,839.000)(3401.201,791.867)
        (3348.805,737.412)(3303.983,676.569)(3267.505,610.386)
        (3240.000,540.000)
\dottedline{45}(1582.000,2093.000)(1664.866,2087.811)(1747.292,2097.795)
        (1826.523,2122.620)(1899.910,2161.454)(1965.000,2213.000)
\dottedline{45}(2040.000,2685.000)(2005.511,2749.280)(1959.401,2805.807)
        (1903.361,2852.507)(1839.446,2887.668)(1770.000,2910.000)
\dottedline{45}(1395.000,2835.000)(1337.351,2784.325)(1290.698,2723.376)
        (1256.831,2654.497)(1237.055,2580.334)(1232.127,2503.737)
        (1242.238,2427.651)(1267.000,2355.000)
\dottedline{45}(5782.000,720.000)(5748.013,785.686)(5705.769,846.392)
        (5655.987,901.086)(5599.512,948.838)(5537.304,988.837)
        (5470.421,1020.404)(5400.000,1043.000)
\dottedline{45}(5820.000,2723.000)(5901.560,2709.200)(5984.277,2709.849)
        (6065.611,2724.929)(6143.063,2753.975)(6214.255,2796.097)
        (6277.001,2849.999)
\dottedline{45}(7297.000,1815.000)(7241.495,1761.577)(7197.544,1698.306)
        (7166.854,1627.645)(7150.617,1552.338)(7149.464,1475.308)
        (7163.439,1399.548)(7192.000,1328.000)
\dottedline{45}(7072.000,1733.000)(7095.096,1810.451)(7093.656,1891.260)
        (7067.815,1967.840)(7020.000,2033.000)
\dottedline{45}(6540.000,2265.000)(6480.965,2223.897)(6430.247,2172.885)
        (6389.486,2113.614)(6360.000,2048.000)
\dottedline{45}(6525.000,1590.000)(6584.385,1538.377)(6653.178,1500.180)
        (6728.393,1477.067)(6806.764,1470.041)(6884.891,1479.408)
        (6959.381,1504.761)(7027.000,1545.000)
\path(5265,495)(5940,3195)
\path(5940,3188)(7612,1500)
\path(5257,495)(7605,1500)
\path(5265,503)(6825,1860)
\path(5940,3180)(6817,1845)
\path(1890,3420)(3915,270)
\path(90,1845)(3915,270)
\path(90,1845)(1665,2520)
\path(90,1845)(1890,3420)
\path(1665,2520)(1890,3420)
\path(1665,2520)(3915,270)
\path(6810,1850)(7605,1500)
\dottedline{45}(1665,2520)(2047,2685)
\dottedline{45}(1665,2528)(1380,2835)
\dottedline{45}(1665,2520)(1575,2100)
\dottedline{45}(6810,1853)(7020,2033)
\dottedline{45}(6810,1845)(6360,2040)
\dottedline{45}(6817,1845)(7027,1545)
\put(1890,3420){\blacken\ellipse{100}{100}}
\put(90,1845){\blacken\ellipse{100}{100}}
\put(1665,2520){\blacken\ellipse{100}{100}}
\put(5265,495){\shade\ellipse{100}{100}}
\put(5940,3195){\shade\ellipse{100}{100}}
\put(7605,1500){\shade\ellipse{100}{100}}
\put(6810,1853){\shade\ellipse{100}{100}}
\put(3915,270){\blacken\ellipse{100}{100}}
\put(0,1575){$\Phi_{12}$}
\put(3780,0){$\Phi_{23}$}
\put(1800,3555){$\Phi_{13}$}
\put(5175,225){$\Phi_1$}
\put(7515,1260){$\Phi_2$}
\put(5805,3325){$\Phi_3$}
\put(1890,2385){$\Phi$}
\put(6190,1805){$\Phi_{123}$}
\put(3375,540){$\alpha$}
\put(360,1800){$\gamma$}
\put(1850,3060){$\beta$}
\put(1745,2645){$\alpha'$}
\put(1300,2475){$\beta'$}
\put(1640,2185){$\gamma'$}
\put(5405,790){$\alpha'$}
\put(5890,2790){$\gamma'$}
\put(6480,2025){$\alpha$}
\put(6705,1575){$\beta$}
\put(6930,1845){$\gamma$}
\end{picture}
\end{center}
\caption{Reciprocal triangles}
\label{reciprocal}
\end{figure}
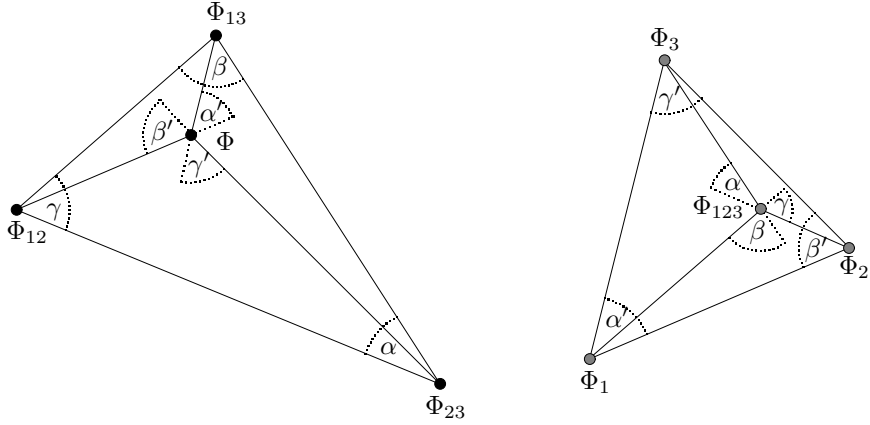

In this section, we investigate in detail the geometric and algebraic
properties of the simplest reciprocal figures, namely reciprocal
triangles. In order to reveal connections with both inversive geometry and 
soliton theory, it proves convenient to complement Maxwell's geometric
construction with a purely algebraic proof of the existence of reciprocal 
triangles. In fact, the proof given in 
\cite{Ped70} and recited below makes use of both a nonlinear algebraic system 
and its associated `linear representation'. Thus, consider a triangle
$\Delta(\Phi_{23},\Phi_{13},\Phi_{12})$ with vertices 
$\Phi_{23},\Phi_{13},\Phi_{12}$ and an additional point $\Phi$ on the
plane. If $\Delta(\Phi_1,\Phi_{2},\Phi_3)$ constitutes a triangle
with edges $(\Phi_1,\Phi_2)$, $(\Phi_2,\Phi_3)$, $(\Phi_3,\Phi_1)$ being
parallel to the line segments $(\Phi,\Phi_{12})$, $(\Phi,\Phi_{23})$,
$(\Phi,\Phi_{13})$ respectively then the two triangles are 
reciprocal\footnote{It is noted that it is justified to use the term {\em
reciprocal triangles}
without referring to the points $\Phi$ and $\Phi_{123}$ since the
latter are uniquely determined by the edges of the two triangles.} if
there exists a point $\Phi_{123}$ such that the line segments
$(\Phi_1,\Phi_{123})$, $(\Phi_2,\Phi_{123})$, $(\Phi_3,\Phi_{123})$ are
parallel to the edges $(\Phi_{13},\Phi_{12})$, $(\Phi_{12},\Phi_{23})$,
$(\Phi_{23},\Phi_{13})$ respectively as indicated in Figure \ref{reciprocal}.
It turns out convenient to identify the plane with the complex plane and 
regard $\Phi,\ldots,\Phi_{123}$ as complex numbers. \mbox{Accordingly}, by 
assumption, there exist real dilation coefficients $a,b,c$ such that
\bela{E1}
  \bear{rl}
   \Phi_{12} - \Phi = & c(\Phi_1 - \Phi_2)\as
   \Phi_{23} - \Phi = & a(\Phi_2 - \Phi_3)\as
   \Phi_{13} - \Phi = & b(\Phi_3 - \Phi_1)
  \ear
\ela
and the reciprocity condition amounts to the existence of a complex
number $\Phi_{123}\in\mathbb{C}$ and three real dilations $a_1,b_2,c_3\in
\mathbb{R}$ obeying the algebraic system
\bela{E2}
  \bear{rl}
   \Phi_{123} - \Phi_3 = & c_3(\Phi_{13} - \Phi_{23})\as
   \Phi_{123} - \Phi_1 = & a_1(\Phi_{12} - \Phi_{13})\as
   \Phi_{123} - \Phi_2 = & b_2(\Phi_{23} - \Phi_{12}).
  \ear
\ela
Now, any of the relations (\ref{E2}) may be regarded as a definition for
$\Phi_{123}$ and elimination of $\Phi_{123}$ from (\ref{E2})$_{1,2}$ and
(\ref{E2})$_{2,3}$ leads, on use of (\ref{E1}), to the necessary and
sufficient conditions
\bela{E3}
  \bear{rl}
   (1+c_3b+a_1b+c_3a)(\Phi_1-\Phi_3) = & (c_3a-a_1c)(\Phi_1-\Phi_2)\as
   (1+a_1c+b_2c+a_1b)(\Phi_2-\Phi_1) = & (a_1b-b_2a)(\Phi_2-\Phi_3).
  \ear
\ela
Since the triangles are assumed to be non-degenerate, the real 
coefficients in~(\ref{E3}) must vanish. If these four conditions on
the dilations were independent then there would exist a constraint on
the dilations $a,b,c$ and therefore the second triangle. However, Maxwell's 
geometric construction implies that this is not the case and, indeed, one 
condition turns out to
be redundant. The remaining three conditions define the dilations 
$a_1,b_2,c_3$. Thus, the existence of the reciprocal triangle is guaranteed
and, as a consequence, we may formulate the following theorem:

\begin{theorem} {\bf (Algebraic description of reciprocal triangles).} 
\label{T1}
\newline
If $\Delta(\Phi_{23},\Phi_{13},\Phi_{12})$ and $\Delta(\Phi_1,\Phi_2,\Phi_3)$
constitute reciprocal triangles with `interior points' $\Phi$ and $\Phi_{123}$
respectively then the dilations $a,b,c$ and $a_1,b_2,c_3$ as defined by
(\ref{E1}) and (\ref{E2}) respectively are related by 
\bela{E4}
  a_1 = -\frac{a}{ab+bc+ca},\quad b_2 = -\frac{b}{ab+bc+ca},\quad
  c_3 = -\frac{c}{ab+bc+ca}.
\ela
Conversely, let $\Phi_1,\Phi_2,\Phi_3$ and $\Phi$ be four generic points on the
complex plane, $a,b,c\in\mathbb{R}$ be arbitrary non-vanishing real 
numbers and $a_1,b_2,c_3$ be given by~(\ref{E4}). Then, the unique solution
of the compatible linear system 
(\ref{E1}), (\ref{E2}) gives rise to the reciprocal triangles  
$\Delta(\Phi_{23},\Phi_{13},\Phi_{12})$ and $\Delta(\Phi_1,\Phi_2,\Phi_3)$
with `interior points' $\Phi$ and $\Phi_{123}$ respectively.
\end{theorem} 

The second part of the above theorem expresses the fact that reciprocal
triangles may be constructed in the following way: Given four points 
$\Phi_1,\Phi_2,\Phi_3$, $\Phi$ on the complex plane and three real numbers 
$a,b,c$, we draw three line
segments $(\Phi,\Phi_{23})$, $(\Phi,\Phi_{13})$, $(\Phi,\Phi_{12})$ which are
parallel to the edges $(\Phi_2,\Phi_3)$, $(\Phi_3,\Phi_1)$, $(\Phi_1,\Phi_2)$
and their lengths are determined by the dilations $a,b,c$ respectively. 
The lines which pass through the vertices $\Phi_1,\Phi_2,\Phi_3$ and are 
parallel to the line segments $(\Phi_{13},\Phi_{12})$, $(\Phi_{12},\Phi_{23})$,
$(\Phi_{23},\Phi_{13})$ respectively then meet at a point~$\Phi_{123}$ with
associated dilations $a_1,b_2,c_3$ given by (\ref{E4}).

The relations (\ref{E4}) imply that, for instance, $c_3a = a_1c$ 
(cf.\ (\ref{E3})). Hence, on expressing the dilations $a,c,a_1,c_3$ in terms of
$\Phi,\ldots,\Phi_{123}$ by means of (\ref{E1}),~(\ref{E2}) and rearranging 
terms, we are led to the following observation:

\begin{corollary} {\bf (An 8-point relation).} 
\label{cor1}
The quadruplets  
$\Phi_1,\Phi_2,\Phi_3,\Phi_{123}$ and $\Phi_{23},\Phi_{13},\Phi_{12},\Phi$ 
associated with two reciprocal triangles satisfy the 8-point relation
\bela{E5}
  \frac{(\Phi_1-\Phi_2)(\Phi_3-\Phi_{123})}{(\Phi_2-\Phi_3)
        (\Phi_{123}-\Phi_1)} = 
  \frac{(\Phi_{23}-\Phi_{13})(\Phi_{12}-\Phi)}{(\Phi_{13}-\Phi_{12})
        (\Phi-\Phi_{23})},
\ela
that is
\bela{E6}
  Q(\Phi_1,\Phi_2,\Phi_3,\Phi_{123}) = Q(\Phi_{23},\Phi_{13},\Phi_{12},\Phi),
\ela
where the cross-ratio $Q$ of four points on the complex plane is defined 
as usual~by
\bela{E7}
  Q(P_1,P_2,P_3,P_4) = \frac{(P_1-P_2)(P_3-P_4)}{(P_2-P_3)(P_4-P_1)}.
\ela
\end{corollary}
It is noted that, by construction and due to the symmetries of the cross-ratio,
the above 8-point relation is invariant under any simultaneous permutation of 
the arguments of the two cross-ratios. 

\section{Inversive geometry of reciprocal triangles}
\setcounter{equation}{0}

In the preceding, it has been shown that reciprocal triangles provide 
particular solutions of the 8-point relation (\ref{E5}). This observation
suggests that there should exist a complete geometric characterization
of the 8-point relation in terms of reciprocal triangles if the latter 
are appropriately generalized. 
To this end, it is natural to adopt a
wider definition of reciprocal figures (as suggested by Maxwell). Thus, 
two triangles are
reciprocally related if corresponding pairs of edges are not necessarily 
parallel but meet at the same angle.
In other words, relative rotations of the triangles are admissible and hence
reciprocal triangles are characterized by the equality of corresponding
angles as shown in Figure \ref{reciprocal}.

Since the cross-ratio of four points on the complex plane
is preserved by M\"obius transformations
\bela{E8}
  \Psi \rightarrow \Psi' = \frac{\tilde{a}\Psi + \tilde{b}}{\tilde{c}\Psi
                    + \tilde{d}}
\ela
with $\tilde{a}\tilde{d}-\tilde{b}\tilde{c}\neq0$, it is evident that
the 8-point relation is invariant under what may be referred to as
{\em local} M\"obius transformations 
acting independently on the quadruplets $\Phi_1,\Phi_2,\Phi_3,\Phi_{123}$ and 
$\Phi_{23},\Phi_{13},\Phi_{12},\Phi$. Consequently, the \mbox{8-point}
relation is preserved by {\em global} M\"obius transformations
which act on the complex plane and therefore simultaneously on both
quadruplets. In addition, the \mbox{8-point} relation is seen to be
invariant under complex conjugation. Thus, the 
8-point relation turns out to be preserved by the group of inversive 
transformations which consists of M\"obius transformations and complex 
conjugation and maps (generalized) circles to (generalized) circles 
\cite{Ped70,BraEspGra00}.
Particular inversive transformations include translations, scalings, rotations
and inversions in a line or circle. The latter are represented by
\bela{E9}
  \Psi \rightarrow \Psi' = \Psi_* + \frac{r^2}{\bar{\Psi}-\bar{\Psi}_*},
\ela
where $r$ and $\Psi_*$ denote the radius and centre of a circle respectively.
It is noted that inversive transformations may be decomposed into inversions.
Accordingly, the 8-point relation constitutes an object of
inversive geometry which, in the spirit of Klein,
is concerned with those properties of figures on the plane which are preserved
by inversive transformations \cite{Ped70}-\cite{MorMor54}.

The above analysis of the symmetry group of the 8-point relation shows 
that a canonical definition of 
reciprocity is required to allow for the global group of inversive 
transformations and local group of M\"obius transformations. It is therefore
natural to adopt the following definition:

\begin{definition} {\bf \boldmath
(Reciprocal $(4,6)$ configurations).} A configuration
consisting of four points on the complex plane which are linked by six 
circular arcs is termed a {\em $(4,6)$ configuration} if the circular 
extensions of the arcs meet at a point (cf.~Figure \ref{foursix}). 
Two $(4,6)$ configurations are said to be {\em reciprocally related} if the 
six angles made by the circular arcs of one $(4,6)$ configuration equal
those of the other $(4,6)$ configuration in the manner indicated in Figures 
\ref{foursix} and \ref{pireciprocal}.
\end{definition}
\begin{figure}[h]
\begin{center}
\setlength{\unitlength}{0.00050489in}
\begin{picture}(9180,3638)(0,-10)
\put(0,653){$\Phi_{12}$}
\put(4050,1103){$\Phi_{23}$}
\put(2295,3488){$\Phi_{13}$}
\put(1705,1508){$\Phi$}
\put(7765,558){$\Phi_{*}$}
\put(725,1128){$\tilde{\gamma}$}
\put(495,1373){$\gamma$}
\put(1980,2948){$\tilde{\beta}$}
\put(2340,2813){$\beta$}
\put(3420,1598){$\tilde{\alpha}$}
\put(3150,1373){$\alpha$}
\put(2050,1498){$\gamma'$}
\put(2050,1993){$\alpha'$}
\put(1440,1693){$\beta'$}
\dottedline{45}(960.000,1096.000)(937.388,1181.345)(907.296,1263.897)
        (865.000,1343.000)
\dottedline{45}(1042.000,1111.000)(1024.473,1185.587)(1000.508,1258.361)
        (970.288,1328.768)(934.042,1396.271)(892.046,1460.356)
        (844.622,1520.534)(792.130,1576.346)(734.971,1627.368)
        (673.579,1673.210)(608.423,1713.523)(540.000,1748.000)
\dottedline{45}(1770.000,3053.000)(1806.922,2987.746)(1851.531,2927.485)
        (1903.160,2873.119)(1961.038,2825.459)(2024.300,2785.219)
        (2092.000,2753.000)
\dottedline{45}(1687.000,2986.000)(1727.964,2924.038)(1774.383,2866.050)
        (1825.872,2812.513)(1882.007,2763.870)(1942.325,2720.522)
        (2006.328,2682.827)(2073.487,2651.095)(2143.249,2625.589)
        (2215.038,2606.519)(2288.262,2594.042)(2362.315,2588.262)
        (2436.587,2589.227)(2510.466,2596.927)(2583.340,2611.300)
        (2654.610,2632.227)(2723.687,2659.536)(2790.000,2693.000)
\dottedline{45}(9000.000,338.000)(8945.768,385.167)(8891.078,431.801)
        (8835.933,477.898)(8780.341,523.454)(8724.306,568.463)
        (8667.833,612.922)(8610.927,656.827)(8553.596,700.173)
        (8495.842,742.956)(8437.673,785.172)(8379.094,826.817)
        (8320.111,867.887)(8260.728,908.378)(8200.952,948.286)
        (8140.789,987.608)(8080.244,1026.339)(8019.323,1064.477)
        (7958.032,1102.016)(7896.377,1138.954)(7834.363,1175.287)
        (7771.997,1211.012)(7709.284,1246.125)(7646.231,1280.623)
        (7582.843,1314.503)(7519.127,1347.761)(7455.089,1380.394)
        (7390.734,1412.399)(7326.070,1443.773)(7261.102,1474.513)
        (7195.836,1504.616)(7130.279,1534.080)(7064.437,1562.900)
        (6998.316,1591.075)(6931.922,1618.602)(6865.263,1645.478)
        (6798.344,1671.701)(6731.171,1697.268)(6663.752,1722.177)
        (6596.092,1746.425)(6528.198,1770.010)(6460.077,1792.930)
        (6391.735,1815.182)(6323.179,1836.765)(6254.414,1857.676)
        (6185.449,1877.914)(6116.288,1897.476)(6046.940,1916.360)
        (5977.410,1934.566)(5907.706,1952.090)(5837.833,1968.931)
        (5767.799,1985.089)(5697.610,2000.560)(5627.274,2015.344)
        (5556.796,2029.440)(5486.183,2042.845)(5415.443,2055.559)
        (5344.582,2067.580)(5273.607,2078.908)(5202.524,2089.541)
        (5131.341,2099.478)(5060.064,2108.718)(4988.699,2117.261)
        (4917.255,2125.105)(4845.738,2132.250)(4774.153,2138.695)
        (4702.510,2144.439)(4630.813,2149.483)(4559.071,2153.825)
        (4487.290,2157.464)(4415.476,2160.402)(4343.637,2162.637)
        (4271.780,2164.169)(4199.911,2164.998)(4128.038,2165.124)
        (4056.166,2164.547)(3984.304,2163.267)(3912.458,2161.284)
        (3840.634,2158.599)(3768.841,2155.211)(3697.084,2151.121)
        (3625.370,2146.329)(3553.706,2140.836)(3482.100,2134.642)
        (3410.558,2127.748)(3339.087,2120.155)(3267.693,2111.863)
        (3196.384,2102.872)(3125.166,2093.185)(3054.046,2082.802)
        (2983.032,2071.723)(2912.129,2059.950)(2841.345,2047.485)
        (2770.686,2034.327)(2700.159,2020.479)(2629.771,2005.942)
        (2559.528,1990.717)(2489.438,1974.805)(2419.507,1958.209)
        (2349.741,1940.929)(2280.148,1922.968)(2210.734,1904.327)
        (2141.505,1885.008)(2072.469,1865.013)(2003.632,1844.343)
        (1935.000,1823.001)
\path(1935.000,1823.000)(2004.629,1805.761)(2074.308,1788.726)
        (2144.036,1771.894)(2213.813,1755.267)(2283.639,1738.843)
        (2353.512,1722.624)(2423.433,1706.609)(2493.400,1690.798)
        (2563.412,1675.192)(2633.471,1659.790)(2703.573,1644.594)
        (2773.720,1629.602)(2843.911,1614.816)(2914.144,1600.235)
        (2984.420,1585.859)(3054.738,1571.688)(3125.096,1557.724)
        (3195.495,1543.964)(3265.934,1530.411)(3336.413,1517.064)
        (3406.930,1503.923)(3477.485,1490.988)(3548.078,1478.259)
        (3618.707,1465.736)(3689.373,1453.421)(3760.075,1441.311)
        (3830.811,1429.409)(3901.583,1417.713)(3972.388,1406.224)
        (4043.226,1394.942)(4114.097,1383.868)(4185.000,1373.000)
\dottedline{45}(1927.000,1816.000)(1927.045,1740.718)(1928.888,1665.458)
        (1932.528,1590.264)(1937.963,1515.179)(1945.189,1440.244)
        (1954.204,1365.504)(1965.000,1291.000)
\dottedline{45}(7860.000,61.000)(7897.980,122.692)(7934.518,185.248)
        (7969.595,248.636)(8003.191,312.820)(8035.288,377.767)
        (8065.869,443.442)(8094.918,509.808)(8122.419,576.831)
        (8148.357,644.474)(8172.718,712.701)(8195.489,781.474)
        (8216.658,850.758)(8236.213,920.514)(8254.144,990.706)
        (8270.441,1061.294)(8285.095,1132.242)(8298.099,1203.511)
        (8309.445,1275.063)(8319.128,1346.858)(8327.142,1418.859)
        (8333.482,1491.026)(8338.146,1563.322)(8341.131,1635.706)
        (8342.436,1708.139)(8342.058,1780.584)(8340.000,1853.000)
\dottedline{45}(8445.000,8.000)(8436.223,79.587)(8426.215,151.013)
        (8414.981,222.256)(8402.522,293.295)(8388.843,364.110)
        (8373.948,434.678)(8357.841,504.980)(8340.526,574.994)
        (8322.010,644.701)(8302.298,714.078)(8281.395,783.106)
        (8259.307,851.764)(8236.041,920.032)(8211.605,987.889)
        (8186.005,1055.316)(8159.248,1122.293)(8131.343,1188.799)
        (8102.298,1254.816)(8072.122,1320.323)(8040.823,1385.301)
        (8008.411,1449.731)(7974.896,1513.595)(7940.286,1576.872)
        (7904.593,1639.544)(7867.828,1701.593)(7830.000,1763.000)
\dottedline{45}(4185.000,1373.000)(4255.674,1361.236)(4326.371,1349.608)
        (4397.089,1338.114)(4467.830,1326.757)(4538.592,1315.534)
        (4609.375,1304.447)(4680.180,1293.495)(4751.005,1282.679)
        (4821.851,1271.998)(4892.717,1261.453)(4963.603,1251.043)
        (5034.509,1240.770)(5105.435,1230.631)(5176.380,1220.629)
        (5247.343,1210.762)(5318.326,1201.031)(5389.327,1191.436)
        (5460.346,1181.976)(5531.383,1172.653)(5602.438,1163.465)
        (5673.511,1154.414)(5744.600,1145.498)(5815.707,1136.719)
        (5886.830,1128.075)(5957.970,1119.568)(6029.125,1111.197)
        (6100.297,1102.961)(6171.484,1094.862)(6242.687,1086.900)
        (6313.905,1079.073)(6385.137,1071.383)(6456.384,1063.829)
        (6527.646,1056.411)(6598.921,1049.130)(6670.210,1041.985)
        (6741.513,1034.977)(6812.829,1028.105)(6884.159,1021.369)
        (6955.500,1014.770)(7026.855,1008.307)(7098.222,1001.981)
        (7169.600,995.791)(7240.990,989.738)(7312.392,983.822)
        (7383.805,978.042)(7455.229,972.399)(7526.664,966.892)
        (7598.109,961.523)(7669.564,956.290)(7741.029,951.193)
        (7812.503,946.233)(7883.987,941.410)(7955.480,936.724)
        (8026.982,932.175)(8098.493,927.762)(8170.011,923.487)
        (8241.538,919.348)(8313.073,915.346)(8384.615,911.480)
        (8456.164,907.752)(8527.721,904.161)(8599.284,900.706)
        (8670.853,897.389)(8742.429,894.208)(8814.011,891.164)
        (8885.598,888.257)(8957.191,885.488)(9028.789,882.855)
        (9100.392,880.359)(9172.000,878.000)
\path(2385.000,3398.000)(2348.620,3335.412)(2313.642,3272.031)
        (2280.082,3207.886)(2247.958,3143.012)(2217.284,3077.438)
        (2188.077,3011.199)(2160.351,2944.326)(2134.119,2876.853)
        (2109.395,2808.813)(2086.190,2740.240)(2064.516,2671.168)
        (2044.385,2601.631)(2025.805,2531.663)(2008.787,2461.299)
        (1993.338,2390.573)(1979.466,2319.522)(1967.178,2248.180)
        (1956.480,2176.582)(1947.378,2104.764)(1939.875,2032.761)
        (1933.976,1960.608)(1929.684,1888.343)(1927.000,1816.000)
\dottedline{45}(4177.000,1381.000)(4238.041,1343.558)(4299.641,1307.042)
        (4361.786,1271.462)(4424.461,1236.824)(4487.652,1203.137)
        (4551.345,1170.408)(4615.525,1138.646)(4680.178,1107.857)
        (4745.289,1078.048)(4810.842,1049.226)(4876.823,1021.398)
        (4943.218,994.569)(5010.009,968.747)(5077.184,943.937)
        (5144.726,920.145)(5212.619,897.375)(5280.848,875.634)
        (5349.399,854.927)(5418.254,835.257)(5487.399,816.630)
        (5556.817,799.049)(5626.492,782.519)(5696.410,767.044)
        (5766.553,752.627)(5836.907,739.271)(5907.453,726.980)
        (5978.178,715.755)(6049.064,705.601)(6120.095,696.518)
        (6191.256,688.510)(6262.529,681.578)(6333.899,675.723)
        (6405.349,670.947)(6476.864,667.252)(6548.426,664.637)
        (6620.019,663.103)(6691.627,662.651)(6763.234,663.282)
        (6834.823,664.993)(6906.379,667.787)(6977.883,671.661)
        (7049.321,676.614)(7120.677,682.647)(7191.932,689.757)
        (7263.073,697.942)(7334.081,707.202)(7404.942,717.533)
        (7475.638,728.933)(7546.154,741.400)(7616.474,754.931)
        (7686.581,769.523)(7756.460,785.172)(7826.094,801.876)
        (7895.468,819.629)(7964.566,838.429)(8033.372,858.270)
        (8101.871,879.148)(8170.046,901.059)(8237.883,923.998)
        (8305.365,947.958)(8372.477,972.936)(8439.204,998.924)
        (8505.532,1025.918)(8571.443,1053.910)(8636.925,1082.895)
        (8701.961,1112.866)(8766.537,1143.817)(8830.637,1175.739)
        (8894.249,1208.626)(8957.356,1242.470)(9019.944,1277.264)
        (9082.000,1313.000)
\path(2385.000,3398.000)(2415.475,3332.497)(2446.984,3267.485)
        (2479.517,3202.979)(2513.068,3138.997)(2547.628,3075.554)
        (2583.187,3012.666)(2619.738,2950.349)(2657.271,2888.619)
        (2695.777,2827.490)(2735.245,2766.979)(2775.666,2707.100)
        (2817.031,2647.868)(2859.328,2589.299)(2902.547,2531.407)
        (2946.677,2474.206)(2991.707,2417.711)(3037.626,2361.937)
        (3084.422,2306.896)(3132.083,2252.603)(3180.598,2199.071)
        (3229.955,2146.314)(3280.141,2094.345)(3331.143,2043.177)
        (3382.949,1992.823)(3435.545,1943.295)(3488.919,1894.607)
        (3543.058,1846.769)(3597.946,1799.795)(3653.572,1753.695)
        (3709.920,1708.482)(3766.978,1664.166)(3824.729,1620.760)
        (3883.161,1578.273)(3942.258,1536.717)(4002.005,1496.101)
        (4062.388,1456.437)(4123.392,1417.733)(4185.000,1380.000)
\dottedline{45}(9142.000,698.000)(9073.004,717.545)(9003.929,736.809)
        (8934.775,755.791)(8865.545,774.490)(8796.240,792.907)
        (8726.859,811.041)(8657.406,828.892)(8587.880,846.460)
        (8518.283,863.744)(8448.616,880.744)(8378.880,897.460)
        (8309.077,913.891)(8239.207,930.037)(8169.272,945.899)
        (8099.273,961.474)(8029.211,976.765)(7959.088,991.769)
        (7888.903,1006.487)(7818.659,1020.919)(7748.357,1035.064)
        (7677.998,1048.923)(7607.583,1062.494)(7537.113,1075.778)
        (7466.589,1088.775)(7396.013,1101.483)(7325.386,1113.904)
        (7254.709,1126.037)(7183.983,1137.881)(7113.209,1149.437)
        (7042.389,1160.704)(6971.523,1171.682)(6900.613,1182.371)
        (6829.660,1192.771)(6758.665,1202.881)(6687.630,1212.702)
        (6616.555,1222.232)(6545.442,1231.473)(6474.291,1240.424)
        (6403.105,1249.084)(6331.884,1257.454)(6260.630,1265.534)
        (6189.343,1273.323)(6118.025,1280.820)(6046.677,1288.028)
        (5975.300,1294.944)(5903.896,1301.568)(5832.465,1307.902)
        (5761.009,1313.944)(5689.529,1319.695)(5618.026,1325.154)
        (5546.501,1330.321)(5474.956,1335.197)(5403.392,1339.781)
        (5331.809,1344.073)(5260.210,1348.073)(5188.595,1351.781)
        (5116.965,1355.196)(5045.322,1358.320)(4973.667,1361.151)
        (4902.001,1363.691)(4830.325,1365.937)(4758.641,1367.892)
        (4686.949,1369.554)(4615.251,1370.924)(4543.548,1372.001)
        (4471.841,1372.786)(4400.132,1373.278)(4328.421,1373.478)
        (4256.710,1373.385)(4185.000,1373.000)
\path(4185.000,1373.000)(4113.370,1372.723)(4041.741,1372.173)
        (3970.115,1371.349)(3898.493,1370.253)(3826.875,1368.883)
        (3755.263,1367.240)(3683.658,1365.324)(3612.060,1363.135)
        (3540.472,1360.673)(3468.893,1357.938)(3397.326,1354.930)
        (3325.770,1351.649)(3254.227,1348.095)(3182.699,1344.268)
        (3111.186,1340.169)(3039.688,1335.797)(2968.208,1331.152)
        (2896.746,1326.234)(2825.304,1321.044)(2753.882,1315.582)
        (2682.481,1309.847)(2611.102,1303.840)(2539.747,1297.561)
        (2468.417,1291.010)(2397.112,1284.186)(2325.833,1277.091)
        (2254.582,1269.724)(2183.360,1262.085)(2112.167,1254.175)
        (2041.005,1245.993)(1969.875,1237.540)(1898.777,1228.816)
        (1827.713,1219.820)(1756.684,1210.554)(1685.691,1201.017)
        (1614.735,1191.209)(1543.817,1181.130)(1472.938,1170.781)
        (1402.098,1160.162)(1331.300,1149.273)(1260.544,1138.114)
        (1189.831,1126.685)(1119.162,1114.986)(1048.538,1103.018)
        (977.960,1090.781)(907.429,1078.275)(836.947,1065.500)
        (766.514,1052.456)(696.131,1039.143)(625.799,1025.562)
        (555.520,1011.713)(485.294,997.596)(415.123,983.212)
        (345.006,968.560)(274.946,953.640)(204.944,938.453)
        (135.000,923.000)
\path(1928.000,1815.000)(1859.941,1792.223)(1792.120,1768.748)
        (1724.543,1744.577)(1657.218,1719.713)(1590.152,1694.160)
        (1523.353,1667.918)(1456.826,1640.992)(1390.580,1613.383)
        (1324.620,1585.096)(1258.955,1556.133)(1193.590,1526.496)
        (1128.534,1496.190)(1063.792,1465.216)(999.372,1433.580)
        (935.281,1401.283)(871.524,1368.330)(808.109,1334.723)
        (745.043,1300.467)(682.332,1265.564)(619.984,1230.020)
        (558.003,1193.836)(496.397,1157.018)(435.173,1119.569)
        (374.337,1081.493)(313.895,1042.794)(253.854,1003.476)
        (194.220,963.543)(135.000,923.000)
\path(2385.000,3413.000)(2320.039,3382.312)(2255.602,3350.538)
        (2191.708,3317.686)(2128.375,3283.766)(2065.620,3248.787)
        (2003.461,3212.759)(1941.917,3175.692)(1881.003,3137.597)
        (1820.738,3098.484)(1761.138,3058.365)(1702.220,3017.250)
        (1644.001,2975.152)(1586.497,2932.082)(1529.725,2888.053)
        (1473.700,2843.076)(1418.438,2797.165)(1363.955,2750.333)
        (1310.266,2702.591)(1257.387,2653.956)(1205.331,2604.438)
        (1154.115,2554.054)(1103.752,2502.817)(1054.256,2450.741)
        (1005.642,2397.841)(957.924,2344.133)(911.113,2289.630)
        (865.225,2234.349)(820.272,2178.306)(776.266,2121.515)
        (733.220,2063.993)(691.146,2005.757)(650.056,1946.822)
        (609.961,1887.206)(570.874,1826.924)(532.804,1765.995)
        (495.763,1704.435)(459.760,1642.261)(424.807,1579.492)
        (390.913,1516.145)(358.088,1452.237)(326.340,1387.787)
        (295.679,1322.813)(266.114,1257.333)(237.653,1191.366)
        (210.303,1124.931)(184.073,1058.045)(158.969,990.729)
        (135.000,923.000)
\dottedline{45}(1965.000,1291.000)(2040.890,1299.337)(2114.717,1318.788)
        (2184.863,1348.925)(2249.791,1389.089)(2308.077,1438.399)
        (2358.444,1495.774)(2399.788,1559.957)(2431.203,1629.541)
        (2452.000,1703.000)
\dottedline{45}(2452.000,1973.000)(2424.397,2048.119)(2385.785,2118.217)
        (2337.047,2181.694)(2279.296,2237.098)(2213.853,2283.163)
        (2142.214,2318.836)(2066.016,2343.301)(1987.000,2356.000)
\dottedline{45}(1402.000,1951.000)(1386.953,1871.945)(1385.818,1791.479)
        (1398.629,1712.031)(1425.000,1636.000)
\dottedline{45}(3292.000,2063.000)(3245.332,2001.073)(3203.240,1935.949)
        (3165.942,1867.965)(3133.633,1797.473)(3106.480,1724.840)
        (3084.623,1650.441)(3068.175,1574.663)(3057.222,1497.897)
        (3051.821,1420.543)(3052.000,1343.000)
\dottedline{45}(3352.000,2011.000)(3309.831,1951.951)(3271.489,1890.348)
        (3237.127,1826.439)(3206.885,1760.481)(3180.882,1692.740)
        (3159.225,1623.486)(3142.000,1553.000)
\dottedline{45}(1927,1823)(1410,1951)
\put(135,923){\blacken\ellipse{100}{100}}
\put(2385,3398){\blacken\ellipse{100}{100}}
\put(1935,1823){\blacken\ellipse{100}{100}}
\put(4185,1373){\blacken\ellipse{100}{100}}
\put(8235,923){\whiten\ellipse{100}{100}}
\end{picture}
\end{center}
\caption{A $(4,6)$ configuration}
\label{foursix}
\end{figure}
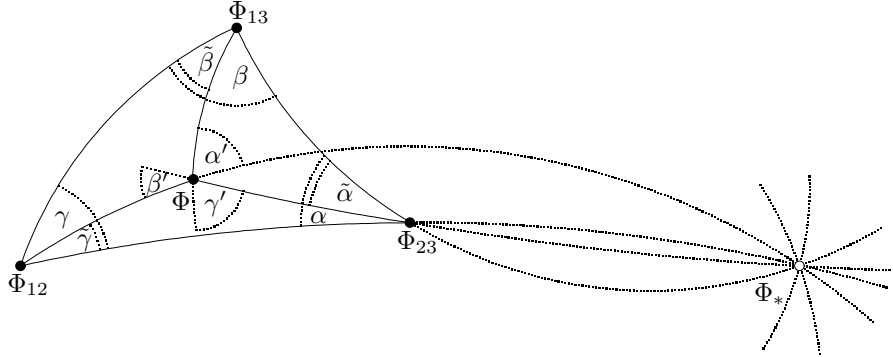
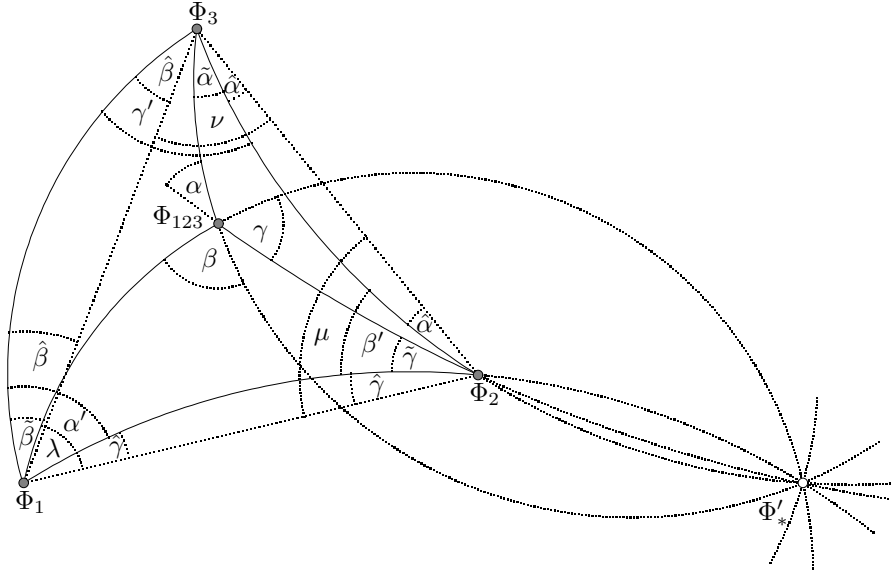
\begin{figure}[h]
\begin{center}
\setlength{\unitlength}{0.00050489in}
\begin{picture}(9191,5865)(0,-10)
\put(586,1445){$\alpha'$}
\put(3686,2265){$\beta'$}
\put(1301,4680){$\gamma'$}
\put(121,1295){$\tilde{\beta}$}
\put(4116,2135){$\tilde{\gamma}$}
\put(1971,5040){$\tilde{\alpha}$}
\put(3191,2385){$\mu$}
\put(2111,4590){$\nu$}
\put(86,630){$\Phi_1$}
\put(4811,1755){$\Phi_2$}
\put(1886,5715){$\Phi_3$}
\put(1511,3595){$\Phi_{123}$}
\put(1061,1210){$\hat{\gamma}$}
\put(266,2115){$\hat{\beta}$}
\put(3771,1845){$\hat{\gamma}$}
\put(4256,2475){$\hat{\alpha}$}
\put(2251,4955){$\hat{\alpha}$}
\put(1571,5085){$\hat{\beta}$}
\put(1856,3915){$\alpha$}
\put(2021,3150){$\beta$}
\put(2561,3465){$\gamma$}
\put(401,1170){$\lambda$}
\put(7806,520){$\Phi_*'$}
\path(1976.000,5625.000)(1915.257,5585.650)(1855.194,5545.271)
        (1795.829,5503.872)(1737.178,5461.468)(1679.258,5418.069)
        (1622.087,5373.689)(1565.682,5328.341)(1510.057,5282.037)
        (1455.231,5234.792)(1401.218,5186.618)(1348.035,5137.530)
        (1295.696,5087.543)(1244.218,5036.670)(1193.614,4984.926)
        (1143.901,4932.328)(1095.091,4878.889)(1047.200,4824.625)
        (1000.241,4769.553)(954.228,4713.688)(909.175,4657.046)
        (865.094,4599.645)(821.998,4541.499)(779.901,4482.628)
        (738.813,4423.047)(698.747,4362.774)(659.715,4301.827)
        (621.728,4240.223)(584.797,4177.980)(548.933,4115.116)
        (514.146,4051.650)(480.447,3987.599)(447.845,3922.984)
        (416.349,3857.822)(385.969,3792.132)(356.714,3725.934)
        (328.592,3659.246)(301.611,3592.088)(275.779,3524.481)
        (251.105,3456.442)(227.594,3387.993)(205.253,3319.152)
        (184.090,3249.941)(164.110,3180.379)(145.320,3110.486)
        (127.724,3040.283)(111.328,2969.790)(96.136,2899.028)
        (82.153,2828.017)(69.384,2756.778)(57.831,2685.331)
        (47.498,2613.698)(38.388,2541.899)(30.504,2469.955)
        (23.848,2397.887)(18.421,2325.716)(14.227,2253.463)
        (11.265,2181.149)(9.537,2108.795)(9.043,2036.422)
        (9.783,1964.052)(11.757,1891.704)(14.965,1819.400)
        (19.405,1747.162)(25.077,1675.010)(31.978,1602.965)
        (40.106,1531.049)(49.460,1459.281)(60.037,1387.683)
        (71.833,1316.276)(84.845,1245.081)(99.069,1174.118)
        (114.501,1103.408)(131.137,1032.971)(148.971,962.828)
        (167.999,893.000)
\dottedline{45}(7938.000,68.000)(7971.627,131.628)(8004.140,195.834)
        (8035.528,260.596)(8065.782,325.896)(8094.892,391.713)
        (8122.851,458.028)(8149.648,524.821)(8175.277,592.070)
        (8199.730,659.757)(8222.998,727.859)(8245.075,796.357)
        (8265.955,865.230)(8285.630,934.456)(8304.095,1004.014)
        (8321.344,1073.884)(8337.372,1144.045)(8352.175,1214.474)
        (8365.747,1285.150)(8378.084,1356.053)(8389.183,1427.159)
        (8399.041,1498.449)(8407.654,1569.899)(8415.019,1641.489)
        (8421.135,1713.197)(8426.000,1785.000)
\path(1976.000,5625.000)(2001.158,5558.012)(2026.980,5491.277)
        (2053.464,5424.803)(2080.608,5358.594)(2108.409,5292.659)
        (2136.863,5227.004)(2165.969,5161.634)(2195.723,5096.558)
        (2226.123,5031.780)(2257.165,4967.307)(2288.846,4903.146)
        (2321.164,4839.304)(2354.114,4775.786)(2387.694,4712.598)
        (2421.901,4649.747)(2456.731,4587.240)(2492.180,4525.081)
        (2528.246,4463.279)(2564.924,4401.837)(2602.211,4340.764)
        (2640.103,4280.064)(2678.597,4219.743)(2717.688,4159.809)
        (2757.373,4100.266)(2797.649,4041.120)(2838.510,3982.377)
        (2879.953,3924.044)(2921.973,3866.125)(2964.568,3808.627)
        (3007.731,3751.555)(3051.460,3694.915)(3095.750,3638.713)
        (3140.597,3582.954)(3185.995,3527.643)(3231.941,3472.786)
        (3278.430,3418.388)(3325.457,3364.456)(3373.018,3310.993)
        (3421.108,3258.006)(3469.723,3205.499)(3518.857,3153.479)
        (3568.506,3101.949)(3618.664,3050.915)(3669.328,3000.383)
        (3720.491,2950.356)(3772.148,2900.841)(3824.296,2851.841)
        (3876.928,2803.362)(3930.039,2755.409)(3983.624,2707.986)
        (4037.678,2661.099)(4092.196,2614.750)(4147.171,2568.946)
        (4202.599,2523.691)(4258.474,2478.989)(4314.791,2434.844)
        (4371.544,2391.262)(4428.727,2348.246)(4486.335,2305.800)
        (4544.362,2263.929)(4602.802,2222.637)(4661.650,2181.928)
        (4720.899,2141.806)(4780.545,2102.274)(4840.580,2063.338)
        (4901.000,2025.000)
\dottedline{45}(4901.000,2025.000)(4961.269,1986.582)(5021.917,1948.767)
        (5082.940,1911.558)(5144.330,1874.960)(5206.082,1838.975)
        (5268.190,1803.608)(5330.647,1768.862)(5393.448,1734.740)
        (5456.586,1701.246)(5520.054,1668.383)(5583.847,1636.154)
        (5647.958,1604.563)(5712.381,1573.612)(5777.109,1543.306)
        (5842.136,1513.646)(5907.456,1484.636)(5973.061,1456.278)
        (6038.946,1428.576)(6105.104,1401.532)(6171.528,1375.149)
        (6238.212,1349.430)(6305.149,1324.376)(6372.332,1299.992)
        (6439.755,1276.278)(6507.412,1253.237)(6575.294,1230.872)
        (6643.396,1209.185)(6711.711,1188.178)(6780.232,1167.853)
        (6848.952,1148.212)(6917.865,1129.257)(6986.963,1110.990)
        (7056.240,1093.413)(7125.688,1076.527)(7195.302,1060.335)
        (7265.073,1044.837)(7334.995,1030.036)(7405.062,1015.932)
        (7475.266,1002.528)(7545.599,989.824)(7616.056,977.822)
        (7686.629,966.523)(7757.312,955.928)(7828.096,946.038)
        (7898.975,936.854)(7969.943,928.378)(8040.991,920.609)
        (8112.113,913.549)(8183.303,907.199)(8254.551,901.558)
        (8325.853,896.628)(8397.200,892.409)(8468.586,888.902)
        (8540.003,886.106)(8611.445,884.022)(8682.903,882.651)
        (8754.372,881.992)(8825.844,882.046)(8897.312,882.812)
        (8968.768,884.291)(9040.207,886.482)(9111.620,889.385)
        (9183.000,893.000)
\path(4901.000,2025.000)(4829.616,2031.439)(4758.173,2037.198)
        (4686.679,2042.276)(4615.140,2046.674)(4543.563,2050.391)
        (4471.953,2053.425)(4400.318,2055.778)(4328.663,2057.449)
        (4256.996,2058.438)(4185.322,2058.744)(4113.649,2058.368)
        (4041.983,2057.309)(3970.330,2055.568)(3898.697,2053.145)
        (3827.090,2050.041)(3755.516,2046.254)(3683.981,2041.787)
        (3612.492,2036.638)(3541.056,2030.809)(3469.678,2024.301)
        (3398.365,2017.113)(3327.124,2009.246)(3255.961,2000.702)
        (3184.883,1991.480)(3113.895,1981.583)(3043.005,1971.009)
        (2972.219,1959.762)(2901.544,1947.841)(2830.985,1935.248)
        (2760.549,1921.983)(2690.242,1908.048)(2620.071,1893.445)
        (2550.043,1878.175)(2480.163,1862.238)(2410.438,1845.637)
        (2340.874,1828.373)(2271.478,1810.447)(2202.256,1791.861)
        (2133.213,1772.617)(2064.357,1752.717)(1995.694,1732.162)
        (1927.229,1710.955)(1858.970,1689.096)(1790.921,1666.589)
        (1723.090,1643.434)(1655.482,1619.636)(1588.105,1595.194)
        (1520.962,1570.112)(1454.062,1544.392)(1387.410,1518.036)
        (1321.011,1491.047)(1254.873,1463.427)(1189.000,1435.179)
        (1123.399,1406.305)(1058.076,1376.807)(993.037,1346.689)
        (928.288,1315.953)(863.834,1284.602)(799.682,1252.639)
        (735.837,1220.066)(672.305,1186.887)(609.091,1153.105)
        (546.202,1118.723)(483.644,1083.743)(421.421,1048.169)
        (359.539,1012.005)(298.005,975.253)(236.824,937.917)
        (176.000,900.000)
\dottedline{45}(9011.000,383.000)(8954.098,427.771)(8896.777,472.006)
        (8839.044,515.700)(8780.902,558.849)(8722.358,601.451)
        (8663.416,643.501)(8604.081,684.995)(8544.360,725.929)
        (8484.256,766.301)(8423.776,806.106)(8362.924,845.342)
        (8301.706,884.004)(8240.128,922.089)(8178.195,959.594)
        (8115.912,996.515)(8053.285,1032.849)(7990.320,1068.594)
        (7927.021,1103.745)(7863.395,1138.301)(7799.447,1172.256)
        (7735.183,1205.610)(7670.609,1238.358)(7605.729,1270.498)
        (7540.551,1302.028)(7475.079,1332.943)(7409.319,1363.242)
        (7343.278,1392.922)(7276.961,1421.980)(7210.374,1450.413)
        (7143.522,1478.220)(7076.412,1505.398)(7009.050,1531.943)
        (6941.442,1557.855)(6873.593,1583.131)(6805.509,1607.767)
        (6737.197,1631.764)(6668.663,1655.117)(6599.912,1677.825)
        (6530.951,1699.886)(6461.786,1721.299)(6392.422,1742.061)
        (6322.867,1762.170)(6253.126,1781.625)(6183.205,1800.423)
        (6113.110,1818.564)(6042.849,1836.046)(5972.426,1852.867)
        (5901.848,1869.025)(5831.121,1884.519)(5760.252,1899.349)
        (5689.247,1913.512)(5618.112,1927.007)(5546.853,1939.833)
        (5475.476,1951.989)(5403.989,1963.474)(5332.397,1974.287)
        (5260.707,1984.427)(5188.924,1993.893)(5117.056,2002.684)
        (5045.108,2010.799)(4973.087,2018.238)(4901.000,2025.000)
\path(2201.000,3600.000)(2137.807,3564.525)(2075.252,3527.937)
        (2013.353,3490.248)(1952.131,3451.471)(1891.605,3411.616)
        (1831.793,3370.696)(1772.715,3328.725)(1714.388,3285.716)
        (1656.831,3241.681)(1600.062,3196.635)(1544.099,3150.592)
        (1488.960,3103.566)(1434.661,3055.571)(1381.219,3006.624)
        (1328.652,2956.739)(1276.975,2905.932)(1226.205,2854.218)
        (1176.359,2801.615)(1127.450,2748.138)(1079.495,2693.804)
        (1032.509,2638.630)(986.507,2582.634)(941.502,2525.832)
        (897.509,2468.243)(854.542,2409.885)(812.614,2350.776)
        (771.738,2290.935)(731.927,2230.379)(693.194,2169.129)
        (655.550,2107.204)(619.008,2044.622)(583.578,1981.403)
        (549.273,1917.567)(516.102,1853.135)(484.076,1788.126)
        (453.206,1722.560)(423.500,1656.459)(394.968,1589.842)
        (367.619,1522.731)(341.461,1455.147)(316.504,1387.111)
        (292.753,1318.643)(270.218,1249.767)(248.905,1180.502)
        (228.820,1110.871)(209.971,1040.896)(192.362,970.598)
        (176.000,900.000)
\dottedline{45}(8381.000,8.000)(8379.744,79.699)(8377.250,151.365)
        (8373.520,222.978)(8368.555,294.515)(8362.357,365.957)
        (8354.926,437.280)(8346.266,508.465)(8336.379,579.490)
        (8325.267,650.333)(8312.935,720.975)(8299.386,791.393)
        (8284.623,861.567)(8268.652,931.475)(8251.477,1001.098)
        (8233.103,1070.413)(8213.536,1139.402)(8192.781,1208.042)
        (8170.845,1276.315)(8147.734,1344.198)(8123.455,1411.673)
        (8098.016,1478.718)(8071.423,1545.315)(8043.685,1611.442)
        (8014.810,1677.082)(7984.807,1742.213)(7953.684,1806.817)
        (7921.451,1870.874)(7888.118,1934.366)(7853.694,1997.272)
        (7818.189,2059.576)(7781.615,2121.257)(7743.982,2182.298)
        (7705.301,2242.681)(7665.584,2302.387)(7624.843,2361.399)
        (7583.089,2419.700)(7540.335,2477.271)(7496.595,2534.095)
        (7451.880,2590.157)(7406.205,2645.438)(7359.583,2699.923)
        (7312.027,2753.596)(7263.552,2806.440)(7214.173,2858.440)
        (7163.904,2909.579)(7112.759,2959.844)(7060.755,3009.219)
        (7007.907,3057.689)(6954.230,3105.239)(6899.741,3151.857)
        (6844.455,3197.527)(6788.390,3242.237)(6731.561,3285.972)
        (6673.986,3328.720)(6615.682,3370.469)(6556.667,3411.205)
        (6496.957,3450.917)(6436.571,3489.592)(6375.526,3527.220)
        (6313.841,3563.788)(6251.535,3599.287)(6188.625,3633.706)
        (6125.130,3667.033)(6061.070,3699.260)(5996.463,3730.377)
        (5931.329,3760.375)(5865.688,3789.244)(5799.557,3816.976)
        (5732.958,3843.562)(5665.910,3868.996)(5598.434,3893.269)
        (5530.548,3916.374)(5462.274,3938.304)(5393.632,3959.052)
        (5324.641,3978.613)(5255.324,3996.981)(5185.700,4014.150)
        (5115.790,4030.115)(5045.615,4044.871)(4975.196,4058.414)
        (4904.553,4070.739)(4833.709,4081.844)(4762.683,4091.725)
        (4691.497,4100.379)(4620.173,4107.803)(4548.731,4113.995)
        (4477.193,4118.954)(4405.580,4122.677)(4333.914,4125.164)
        (4262.215,4126.414)(4190.505,4126.427)(4118.806,4125.202)
        (4047.138,4122.740)(3975.524,4119.042)(3903.984,4114.109)
        (3832.540,4107.942)(3761.213,4100.543)(3690.025,4091.914)
        (3618.995,4082.058)(3548.147,4070.978)(3477.500,4058.677)
        (3407.076,4045.159)(3336.896,4030.428)(3266.980,4014.488)
        (3197.350,3997.343)(3128.026,3979.000)(3059.029,3959.463)
        (2990.380,3938.739)(2922.098,3916.833)(2854.204,3893.752)
        (2786.719,3869.503)(2719.662,3844.093)(2653.054,3817.530)
        (2586.914,3789.821)(2521.261,3760.975)(2456.117,3731.001)
        (2391.499,3699.907)(2327.428,3667.702)(2263.922,3634.397)
        (2201.000,3600.001)
\dottedline{45}(4901.000,2025.000)(4966.283,1994.533)(5031.709,1964.372)
        (5097.274,1934.516)(5162.978,1904.968)(5228.820,1875.727)
        (5294.798,1846.795)(5360.910,1818.171)(5427.156,1789.857)
        (5493.533,1761.852)(5560.040,1734.159)(5626.676,1706.776)
        (5693.440,1679.706)(5760.329,1652.948)(5827.342,1626.503)
        (5894.479,1600.371)(5961.737,1574.554)(6029.115,1549.051)
        (6096.611,1523.864)(6164.225,1498.993)(6231.954,1474.438)
        (6299.797,1450.200)(6367.753,1426.279)(6435.819,1402.677)
        (6503.996,1379.393)(6572.280,1356.428)(6640.671,1333.782)
        (6709.168,1311.457)(6777.767,1289.452)(6846.469,1267.768)
        (6915.272,1246.405)(6984.174,1225.365)(7053.173,1204.647)
        (7122.269,1184.251)(7191.459,1164.179)(7260.742,1144.431)
        (7330.117,1125.007)(7399.582,1105.907)(7469.135,1087.132)
        (7538.776,1068.683)(7608.502,1050.560)(7678.311,1032.762)
        (7748.204,1015.292)(7818.177,998.148)(7888.230,981.332)
        (7958.360,964.843)(8028.567,948.683)(8098.849,932.851)
        (8169.204,917.348)(8239.630,902.173)(8310.127,887.329)
        (8380.692,872.814)(8451.325,858.629)(8522.023,844.774)
        (8592.785,831.251)(8663.610,818.058)(8734.495,805.197)
        (8805.440,792.667)(8876.443,780.469)(8947.501,768.603)
        (9018.615,757.069)(9089.782,745.868)(9161.000,735.000)
\path(2193.000,3600.000)(2252.352,3557.556)(2311.893,3515.379)
        (2371.622,3473.468)(2431.538,3431.825)(2491.640,3390.451)
        (2551.927,3349.346)(2612.397,3308.512)(2673.050,3267.949)
        (2733.883,3227.657)(2794.896,3187.639)(2856.088,3147.894)
        (2917.457,3108.423)(2979.002,3069.228)(3040.722,3030.308)
        (3102.616,2991.666)(3164.682,2953.300)(3226.919,2915.213)
        (3289.326,2877.405)(3351.902,2839.877)(3414.645,2802.629)
        (3477.554,2765.662)(3540.628,2728.978)(3603.865,2692.576)
        (3667.265,2656.458)(3730.826,2620.624)(3794.547,2585.075)
        (3858.427,2549.811)(3922.463,2514.834)(3986.656,2480.144)
        (4051.003,2445.741)(4115.504,2411.628)(4180.156,2377.803)
        (4244.960,2344.268)(4309.913,2311.023)(4375.014,2278.069)
        (4440.262,2245.408)(4505.655,2213.038)(4571.193,2180.962)
        (4636.874,2149.180)(4702.696,2117.692)(4768.659,2086.499)
        (4834.761,2055.601)(4901.000,2025.000)
\path(1976.000,5625.000)(1968.050,5552.089)(1961.264,5479.061)
        (1955.644,5405.934)(1951.191,5332.726)(1947.907,5259.457)
        (1945.791,5186.144)(1944.846,5112.807)(1945.070,5039.465)
        (1946.465,4966.135)(1949.029,4892.837)(1952.762,4819.589)
        (1957.663,4746.410)(1963.730,4673.319)(1970.963,4600.333)
        (1979.359,4527.473)(1988.916,4454.755)(1999.633,4382.199)
        (2011.505,4309.824)(2024.530,4237.646)(2038.705,4165.686)
        (2054.026,4093.962)(2070.489,4022.490)(2088.090,3951.291)
        (2106.825,3880.381)(2126.689,3809.779)(2147.676,3739.503)
        (2169.782,3669.571)(2193.000,3600.000)
\dottedline{45}(2193.000,3600.000)(2217.258,3532.173)(2242.583,3464.738)
        (2268.970,3397.711)(2296.412,3331.109)(2324.903,3264.949)
        (2354.434,3199.246)(2385.000,3134.018)(2416.591,3069.281)
        (2449.200,3005.051)(2482.820,2941.344)(2517.442,2878.175)
        (2553.056,2815.561)(2589.655,2753.518)(2627.229,2692.059)
        (2665.768,2631.202)(2705.264,2570.961)(2745.705,2511.351)
        (2787.084,2452.387)(2829.387,2394.084)(2872.607,2336.455)
        (2916.730,2279.517)(2961.748,2223.282)(3007.647,2167.765)
        (3054.418,2112.980)(3102.047,2058.940)(3150.524,2005.659)
        (3199.837,1953.150)(3249.972,1901.426)(3300.918,1850.500)
        (3352.661,1800.385)(3405.189,1751.093)(3458.489,1702.637)
        (3512.548,1655.028)(3567.351,1608.279)(3622.886,1562.401)
        (3679.138,1517.406)(3736.094,1473.304)(3793.739,1430.108)
        (3852.059,1387.826)(3911.039,1346.471)(3970.665,1306.053)
        (4030.921,1266.580)(4091.793,1228.065)(4153.266,1190.515)
        (4215.324,1153.940)(4277.952,1118.350)(4341.134,1083.753)
        (4404.854,1050.158)(4469.097,1017.574)(4533.847,986.008)
        (4599.086,955.468)(4664.800,925.962)(4730.972,897.497)
        (4797.584,870.081)(4864.621,843.720)(4932.067,818.421)
        (4999.903,794.190)(5068.113,771.032)(5136.681,748.955)
        (5205.588,727.964)(5274.819,708.062)(5344.355,689.257)
        (5414.179,671.552)(5484.274,654.951)(5554.622,639.459)
        (5625.207,625.081)(5696.009,611.818)(5767.012,599.675)
        (5838.198,588.654)(5909.550,578.759)(5981.048,569.992)
        (6052.676,562.355)(6124.416,555.849)(6196.249,550.478)
        (6268.159,546.241)(6340.126,543.140)(6412.133,541.176)
        (6484.162,540.349)(6556.196,540.659)(6628.215,542.107)
        (6700.203,544.692)(6772.141,548.413)(6844.011,553.270)
        (6915.795,559.260)(6987.476,566.384)(7059.036,574.638)
        (7130.456,584.021)(7201.719,594.531)(7272.808,606.165)
        (7343.703,618.919)(7414.389,632.792)(7484.847,647.779)
        (7555.059,663.876)(7625.008,681.080)(7694.677,699.387)
        (7764.049,718.791)(7833.105,739.288)(7901.829,760.873)
        (7970.204,783.540)(8038.212,807.284)(8105.837,832.099)
        (8173.061,857.979)(8239.869,884.917)(8306.243,912.907)
        (8372.166,941.940)(8437.623,972.012)(8502.597,1003.113)
        (8567.072,1035.236)(8631.032,1068.373)(8694.460,1102.516)
        (8757.342,1137.656)(8819.660,1173.785)(8881.401,1210.893)
        (8942.548,1248.971)(9003.086,1288.010)(9063.000,1328.000)
\dottedline{45}(1638.000,3218.000)(1684.607,3156.676)(1737.973,3101.135)
        (1797.387,3052.117)(1862.056,3010.276)(1931.118,2976.170)
        (2003.651,2950.253)(2078.689,2932.872)(2155.230,2924.259)
        (2232.255,2924.528)(2308.734,2933.675)(2383.649,2951.579)
        (2456.000,2978.000)
\dottedline{45}(2748.000,3218.000)(2791.459,3282.373)(2826.498,3351.689)
        (2852.564,3424.854)(2869.245,3500.710)(2876.277,3578.060)
        (2873.548,3655.682)(2861.103,3732.347)(2839.138,3806.846)
        (2808.000,3878.000)
\dottedline{45}(2021.000,4245.000)(1950.122,4224.661)(1882.287,4195.750)
        (1818.527,4158.709)(1759.812,4114.100)(1707.036,4062.603)
        (1661.000,4005.000)
\dottedline{45}(1061.000,1373.000)(1024.423,1436.877)(983.214,1497.869)
        (937.600,1555.640)(887.832,1609.873)(834.182,1660.269)
        (776.946,1706.553)(716.438,1748.469)(652.992,1785.788)
        (586.955,1818.303)(518.690,1845.837)(448.574,1868.237)
        (376.990,1885.382)(304.333,1897.176)(231.002,1903.555)
        (157.400,1904.484)(83.931,1899.957)(11.000,1890.000)
\dottedline{45}(1271.000,1163.000)(1250.207,1254.473)(1217.342,1342.335)
        (1173.000,1425.000)
\dottedline{45}(4338.000,2723.000)(4274.109,2674.272)(4218.651,2616.125)
        (4173.000,2550.000)
\dottedline{45}(4098.000,2408.000)(4062.744,2344.473)(4035.097,2277.285)
        (4015.432,2207.343)(4004.014,2135.592)(4001.000,2063.000)
\dottedline{45}(731.000,2370.000)(663.684,2395.227)(595.214,2417.127)
        (525.754,2435.648)(455.470,2450.746)(384.531,2462.385)
        (313.107,2470.535)(241.370,2475.179)(169.491,2476.305)
        (97.644,2473.910)(26.000,2468.000)
\dottedline{45}(3738.000,3465.000)(3682.404,3418.965)(3628.652,3370.790)
        (3576.826,3320.549)(3527.005,3268.319)(3479.265,3214.180)
        (3433.681,3158.214)(3390.320,3100.508)(3349.251,3041.149)
        (3310.536,2980.229)(3274.233,2917.841)(3240.400,2854.081)
        (3209.086,2789.046)(3180.341,2722.835)(3154.209,2655.550)
        (3130.729,2587.295)(3109.938,2518.173)(3091.867,2448.291)
        (3076.543,2377.755)(3063.991,2306.673)(3054.230,2235.155)
        (3047.274,2163.310)(3043.135,2091.248)(3041.818,2019.078)
        (3043.325,1946.913)(3047.655,1874.862)(3054.801,1803.035)
        (3064.751,1731.543)(3077.490,1660.495)(3093.000,1590.000)
\dottedline{45}(3573.000,2040.000)(3575.843,1958.523)(3584.323,1877.438)
        (3598.397,1797.135)(3618.000,1718.000)
\dottedline{45}(3768.000,2903.000)(3723.494,2846.140)(3682.283,2786.848)
        (3644.500,2725.316)(3610.265,2661.740)(3579.689,2596.327)
        (3552.870,2529.285)(3529.895,2460.831)(3510.837,2391.184)
        (3495.757,2320.570)(3484.705,2249.213)(3477.715,2177.346)
        (3474.810,2105.197)(3476.000,2033.000)
\dottedline{45}(1338.000,5123.000)(1386.900,5065.802)(1440.900,5013.392)
        (1499.534,4966.224)(1562.296,4924.705)(1628.642,4889.195)
        (1698.000,4860.000)
\dottedline{45}(2486.000,4988.000)(2414.423,4924.011)(2306.000,4875.000)
\dottedline{45}(1946.000,4920.000)(2024.684,4918.199)(2102.944,4926.553)
        (2179.474,4944.922)(2253.000,4973.000)
\dottedline{45}(1541.000,4500.000)(1610.660,4473.432)(1681.883,4451.393)
        (1754.375,4433.975)(1827.836,4421.249)(1901.963,4413.269)
        (1976.449,4410.066)(2050.987,4411.655)(2125.269,4418.029)
        (2198.988,4429.161)(2271.840,4445.005)(2343.523,4465.497)
        (2413.743,4490.551)(2482.208,4520.063)(2548.635,4553.913)
        (2612.751,4591.960)(2674.291,4634.047)(2733.000,4680.000)
\dottedline{45}(993.000,4763.000)(1042.631,4707.738)(1095.362,4655.427)
        (1151.019,4606.240)(1209.417,4560.340)(1270.362,4517.881)
        (1333.652,4479.002)(1399.076,4443.834)(1466.417,4412.493)
        (1535.451,4385.082)(1605.950,4361.694)(1677.679,4342.406)
        (1750.400,4327.281)(1823.871,4316.370)(1897.849,4309.709)
        (1972.088,4307.321)(2046.341,4309.213)(2120.362,4315.379)
        (2193.904,4325.798)(2266.725,4340.436)(2338.581,4359.245)
        (2409.235,4382.161)(2478.451,4409.109)(2546.000,4440.000)
\dottedline{45}(386.000,1553.000)(300.538,1568.733)(213.879,1575.176)
        (127.030,1572.253)(41.000,1560.000)
\dottedline{45}(791.000,1043.000)(771.728,1113.002)(744.017,1180.113)
        (708.287,1243.320)(665.077,1301.669)(615.038,1354.280)
        (558.927,1400.359)(497.588,1439.209)(431.949,1470.246)
        (363.000,1493.000)
\dottedline{45}(2201,3600)(1669,3998)
\dottedline{45}(176,900)(1976,5625)
\dottedline{45}(176,900)(4901,2025)
\dottedline{45}(1976,5625)(4901,2025)
\put(176,900){\shade\ellipse{100}{100}}
\put(2201,3600){\shade\ellipse{100}{100}}
\put(4901,2025){\shade\ellipse{100}{100}}
\put(1976,5625){\shade\ellipse{100}{100}}
\put(8276,900){\whiten\ellipse{100}{100}}
\end{picture}
\end{center}
\caption{A reciprocal $(4,6)$ configuration}
\label{pireciprocal}
\end{figure}

It is evident that, by construction,  $(4,6)$ configurations are images 
under inversive transformations of four points linked by six straight line 
segments. 
As a consequence, the angles in any of the four triangles made by three 
circular arcs add up to $\pi$. If the circular arcs of two reciprocal
$(4,6)$ configurations degenerate to straight line segments then the usual
definition of reciprocal triangles is retrieved. The generalization of
Corollary \ref{cor1} is now the following:

\begin{theorem} 
\label{T2}
{\bf (Inversive geometry of the 8-point relation).}
Two quadruplets $(\Phi_1,\Phi_2,\Phi_3,\Phi_{123})$ and 
$(\Phi_{23},\Phi_{13},\Phi_{12},\Phi)$ of complex numbers 
may be regarded as the vertices of two
reciprocal $(4,6)$ configurations if and only if the
8-point relation
\bela{E10}
  Q(\Phi_1,\Phi_2,\Phi_3,\Phi_{123}) = Q(\Phi_{23},\Phi_{13},\Phi_{12},\Phi)
\ela
holds.
\end{theorem}

\noindent
{\bf Proof.} On the one hand, consider two reciprocal $(4,6)$ configurations 
with vertices
$\Phi_1,\Phi_2,\Phi_3,\Phi_{123}$ and $\Phi_{23},\Phi_{13},\Phi_{12},\Phi$
respectively. If we apply two local M\"obius transformations which map the
two points of intersection $\Phi_*$ and $\Phi_*'$ to infinity (cf.\ Figures
\ref{foursix} and \ref{pireciprocal}) then two reciprocal triangles are 
obtained. Since the cross-ratios associated with the two $(4,6)$ configurations
are preserved by local M\"obius transformations, Corollary \ref{cor1} implies 
that the 8-point relation (\ref{E10}) is satisfied. 

On the other hand, 
let the 8-point relation be satisfied and choose an arbitrary
point $\Phi_*$. Then, by drawing circles through $\Phi_*$ and any two
of the points $\Phi_{23},\Phi_{13},\Phi_{12},\Phi$, a $(4,6)$ configuration
is obtained. The circles passing through the point $\Phi$ meet at some angles
$\alpha',\beta'$ and $\gamma'$ as indicated in Figure \ref{foursix}. Here, we
regard angles as oriented quantities. Another
$(4,6)$ configuration is now constructed in the following way: If 
$\lambda,\mu,\nu$ denote the angles made by the line segments 
$(\Phi_1,\Phi_2)$, $(\Phi_2,\Phi_3)$, $(\Phi_3,\Phi_1)$ then we may draw 
three circles which pass through the pairs
$\{\Phi_2,\Phi_3\}$, $\{\Phi_3,\Phi_1\}$, $\{\Phi_1,\Phi_2\}$ and meet these 
line segments 
at angles $\hat{\alpha}=\alpha'-\lambda$, $\hat{\beta}=\beta'-\mu$, 
$\hat{\gamma} =\gamma'-\nu$ as shown in Figure \ref{pireciprocal}. 
By construction, these circles meet at angles $\alpha',\beta',\gamma'$. 
Since $\alpha'+\beta'+\gamma'=\pi$, the three circles intersect at a point, 
$\Phi_*'$, say. Finally, another three circles $S_1,S_2,S_3$ are defined by 
the requirement
that they pass through the pairs $\{\Phi_1,\Phi_*'\}$, $\{\Phi_2,\Phi_*'\}$,  
$\{\Phi_3,\Phi_*'\}$ and meet the other three circles at angles 
$\tilde{\alpha}$, $\tilde{\beta}$, $\tilde{\gamma}$ which are defined in
Figure~\ref{foursix}. We now apply two local M\"obius transformations
which map the points $\Phi_*$ and $\Phi_*'$ to infinity and the twelve
circles to twelve straight lines. The existence theorem of reciprocal 
triangles then implies
that the images of the circles $S_1,S_2,S_3$ are concurrent and hence the
circles $S_1,S_2,S_3$ intersect at a point $\Phi_{\circ\circ\circ}$, say.
Thus, the quadruplets $(\Phi_1,\Phi_2,\Phi_3,\Phi_{\circ\circ\circ})$ and 
$(\Phi_{23},\Phi_{13},\Phi_{12},\Phi)$ of points constitute the vertices of
two reciprocal $(4,6)$ configurations and hence
\bela{E11}
  Q(\Phi_1,\Phi_2,\Phi_3,\Phi_{\circ\circ\circ}) = 
  Q(\Phi_{23},\Phi_{13},\Phi_{12},\Phi).
\ela
However, since, by assumption, the 8-point relation (\ref{E10}) is likewise 
satisfied, it is concluded that $\Phi_{\circ\circ\circ}=\Phi_{123}$. This 
completes the proof.
\medskip 

Another algebraic description of reciprocal $(4,6)$ configurations is obtained
by introducing {\em complex} dilations $a,b,c$ and $a_1,b_2,c_3$ according to 
\bela{E12}
  \bear{rl}
   \Phi_{12} - \Phi = & c(\Phi_1 - \Phi_2)\as
   \Phi_{23} - \Phi = & a(\Phi_2 - \Phi_3)\as
   \Phi_{13} - \Phi = & b(\Phi_3 - \Phi_1)
  \ear
\ela
and 
\bela{E13}
  \bear{rl}
   \Phi_{123} - \Phi_3 = & c_3(\Phi_{13} - \Phi_{23})\as
   \Phi_{123} - \Phi_1 = & a_1(\Phi_{12} - \Phi_{13})\as
   \Phi_{123} - \Phi_2 = & b_2(\Phi_{23} - \Phi_{12}).
  \ear
\ela
As in the case of reciprocal triangles, the latter system implies that
\bela{E14}
  \bear{rl}
   (1+c_3b+a_1b+c_3a)(\Phi_1-\Phi_3) = & (c_3a-a_1c)(\Phi_1-\Phi_2)\as
   (1+a_1c+b_2c+a_1b)(\Phi_2-\Phi_1) = & (a_1b-b_2a)(\Phi_2-\Phi_3).
  \ear
\ela
However, the 8-point relation (\ref{E10}) is equivalent to any of the
relations $a_1b=b_2a$, $b_2c=c_3b$ or $c_3a=a_1c$. Consequently, the
right-hand sides of (\ref{E14}) vanish and the left-hand sides provide
another two constraints on the complex dilations. Thus, we obtain the
following generalization of Theorem \ref{T1}:

\begin{theorem} {\bf\boldmath (Algebraic description of reciprocal $(4,6)$
configurations).} 
\label{T3}
The complex dilations $a,b,c$ and $a_1,b_2,c_3$ as defined by (\ref{E12}) and
(\ref{E13}) associated with two reciprocal $(4,6)$ configurations
are related by 
\bela{E15}
  a_1 = -\frac{a}{ab+bc+ca},\quad b_2 = -\frac{b}{ab+bc+ca},\quad
  c_3 = -\frac{c}{ab+bc+ca}.
\ela
Conversely, let $\Phi_1,\Phi_2,\Phi_3$ and $\Phi$ be four generic points on the
complex plane, $a,b,c\in\mathbb{C}$ be arbitrary non-vanishing complex 
numbers and $a_1,b_2,c_3$ be given by~(\ref{E15}). Then, the unique solution
of the compatible linear system 
(\ref{E12}), (\ref{E13}) gives rise to reciprocal $(4,6)$ configurations
with vertices $\Phi_1,\Phi_2,\Phi_3,\Phi_{123}$ and 
$\Phi_{23},\Phi_{13},\Phi_{12},\Phi$ respectively.
\end{theorem} 

Theorems \ref{T1} and \ref{T3} are formulated in such a way that a
remarkable connection with soliton theory is readily established. 
Indeed, it is shown below that
the nonlinear system (\ref{E15}) may be identified with a well-known and
distinct discrete integrable equation with (\ref{E12}) being its standard 
linear representation. A similar geometric link with the Schwarzian 
Kadomtsev-Petviashvili hierarchy provided by the ancient Greek Theorem of
Menelaus has been recorded earlier in \cite{KonSch01}.

\section{The discrete BKP equation}
\setcounter{equation}{0}

In this section, we are concerned with (pairs) of lattices which consist
of an infinite number of reciprocal triangles or $(4,6)$ configurations.
In order to show the existence of such lattices, it is convenient to 
introduce a canonical notation associated with three-dimensional
lattices on the complex plane of $\mathbb{Z}^3$ combinatorics, that is maps of
the form
\bela{E16}
  \Phi : \mathbb{Z}^3\rightarrow \mathbb{C},\quad 
  (n_1,n_2,n_3)\mapsto \Phi(n_2,n_2,n_3).
\ela
Thus, we supress the arguments of $\Phi$ and indicate unit increments
of the discrete variables or, equivalently, shifts on the lattice along the
edges by indices:
\bela{E17}
  \Phi = \Phi(n_1,n_2,n_3),\quad \Phi_1 = \Phi(n_1+1,n_2,n_3),\quad
  \Phi_{12} = \Phi(n_1+1,n_2+1,n_3)\ldots.
\ela
Since reciprocal triangles constitute projections of tetrahedral
polyhedra, it is natural to consider lattices on the complex plane which 
represent images of three-dimensional lattices consisting of tetrahedra. A 
canonical construction of the latter is obtained by starting at a vertex of
the $\mathbb{Z}^3$ cubic lattice and drawing diagonals across the faces of the 
cubes. The six diagonals on each cube then form a tetrahedron as shown in
Figure \ref{cell}, while the polyhedron inscribed in eight adjacent cubes is 
readily seen to be the stellated octahedron (stella octangula), that is an 
octahedron which is
enclosed by eight tetrahedra (cf.~Figure~\ref{stella1}). \mbox{Accordingly},
\setlength{\unitlength}{0.00055in}
\begin{figure}
\begin{center}
\begin{picture}(5316,2146)(0,-10)
\path(180,1178)(180,53)(1305,53)
        (1305,1178)(180,1178)(855,2078)
        (1980,2078)(1980,953)(1305,53)
\path(1305,1178)(1980,2078)
\path(3105,1178)(3105,53)(4230,53)
        (4230,1178)(3105,1178)(3780,2078)
        (4905,2078)(4905,953)(4230,53)
\path(4230,1178)(4905,2078)
\dottedline{45}(180,53)(855,953)(1980,953)
\dottedline{45}(855,953)(855,2078)
\dottedline{45}(3105,53)(3780,953)(4905,953)
\dottedline{45}(3780,953)(3780,2078)
\thicklines
\path(1305,53)(180,1178)(855,953)
        (1305,53)(1980,2078)(180,1178)
\path(855,953)(1980,2078)
\path(3105,53)(4905,953)(3780,2078)
        (3105,53)(4230,1178)(3780,2078)
\path(4230,1178)(4905,953)
\thinlines
\put(855,2078){\blacken\ellipse{100}{100}}
\put(1305,1178){\blacken\ellipse{100}{100}}
\put(180,53){\blacken\ellipse{100}{100}}
\put(1980,953){\blacken\ellipse{100}{100}}
\put(3105,53){\blacken\ellipse{100}{100}}
\put(4905,953){\blacken\ellipse{100}{100}}
\put(3780,2078){\blacken\ellipse{100}{100}}
\put(4230,1178){\blacken\ellipse{100}{100}}
\put(3105,1178){\shade\ellipse{100}{100}}
\put(4230,53){\shade\ellipse{100}{100}}
\put(4905,2078){\shade\ellipse{100}{100}}
\put(3780,953){\shade\ellipse{100}{100}}
\put(855,953){\shade\ellipse{100}{100}}
\put(1305,53){\shade\ellipse{100}{100}}
\put(1980,2078){\shade\ellipse{100}{100}}
\put(180,1178){\shade\ellipse{100}{100}}
\put(0,8){$\scriptstyle0$}
\put(1400,8){$\scriptstyle1$}
\put(960,803){$\scriptstyle2$}
\put(0,1110){$\scriptstyle3$}
\put(0,2000){\scriptsize (a)}
\put(2065,863){$\scriptstyle12$}
\put(2060,2010){$\scriptstyle123$}
\put(565,2010){$\scriptstyle23$}
\put(1385,1110){$\scriptstyle13$}
\put(2925,8){$\scriptstyle0$}
\put(5000,863){$\scriptstyle12$}
\put(5000,2010){$\scriptstyle123$}
\put(3490,2010){$\scriptstyle23$}
\put(2925,1110){$\scriptstyle3$}
\put(2925,2000){\scriptsize (b)}
\put(3600,903){$\scriptstyle2$}
\put(4345,1153){$\scriptstyle13$}
\put(4325,8){$\scriptstyle1$}
\end{picture}
\end{center}
\caption{The elementary cells of $G^{(1)}$ and $G^{(0)}$}
\label{cell}
\end{figure}
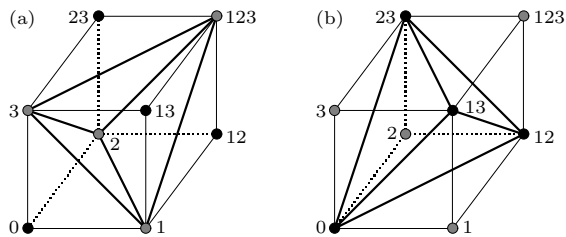
\begin{figure}
 \begin{minipage}[t]{0.49\textwidth}
 \centerline{\includegraphics[angle=270,trim= 50 50 50 50,width=\textwidth,
                                               clip]{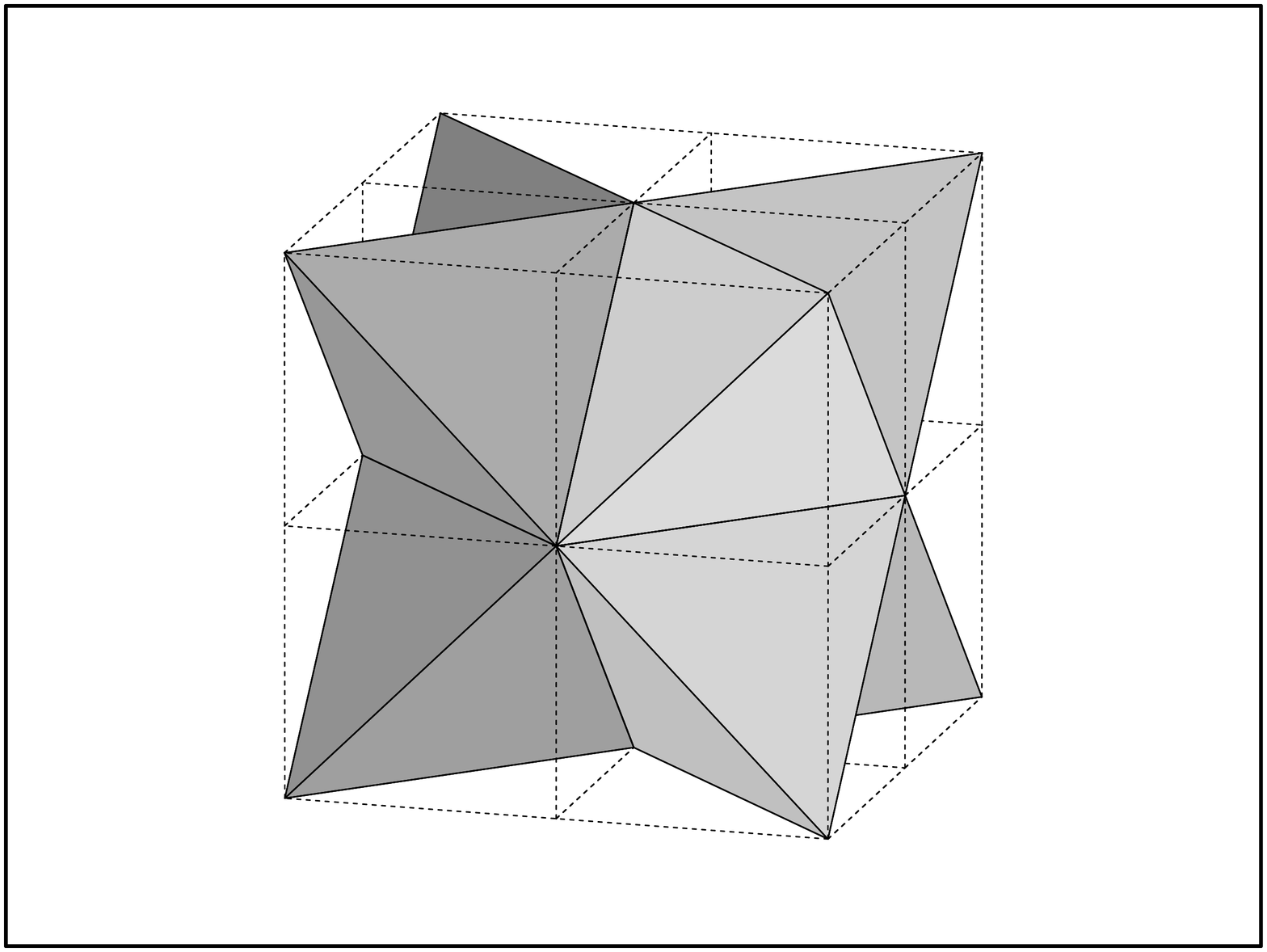}}
 \caption{The stella octangula}
 \label{stella1}
 \end{minipage}
 \begin{minipage}[t]{0.49\textwidth}
 \centerline{\includegraphics[angle=270,trim= 50 50 50 50,width=\textwidth,
                                   clip]{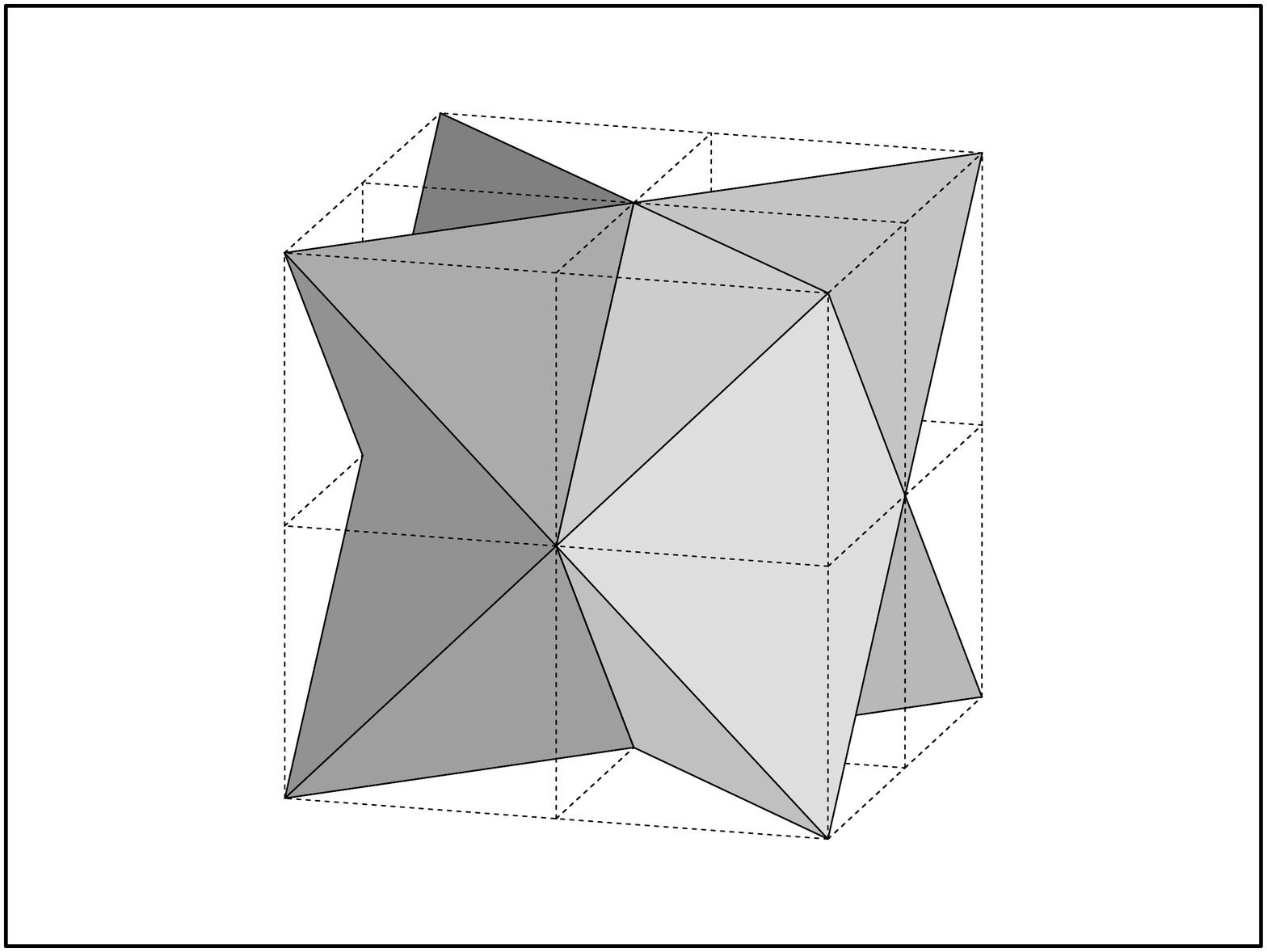}}
 \caption{A modified stella octangula}
 \label{stella2}
 \end{minipage} 
\end{figure}
the complete lattice is obtained by stacking stella octangulae.
This lattice may be interpreted as a face-centred cubic (fcc) lattice or a 
face-centred cubic sphere packing \cite{solid} with the edges linking the
centres of the spheres. Hence, we here consider lattices on the complex plane 
of the form
\bela{E17a}
 \bear{rl}
  \Phi^{(0)} : G^{(0)} &\rightarrow \mathbb{C}\as
  G^{(0)} &= \{(n_1,n_2,n_3)\in\mathbb{Z}^3: n_1+n_2+n_3\, \mbox{ even}\}.
 \ear
\ela
It is our aim to show that two lattices of this type which consist of 
reciprocal triangles or $(4,6)$ configurations are integrable.
To this end, the second lattice of the type (\ref{E17a}) is labelled according
to (cf.\ Figure \ref{cell}(a))
\bela{E17b}
  \bear{rl}
  \Phi^{(1)} : G^{(1)} &\rightarrow \mathbb{C}\as
  G^{(1)} &= \{(n_1,n_2,n_3)\in\mathbb{Z}^3: n_1+n_2+n_3\, \mbox{ odd}\}.
 \ear
\ela
Since the set $G^{(1)}$ is the complement of $G^{(0)}$ with respect to
$\mathbb{Z}^3$, we are now in a position to combine the lattices
$\Phi^{(0)}$ and $\Phi^{(1)}$ to one lattice $\Phi$:
\bela{E17c}
  \Phi(\n) = \Phi^{(i)}(\n)\,\mbox{ if }\,\n\in G^{(i)}.
\ela
This is merely done for reasons of book-keeping. It is interesting to note
that the structure of these interpenetrating fcc lattices is precisely that
of the sodium chloride crystal, where the Na$^+$ ions form one fcc lattice
and the second fcc lattice consists of Cl$^-$ ions. 

It is now evident that the lattices
$\Phi^{(0)}$ and $\Phi^{(1)}$ are composed of reciprocal triangles or
$(4,6)$ configurations if the lattice~$\Phi$ obeys the linear lattice
equations
\bela{E18}
  \bear{rl}
   \Phi_{12} - \Phi = & c(\Phi_1 - \Phi_2)\as
   \Phi_{23} - \Phi = & a(\Phi_2 - \Phi_3)\as
   \Phi_{13} - \Phi = & b(\Phi_3 - \Phi_1),
  \ear
\ela
where $a,b,c$ are as yet unspecified dilation functions. For such lattices to
exist, the compatibility conditions 
$(\Phi_{12})_3=(\Phi_{23})_1=(\Phi_{13})_2$ need to be satisfied. These 
encapsulate the fact that the construction of the vertex $\Phi_{123}$
in the three different ways
\bela{E19}
  \bear{rl}
   \Phi_{123} - \Phi_3 = & c_3(\Phi_{13} - \Phi_{23})\as
   \Phi_{123} - \Phi_1 = & a_1(\Phi_{12} - \Phi_{13})\as
   \Phi_{123} - \Phi_2 = & b_2(\Phi_{23} - \Phi_{12})
  \ear
\ela
must lead to the same result. Since the systems (\ref{E13}) and (\ref{E19})
are identical, the compatibility conditions coincide with
those associated with reciprocal triangles or $(4,6)$ configurations, that is
\bela{E20}
   E^1(\Phi_1-\Phi_3) = E^2(\Phi_1-\Phi_2),\quad
   E^3(\Phi_2-\Phi_3) = E^4(\Phi_2-\Phi_3)
\ela
with the coefficients $E^i$ given by
\bela{E21}
  \bear{rlrl}
   E^1 = & 1+c_3b+a_1b+c_3a, \quad & E^2 = & c_3a-a_1c\as
   E^3 = & 1+a_1c+b_2c+a_1b, \quad & E^4 = & a_1b-b_2a.
  \ear
\ela
Moreover, the results of Section 3 imply that
the coefficients $E^i$ vanish if and only if the dilation functions
satisfy the algebraic system (\ref{E15}) now regarded as lattice equations.
Any real or complex solution of the latter therefore defines
lattices composed of reciprocal triangles or $(4,6)$ configurations 
respectively.

The fact that the {\em nonlinear} system of difference equations (\ref{E15}) 
is obtainable from the compatibility conditions for the {\em linear} 
system (\ref{E18}) suggests that there exists a connection with soliton theory.
Indeed, the relations \mbox{$a_1b=b_2a$}, 
\mbox{$b_2c=c_3b$} and $c_3a=a_1c$ guarantee that
there exists a potential $\tau$ which para\-metrizes the dilations $a,b,c$
according to
\bela{E22}
  a = \frac{\tau_2\tau_3}{\tau\tau_{23}},\quad
  b = \frac{\tau_1\tau_3}{\tau\tau_{13}},\quad
  c = \frac{\tau_1\tau_2}{\tau\tau_{12}}.
\ela
The system (\ref{E15}) then reduces to an integrable single equation known as 
the discrete BKP (dBKP) equation \cite{MIWA} and (\ref{E18}) is nothing but its
standard linear representation (see, e.g., \cite{NimSch97}). Thus, the 
following theorem obtains:

\begin{theorem} {\bf (BKP lattices).} Three-dimensional lattices $\Phi$
on the complex plane with constituent sublattices $\Phi^{(0)}$ and 
$\Phi^{(1)}$ of fcc combinatorics
consist of reciprocal $(4,6)$ configurations (triangles) 
with corresponding quadruplets of vertices
$(\Phi_{23},\Phi_{13},\Phi_{12},\Phi)$ and
$(\Phi_1,\Phi_2,\Phi_3,\Phi_{123})$ if
and only if the complex (real) dilations defined by
\bela{E23}
  \bear{rl}
   \Phi_{12} - \Phi = &\dis
   \frac{\tau_1\tau_2}{\tau\tau_{12}}(\Phi_1 - \Phi_2)\AS
   \Phi_{23} - \Phi = &\dis
   \frac{\tau_2\tau_3}{\tau\tau_{23}}(\Phi_2 - \Phi_3)\AS
   \Phi_{13} - \Phi = &\dis
   \frac{\tau_1\tau_3}{\tau\tau_{13}}(\Phi_3 - \Phi_1)
  \ear
\ela
may be parametrized in terms of solutions of the discrete BKP equation
\bela{E24}
  \tau\tau_{123} + \tau_1\tau_{23} + \tau_2\tau_{13} +\tau_3\tau_{12} = 0.
\ela
Lattices of this type are equivalently described by the 8-point
lattice equation
\bela{E25}
  Q(\Phi_1,\Phi_2,\Phi_3,\Phi_{123}) = Q(\Phi_{23},\Phi_{13},\Phi_{12},\Phi).
\ela  
\end{theorem}

The celebrated Kadomtsev-Petviashvili (KP) hierarchy of integrable equations 
is often referred to as the AKP hierarchy due to its deep connection
with the Lie algebra $A_{\infty}=gl(\infty)$ \cite{JimMiw83}. 
The hierarchy associated with
the Lie algebra $B_{\infty}=so(\infty)$ is known as the BKP hierarchy.
The dBKP equation encodes the complete BKP hierarchy of soliton equations 
\cite{MIWA}. Indeed, by applying appropriate
continuum limits to the dBKP equation, any member of the BKP family may be 
obtained. Proto-typical examples are the 2+1-dimensional 
Sawada-Kotera and Nizhnik-Veselov-Novikov equations \cite{JimMiw83}.

It was
pointed out in \cite{NimSch97} 
that there exists a `natural' continuum limit in 
which the variables $n_i$ are regarded as direct discretizations of some 
continuous variables which leads to the integrable 2+1-dimensional
generalization of the classical sine-Gordon equation set down in 
\cite{KonRog91}.
In fact, the dBKP equation was shown to represent a superposition
principle for eight solutions of the \mbox{2+1-dimensional} sine-Gordon system
generated by the classical Moutard transformation \cite{Mou78}. In subsequent
work \cite{NimSch98}, 
the superposition principle for the associated eigenfunctions 
was set down but it did not occur to the authors that it may be cast into the
form of the 8-point relation (\ref{E25}). By construction, the latter is
equivalent to the dSKP equation. In analogy with the discrete (Schwarzian)
KP equation \cite{Wei83}-\cite{BogKon98b}, it may be termed discrete 
Schwarzian BKP (dSBKP) equation since it is invariant under M\"obius
transformations and constitutes a compact form of the hierarchy of singular
manifold equations associated with the BKP hierarchy.

In summary, it has been shown that, in the setting of inversive geometry, 
reciprocal triangles and, more generally, reciprocal $(4,6)$ configurations 
encapsulate the integrable
dBKP and dSBKP equations and hence the associated semi-discrete and 
continuous hierarchies of integrable equations. It is therefore natural to
refer to the lattices composed of these figures as BKP lattices. BKP lattices
consisting of reciprocal triangles correspond to real solutions of the
BKP equation. Purely imaginary solutions of the BKP equation in the 
form~(\ref{E15}) may be associated with lattices that contain reciprocal 
triangles whose edges meet at right angles.
We observe in passing that BKP lattices consisting of reciprocal triangles
may also be defined in ambient spaces of arbitrary dimension. Indeed, if we 
regard $\Phi$ as a real vector-valued function in $\mathbb{R}^n$ and the
function~$\tau$ constitutes a real solution of the BKP equation then the linear
system~(\ref{E23}) remains compatible and the corresponding
lattice is composed of reciprocal
triangles. Furthermore, since both the vertices 
$\Phi_1,\Phi_2,\Phi_3,\Phi_{123}$ and the vertices 
$\Phi_{23},\Phi_{13},\Phi_{12},\Phi$ are coplanar, the quadrilaterals
$(\Phi,\Phi_i,\Phi_k,\Phi_{ik}),\,i<k$ are planar (but non-embedded). 
Thus, lattices in 
$\mathbb{R}^n$ consisting of reciprocal triangles constitute particular
conjugate lattices \cite{BobSei99}. 
From an algebraic point of view, this is 
evident since (\ref{E23}) represents three coupled discrete Darboux 
equations~\cite{BogKon95,Dol97}.

\section{Particular BKP lattices. Sine-Gordon and \qquad \mbox{ }
Tzitz\'eica lattices}
\setcounter{equation}{0}

There exists a great variety of reductions of both the dBKP and dSBKP equations
each of which may now be analysed with respect to its geometric 
significance. Here, we embark on such a geometric analysis and discuss two 
canonical reductions leading to particular BKP lattices. Firstly, since the
\mbox{dSBKP} equation is invariant under inversion, it is natural to focus on 
lattices
which constitute fixed points of inversive transformations. For instance, if 
we consider the reduction $\Phi = 1/\bar{\Phi}$ 
then the parametrization
\bela{E26}
  \Phi = e^{2\ii\omega},
\ela
where $\omega$ denotes a real function, reduces the dSBKP equation to
\bela{E27}
  \frac{\sin(\omega_1-\omega_2)\sin(\omega_3-\omega_{123})}{
        \sin(\omega_2-\omega_3)\sin(\omega_{123}-\omega_1)} = 
  \frac{\sin(\omega_{23}-\omega_{13})\sin(\omega_{12}-\omega)}{
        \sin(\omega_{13}-\omega_{12})\sin(\omega-\omega_{23})}.
\ela
As shown in \cite{NimSch98}, 
this equation constitutes an integrable discretization 
of the 2+1-dimensional sine-Gordon equation 
\bela{E27a}
  \omega_{xyz} = \omega_x\omega_{yz}\cot\omega
  - \omega_y\omega_{xz}\tan\omega
\ela
which was first set down by Darboux 
\cite{Dar10} 
in connection with triply orthogonal systems of surfaces and later 
rediscovered in~\cite{KonRog91,KonSchRog92} 
as a canonical reduction of the 2+1-dimensional 
sine-Gordon system alluded to in the preceding section. It has also been
derived in the context of three-dimensional discrete `curvature' lattices
on the plane \cite{KonSch98}.

Secondly, it is natural to consider BKP lattices which are mapped to itself
by discrete rotations, scalings or both. For instance, it may be assumed 
that $k$ shifts along the edges in the $n_3$-direction amount to a combination
of a scaling and a rotation of the lattice, viz
\bela{E28}
  \Phi(n_3+k) = c\Phi(n_3),\quad c\in\mathbb{C},\,k\in\mathbb{N}.
\ela
This corresponds to $k$-periodic reductions of the dBKP equation. It turns out
that the dBKP equation may be solved explicitly if $k=1$ and reduces to
a linear equation if the period is $k=2$. The first nontrivial case is 
given by $k=3$, that is
\bela{E29}
   \Phi_{333}=c\Phi,\quad \tau_{333}=\tau.
\ela
Thus, if we introduce the quantities
\bela{E30}
  \rho = \tau_3,\quad \sigma=\tau_{33}
\ela
then the dBKP equation (\ref{E24}) turns into the coupled system of three
two-dimensional equations 
\bela{E31}
  \bear{rl}
    \tau\rho_{12} + \tau_1\rho_2 + \tau_2\rho_1 +\rho\tau_{12} = &0\as
    \rho\sigma_{12} + \rho_1\sigma_2 + \rho_2\sigma_1 +\sigma\rho_{12} = &0\as
    \sigma\tau_{12} + \sigma_1\tau_2 + \sigma_2\tau_1 +\tau\sigma_{12} = &0.
  \ear
\ela
Inspection of this system shows that it is convenient to define quantities
$\phi,\psi$ and $H$ according to
\bela{E32}
  \rho = \tau\phi,\quad \sigma = \tau\psi,\quad 
  H = -\frac{\tau_1\tau_2}{\tau\tau_{12}}
\ela
so that we obtain the two {\em linear} equations
\bela{E33}
  \phi_{12}+\phi = H(\phi_1+\phi_2),\quad \psi_{12}+\psi = H(\psi_1 + \psi_2)
\ela
subject to the {\em quadratic} constraint
\bela{E34}
  H[\psi(\phi_1+\phi_2)+(\psi_1+\psi_2)\phi-\psi_2\phi_1-\psi_1\phi_2] = 
  2\psi\phi.
\ela
If we shift this constraint in the $n_1$- and $n_2$-directions then we
obtain two equations which may be separated into four {\em linear}
equations by introducing functions of separation $A$ and $B$. These read 
\bela{E35}
  \bear{rl}
    \phi_{11} - \phi_1 = &\dis\frac{H_1-1}{H_1(H-1)}(\phi_1-\phi)
    + \frac{A}{H-1}(\phi_{12}-\phi_1)\AS
    \phi_{22} - \phi_2 = &\dis\frac{H_2-1}{H_2(H-1)}(\phi_2-\phi)
    + \frac{B}{H-1}(\phi_{12}-\phi_2)\AS    
    \psi_{11} - \psi_1 = &\dis\frac{H_1-1}{H_1(H-1)}(\psi_1-\psi)
    - \frac{A}{H-1}(\psi_{12}-\psi_1)\AS
    \psi_{22} - \psi_2 = &\dis\frac{H_2-1}{H_2(H-1)}(\psi_2-\psi)
    - \frac{B}{H-1}(\psi_{12}-\psi_2).
  \ear
\ela
We observe that the two systems (\ref{E35})$_{1,2}$ and (\ref{E35})$_{3,4}$
differ only by the signs in front of $A$ and $B$. In fact, the two systems
may be regarded as mutually adjoint with (\ref{E34}) being an admissible
constraint. This is reflected by the fact that
the compatibility conditions \mbox{$(\phi_{11})_2=(\phi_{12})_1,\,
(\phi_{22})_1=(\phi_{12})_2$} and \mbox{$(\psi_{11})_2=(\psi_{12})_1,\,
(\psi_{22})_1=(\psi_{12})_2$} produce the same nonlinear difference equations
for $A,B$ and $H$, namely
\bela{E36}
  \bear{c}\dis
    A_2 = \frac{H_1}{H}A,\quad B_1 = \frac{H_2}{H}B\AS\dis
   H_{12} = \frac{H(H-1)}{H^2(H_1+H_2-H_1H_2)-H+ABH_1H_2}.
 \ear
\ela
The latter 
system constitutes an integrable discretization of the gauge-invariant 
form of the classical Tzitz\'eica equation \cite{Tzi07,Tzi10}
\bela{E37}
  (\ln h)_{xy} = h - h^{-2}.  
\ela
It is interesting to note that the discrete Tzitz\'eica 
system has been shown to govern canonical 
discrete analogues of affine spheres which have been defined in a purely
geometric manner in \cite{BobSch99a,BobSch99b}. 

A single equation may be obtained by
parametrizing $A,B$ and $H$ according~to
\bela{E38}
  A = \frac{\tilde{\tau}_1^2}{\tilde{\tau}\tilde{\tau}_{11}},\quad
  H = \frac{\tilde{\tau}_1\tilde{\tau}_2}{\tilde{\tau}\tilde{\tau}_{12}},\quad
  B = \frac{\tilde{\tau}_2^2}{\tilde{\tau}\tilde{\tau}_{22}}
\ela
so that (\ref{E36})$_{1,2}$ are identically satisfied. The remaining relation
then reduces to the discrete Tzitz\'eica equation
\bela{E39}
  \left|\bear{ccc}
   \tilde{\tau}&\tilde{\tau}_1&\tilde{\tau}_{11}\\
   \tilde{\tau}_2&\tilde{\tau}_{12}&\tilde{\tau}_{112}\\
   \tilde{\tau}_{22}&\tilde{\tau}_{122}&\tilde{\tau}_{1122}\ear\right|
  +\tilde{\tau}_{12}^3 = 0.
\ela
Since $\tilde{\tau} = (-1)^{n_1n_2}\tau$ without loss of generality, it is
evident that the quantities $\tilde{\rho} = (-1)^{n_1n_2}\rho$ and
$\tilde{\sigma} = (-1)^{n_1n_2}\sigma$ constitute another two solutions
of the discrete Tzitz\'eica equation. In fact, the relations 
(\ref{E32})$_{1,2}$ imply that these three solutions of the Tzitz\'eica 
equation
are related by the discrete version of the classical Tzitz\'eica transformation
set down in \cite{Sch99,BobSch99a}. 

The above link with the discrete Tzitz\'eica transformation may be further 
investigated by considering the linear system (\ref{E23}) for the lattice
function $\Phi$. Thus, if we introduce the notation
\bela{E40}
  \Phi_3 = \Psi,\quad \Phi_{33} = \Xi
\ela
then we obtain nine equations with coefficients depending on $\tau,\rho$
and $\sigma$. For brevity, we merely state that this system may be decoupled
into
\bela{E41}
 \bear{rl}
  \Phi_{11} - \Phi_1 = &\dis\frac{H_1-1}{H_1(H-1)}(\Phi_1-\Phi)
    - \lambda\frac{A}{H-1}(\Phi_{12}+\Phi_1)\AS
  \Phi_{12} - \Phi = & -H(\Phi_1-\Phi_2)\as  
  \Phi_{22} + \Phi_2 = &-\dis\frac{H_2-1}{H_2(H-1)}(\Phi_2+\Phi)
    - \frac{1}{\lambda}\frac{B}{H-1}(\Phi_{12}-\Phi_2),
 \ear
\ela
where $\lambda = (1-c)/(1+c)$, and 
\bela{E42}
 \bear{rl}
   \Psi = &c\{\dis \Phi + \frac{H}{(\lambda-1)(H-1)\phi}
    [\lambda(\phi_1-\phi)(\Phi_2+\Phi)+(\phi_2-\phi)(\Phi_1-\Phi)]\}\AS
   \Xi = &\dis \Phi + \frac{H}{(\lambda+1)(H-1)\psi}
    [\lambda(\psi_1-\psi)(\Phi_2+\Phi)-(\psi_2-\psi)(\Phi_1-\Phi)].
 \ear
\ela
The substitution $\Phi\rightarrow (-1)^{n_2}\Phi$ shows that (\ref{E41}) is
nothing but another copy of the linear representation (\ref{E33})$_1$,
(\ref{E35})$_{1,2}$ of the discrete Tzitz\'eica system if the invariance
$A \rightarrow \lambda A,\,B\rightarrow B/\lambda$ of (\ref{E36}) is
taken into account. Hence, the system~(\ref{E41}) represents the
standard parameter-dependent linear representation of the discrete
Tzitz\'eica system \cite{Sch99}-\cite{BobSch99b}. 
Furthermore, for symmetry reasons, the quantities 
$\Psi$ and~$\Xi$ obey analogous linear systems associated with the solutions
$\tilde{\rho}$ and $\tilde{\sigma}$ of the discrete Tzitz\'eica 
equation. Indeed,
it turns out that $\Psi$ and~$\Xi$ as given by (\ref{E42}) are discrete
Tzitz\'eica transforms \cite{Sch99,BobSch99a} of $\Phi$. We have therefore
established that BKP 
lattices subject to the symmetry constraint $\Phi_{333}=c\Phi$ may be 
decomposed into two-dimensional `Tzitz\'eica' 
lattices defined by (\ref{E41}) and
their discrete Tzitz\'eica transforms $\Psi$ and $\Xi$.

We conclude this section with the remark that the dBKP equation may also be 
regarded as a fully discrete version of the  $B_{\infty}$ Toda lattice.
Accordingly, the discrete Tzitz\'eica equation constitutes the
period 3 reduction of the fully discrete Toda lattice. This is in
harmony with the fact that the Tzitz\'eica equation not only
appears in the context of affine differential geometry \cite{Bla67} but also 
constitutes the period 3 reduction of the $B_{\infty}$ Toda lattice
\cite{TZITODA}. 

\section{Continuum limits. Quasi-conformal mappings}
\setcounter{equation}{0}

As mentioned earlier, any member of
the (S)BKP hierarchy may be retrieved from the d(S)BKP equation
by application of an appropriate continuum limit.
Furthermore, if we make the substitution (\ref{E26}), where the function
$\omega$ is now regarded to be complex, then the dSBKP equation assumes the
trigonometric form (\ref{E27}). The connection with Darboux's (complex)
sine-Gordon-type equation (\ref{E27a}) is established by relabelling the
lattice according to
\bela{G1}
  \omega \rightarrow \frac{(-1)^{n_3}}{2}\omega + n_2\frac{\pi}{2},
\ela
leading to 
\bela{G1a}
  \frac{\cos[\frac{1}{2}(\omega_1-\omega_2)]
        \cos[\frac{1}{2}(\omega_3-\omega_{123})]}{
        \cos[\frac{1}{2}(\omega_2+\omega_3)]
        \cos[\frac{1}{2}(\omega_{123}+\omega_1)]} = 
  \frac{\cos[\frac{1}{2}(\omega_{23}-\omega_{13})]
        \cos[\frac{1}{2}(\omega_{12}-\omega)]}{
        \cos[\frac{1}{2}(\omega_{13}+\omega_{12})]
        \cos[\frac{1}{2}(\omega+\omega_{23})]},
\ela
and then performing the limit
\bela{G2}
  x = \epsilon n_1,\quad y = \epsilon n_2,\quad z = \epsilon n_3,\quad
  \epsilon\rightarrow0.
\ela
An equation of second order is obtained by considering the natural
continuum limit in which the polygons $\Phi(n_i=\mbox{const},n_k=\mbox{const})$
become coordinate lines on the complex plane. Thus, if we set
\bela{G3}
  \Phi_i = \Phi + \epsilon\Phi_{x_i} + O(\epsilon^2),\quad i=1,2,3,
\ela
where $\epsilon$ denotes a lattice parameter and $\Phi_{x_i} = \del\Phi/\del
x_i$, then, in the limit $\epsilon\rightarrow0$, the dSBKP equation reduces to
\bela{G4}
    \left[\ln\left(\frac{\Phi_y+\Phi_t}{\Phi_y-\Phi_t}\right)\right]_x
  = \left[\ln\left(\frac{\Phi_t+\Phi_x}{\Phi_t-\Phi_x}\right)\right]_y
\ela
with $(x_1,x_2,x_3)=(x,y,t)$, while the linear system (\ref{E18}) becomes
\bela{G5}
  \Phi_y = \frac{c-1}{c+1}\Phi_x,\quad
  \Phi_t = \frac{a-1}{a+1}\Phi_y,\quad
  \Phi_x = \frac{b-1}{b+1}\Phi_t.
\ela
The equation (\ref{G4}) is invariant under transformations
of the form $\Phi\rightarrow F(\Phi)$, where $F$ is an arbitrary differentiable
function, and complex conjugation. Its solutions constitute mappings
$\Phi : \mathbb{R}^3\rightarrow\mathbb{C}$ to which we shall refer as SBKP
mappings. 

In order to investigate the nature of SBKP mappings, it is first noted that
the system (\ref{G5}) gives rise to the consistency condition 
\bela{G6}
  \frac{a-1}{a+1}\frac{b-1}{b+1}\frac{c-1}{c+1} = 1.
\ela
Accordingly, if we set $\gamma = (c-1)/(c+1)$ and $\delta = (b+1)/(b-1)$ then
SBKP mappings are obtained by integrating the linear pair
\bela{G7}
  \Phi_y = \gamma\Phi_x,\quad \Phi_t = \delta\Phi_x,
\ela
where $\gamma,\delta$ are solutions of the coupled nonlinear system
\bela{G8}
  \gamma_t+\gamma\delta_x=\delta_y+\delta\gamma_x,\quad
  (1-\delta^2)\gamma_t=(1-\gamma^2)\delta_y.
\ela
The latter system comprises (\ref{G4}) and the compatibility condition for
(\ref{G5}). In terms of the variables
\bela{G9}
  \mu = \frac{\i-\gamma}{\i+\gamma},\quad z=x+\i y,
\ela
the linear system (\ref{G7}) reads
\bela{G10}
  \Phi_{\bar{z}}=\mu\Phi_z,\quad \Phi_t=\delta(\Phi_z+\Phi_{\bar{z}}).
\ela
Thus, SBKP mappings are descriptive of particular evolutions of quasi-conformal
mappings if the variable $t$ is interpreted as time and the complex 
dilation $\mu$ in the Beltrami equation (\ref{G10})$_1$ is assumed to be
bounded \cite{Ahl66}-\cite{Car47}. 
In other words, SBKP mappings consist of one-parameter
families of quasi-conformal mappings $\Phi(t):\mathbb{C}\rightarrow\mathbb{C}$.
The solutions of the dSBKP equation therefore constitute integrable 
discretizations of particular families of quasi-conformal mappings.

\section{`Integrable' \boldmath reciprocal $(6,10)$ figures, configurations
and associated lattices}
\setcounter{equation}{0}

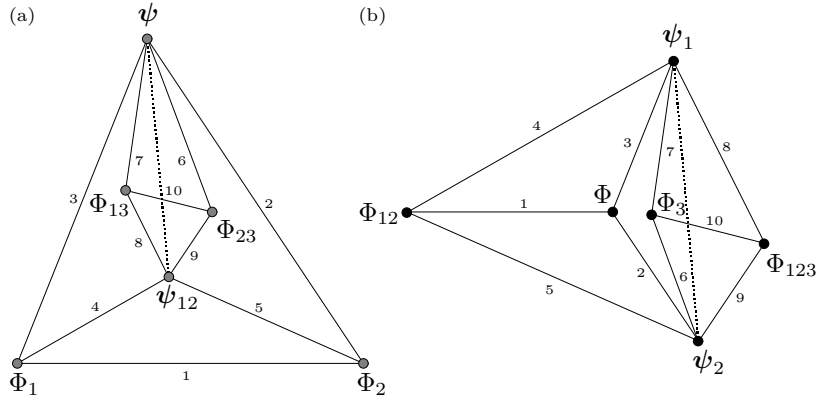
\begin{figure}
\begin{center}
\setlength{\unitlength}{0.00050489in}
\begin{picture}(8259,3885)(0,-10)
\dottedline{45}(7170,500)(6915,3415)
\dottedline{45}(1440,3645)(1665,1170)
\path(1440,3645)(1215,2070)
\path(90,270)(3690,270)
\path(1215,2070)(2115,1845)
\path(1440,3645)(2115,1845)
\path(1215,2070)(1665,1170)
\path(1665,1170)(2115,1845)
\path(1665,1170)(3690,270)
\path(90,270)(1665,1170)
\path(4140,1840)(6915,3420)
\path(4140,1845)(6280,1845)
\path(6910,3425)(6275,1845)
\path(4140,1845)(7180,500)
\path(6280,1840)(7175,500)
\path(90,270)(1440,3645)
\path(1440,3645)(3690,270)
\path(6915,3420)(6685,1815)
\path(6685,1815)(7170,505)
\path(7175,500)(7860,1520)
\path(6910,3420)(7860,1520)
\path(6685,1820)(7855,1520)
\put(1440,3645){\shade\ellipse{100}{100}}
\put(1215,2070){\shade\ellipse{100}{100}}
\put(2115,1845){\shade\ellipse{100}{100}}
\put(90,270){\shade\ellipse{100}{100}}
\put(1665,1170){\shade\ellipse{100}{100}}
\put(3690,270){\shade\ellipse{100}{100}}
\put(4140,1840){\blacken\ellipse{100}{100}}
\put(6280,1845){\blacken\ellipse{100}{100}}
\put(6685,1815){\blacken\ellipse{100}{100}}
\put(7170,500){\blacken\ellipse{100}{100}}
\put(6915,3415){\blacken\ellipse{100}{100}}
\put(7855,1515){\blacken\ellipse{100}{100}}
\put(0,-20){$\Phi_1$}
\put(0,3835){\scriptsize (a)}
\put(3615,-20){$\Phi_2$}
\put(1345,3835){$\Psi$}
\put(835,1845){$\Phi_{13}$}
\put(2160,1575){$\Phi_{23}$}
\put(7110,225){$\Psi_2$}
\put(6795,3610){$\Psi_1$}
\put(1530,900){$\Psi_{12}$}
\put(3635,1755){$\Phi_{12}$}
\put(3635,3835){\scriptsize (b)}
\put(6075,1935){$\Phi$}
\put(6735,1855){$\Phi_3$}
\put(7875,1215){$\Phi_{123}$}
\put(1800,100){$\scriptscriptstyle1$}
\put(2660,1890){$\scriptscriptstyle2$}
\put(630,1935){$\scriptscriptstyle3$}
\put(855,815){$\scriptscriptstyle4$}
\put(2555,810){$\scriptscriptstyle5$}
\put(1755,2340){$\scriptscriptstyle6$}
\put(1310,2340){$\scriptscriptstyle7$}
\put(1620,2025){$\scriptscriptstyle10$}
\put(1305,1485){$\scriptscriptstyle8$}
\put(1885,1350){$\scriptscriptstyle9$}
\put(5310,1890){$\scriptscriptstyle1$}
\put(6525,1170){$\scriptscriptstyle2$}
\put(6390,2520){$\scriptscriptstyle3$}
\put(5445,2695){$\scriptscriptstyle4$}
\put(5580,995){$\scriptscriptstyle5$}
\put(6970,1125){$\scriptscriptstyle6$}
\put(6835,2430){$\scriptscriptstyle7$}
\put(7420,2475){$\scriptscriptstyle8$}
\put(7560,905){$\scriptscriptstyle9$}
\put(7245,1710){$\scriptscriptstyle10$}
\end{picture}
\end{center}
\caption{Reciprocal $(6,10)$ figures}
\label{sixten}
\end{figure}

It turns out that the dBKP equation is but the simplest integrable equation
which may be derived from reciprocal figures. Another example is obtained if
we remove two edges from Maxwell's `octahedral' figure as displayed in
Figure~\ref{indeterminate}(a) in order to achieve determinacy. 
Indeed, Figure \ref{sixten}(a) now consists of six vertices, six polygons 
and ten edges and thus admits a reciprocal figure with the same number of
vertices and edges which is uniquely determined up to an arbitrary scaling.
We shall refer to reciprocal figures of this kind as reciprocal~$(6,10)$
figures. 

\subsection{\boldmath Reciprocal $(6,10)$ figures and configurations}

We first exploit the fact that Maxwell's theorem guarantees the
existence of reciprocal
$(6,10)$ \mbox{figures} and label the vertices by 
$\Psi,\Psi_{12},\Phi_1,\Phi_2,\Phi_{13},
\Phi_{23}$ and $\Psi_1,\Psi_2,\Phi,\Phi_3,\Phi_{12},\Phi_{123}$ respectively.
Here, this particular labelling is of no relevance but will turn out to be
of importance in connection with $(6,10)$ lattices. It
proves convenient to group the conditions of parallelism into two sets of
linear equations, namely
\bela{E43}
 \bear{rlrl}
  \Phi_{12}-\Phi = & a(\Phi_1-\Phi_2),\quad & \Psi_1-\Phi = & d(\Phi_1-\Psi)\as
  \Phi_{13}-\Psi = & b(\Psi_1-\Phi_3),\quad & \Psi_2-\Phi = & e(\Phi_2-\Psi)\as
  \Phi_{23}-\Psi = & c(\Psi_2-\Phi_3)&&
 \ear
\ela
and
\bela{E44}
 \bear{rlrl}
  \Phi_{123}-\Phi_3 = & a_3(\Phi_{13}-\Phi_{23}),
                \quad & \Psi_{12}-\Phi_2 = & d_2(\Phi_{12}-\Psi_2)\as
  \Phi_{123}-\Psi_2 = & b_2(\Psi_{12}-\Phi_{23}),
                \quad & \Psi_{12}-\Phi_1 = & e_1(\Phi_{12}-\Psi_1)\as
  \Phi_{123}-\Psi_1 = & c_1(\Psi_{12}-\Phi_{13})&&
 \ear
\ela
with associated real dilation coefficients 
$a,b,c,d,e$ and $a_3,b_2,c_1,d_2,e_1$. Elimination of $\Psi_{12}$ 
between (\ref{E44})$_{2,4}$ on the one hand
and $\Psi_{12},\Phi_{123}$ between (\ref{E44})$_{1,3,5}$ on the other
hand results in
\bela{E45}
 \bear{rl}
 [d_2a + e_1(d-a) - 1](\Phi_1-\Psi) + [e_1a - d_2(e+a) + 1](\Phi_2-\Psi)=&0\as
 [(c_1a_3-b_2c_1-a_3b_2)b+b_2](\Psi_1-\Phi_3)\qquad\qquad\qquad\qquad
 \qquad\qquad\!&\as 
 \mbox{}+[(b_2a_3-a_3c_1+c_1b_2)c-c_1](\Psi_2-\Phi_3)=&0.
 \ear
\ela
Since we consider the generic case, neither the vertices $\Psi,\Phi_1,\Phi_2$
nor the vertices $\Psi_1,\Psi_2,\Phi_3$ are collinear. Hence, the 
relations (\ref{E45}) may be solved for $b_2,c_1,d_2,e_1$. The remaining
consistency condition $\mbox{(\ref{E44})$_3$} - \mbox{(\ref{E44})$_5$}$ 
then reduces~to
\bela{E46}
  \{a_3[bc(ad+ed-ea)+a(b-c)]-a\}(\Psi_1-\Psi_2) = 0
\ela
so that $a_3$ is likewise determined. 

An algebraic proof of the existence of reciprocal $(6,10)$ figures is now 
readily obtained.
Thus, given a generic $(6,10)$ figure with vertices 
$\Psi,\Psi_{12},\Phi_1,\Phi_2$,
$\Phi_{13},\Phi_{23}$, we draw a line parallel to the edge $(\Phi_1,\Phi_2)$ 
and choose two points $\Phi$ and $\Phi_{12}$ on this line. The vertices 
$\Psi_1$ and $\Psi_2$ are then given by the points of intersection of the
pairs of lines which pass through $\Phi$ and $\Phi_{12}$ and are parallel
to the corresponding edges of the original figure 
(cf.\ Figure \ref{sixten}). The vertex $\Phi_3$ is constructed in a 
similar manner. In algebraic terms,
this implies that the linear equations (\ref{E43}), (\ref{E44})$_{2,4}$ 
are satisfied with some real dilations $a,b,c,d,e$ and $d_2,e_1$. 
The latter two
dilations are determined by (\ref{E45})$_1$. Finally, the three lines 
which pass through $\Psi_1,\Psi_2,\Phi_3$ and are parallel to the
edges $(\Psi_{12},\Phi_{13})$, $(\Psi_{12},\Phi_{23})$, 
$(\Phi_{13},\Phi_{23})$ respectively are
concurrent since the remaining linear equations (\ref{E44})$_{1,3,5}$
are compatible if the coefficients $a_3,b_2,c_1$ are defined by the
compatibility conditions (\ref{E45})$_2$ and (\ref{E46}). Accordingly, the 
reciprocal $(6,10)$ figure exists and is defined up to an arbitrary scaling.
An alternative method of constructing reciprocal $(6,10)$ figures is stated
in the following theorem:

\begin{theorem} {\bf (\boldmath Reciprocal $(6,10)$ figures).} 
\label{T4}
The real
dilations $a,b,c,d,e$ and $a_3,b_2,c_1,d_2,e_1$ as defined by (\ref{E43}) and
(\ref{E44}) associated with two reciprocal $(6,10)$ configurations are
related by
\bela{E47}
  \bear{c}\dis
     a_3  = \frac{a}{bc(ad+ed-ea)+a(b-c)}\AS\dis
     b_2 = \frac{a_3(c-b)+1}{c},\quad
     c_1 = \frac{a_3(c-b)+1}{b}\AS\dis
     d_2 = \frac{d}{ad+ed-ea},\quad e_1 = \frac{e}{ad+ed-ea}.
  \ear
\ela
Conversely, let $\Phi,\Phi_1,\Phi_2,\Phi_3,\Psi$ be five generic points on
the complex plane, $a,b,c,d,e\in\mathbb{R}$ be five arbitrary non-vanishing
real numbers and $a_3,b_2,c_1,d_2,e_1$ be given by (\ref{E47}). Then, the
linear system (\ref{E43}), (\ref{E44}) is compatible and the points
$\Psi,\Psi_{12},\Phi_1,\Phi_2,\Phi_{13},
\Phi_{23}$ and $\Psi_1,\Psi_2,\Phi,\Phi_3,\Phi_{12},\Phi_{123}$ constitute the
vertices of two reciprocal $(6,10)$ figures. These obey the cross-ratio
relations
\bela{E48}
 \bear{rl}
  Q(\Phi_1,\Phi_2,\Psi,\Psi_{12}) = & Q(\Psi_2,\Psi_1,\Phi_{12},\Phi)\as
  Q(\Psi_1,\Psi_2,\Phi_3,\Phi_{123}) = & Q(\Phi_{23},\Phi_{13},\Psi_{12},\Psi).
 \ear
\ela
\end{theorem}

The above cross-ratio relations are a consequence of the identities
$cb_2=bc_1$ and $ed_2=de_1$. They express the fact that the quadruplets
$(\Phi_1,\Phi_2,\Psi,\Psi_{12})$ and $(\Psi_2,\Psi_1,\Phi_{12},\Phi)$ on the
one hand and  $(\Phi_{23},\Phi_{13},\Psi_{12},\Psi)$ and 
$(\Psi_1,\Psi_2,\Phi_3,\Phi_{123})$ on the other hand give rise to two
pairs of reciprocal triangles if the parallel line segments 
$(\Psi,\Psi_{12})$ and $(\Psi_1,\Psi_2)$ are drawn (cf.\ Figure \ref{sixten}).
The occurrence of cross-ratios once again
suggests that reciprocal $(6,10)$ figures should be regarded as degenerate
reciprocal $(6,10)$ configurations. Indeed, if, in a $(6,10)$ figure, we
replace the straight edges by circular arcs whose extensions meet at a point
then two such $(6,10)$ configurations are reciprocally related if 
corresponding angles are equal. As in the case of reciprocal $(4,6)$ 
configurations, it is now straight forward to show that 
reciprocal $(6,10)$ configurations are governed by the cross-ratio
relations (\ref{E48}) and the associated complex dilations satisfy the
algebraic relations (\ref{E47}). 

\subsection{\boldmath Reciprocal $(6,10)$ lattices}

\begin{figure}
\begin{center}
\setlength{\unitlength}{0.00055in}
\begin{picture}(5859,3308)(0,-10)
\put(1415,2320){$\scriptstyle144$}
\path(225,2340)(225,90)(1350,90)
        (1350,2340)(225,2340)(900,3240)
        (2025,3240)(2025,990)(1350,90)
\path(225,1215)(1350,1215)(2025,2115)
\path(1350,2340)(2025,3240)
\path(3600,2340)(3600,90)(4725,90)
        (4725,2340)(3600,2340)(4275,3240)
        (5400,3240)(5400,990)(4725,90)
\path(3600,1215)(4725,1215)(5400,2115)
\path(4725,2340)(5400,3240)
\dottedline{45}(225,90)(900,990)(900,3240)
\dottedline{45}(900,990)(2025,990)
\dottedline{45}(225,1215)(900,2115)(2025,2115)
\dottedline{45}(3600,90)(4275,990)(4275,3240)
\dottedline{45}(4275,990)(5400,990)
\dottedline{45}(3600,1215)(4275,2115)(5400,2115)
\thicklines
\path(225,1215)(1350,90)(900,990)
        (225,1215)(900,3240)(2025,2115)(1350,90)
\path(900,990)(2025,2115)(1350,2340)(900,3240)
\path(1350,2340)(225,1215)
\path(3600,90)(4275,2115)(3600,2340)
        (5400,3240)(4725,1215)(5400,990)(3600,90)
\path(3600,90)(4725,1215)(3600,2340)
\path(5400,3240)(4275,2115)(5400,990)
\thinlines
\put(225,90){\blacken\ellipse{100}{100}}
\put(225,2340){\blacken\ellipse{100}{100}}
\put(900,2115){\blacken\ellipse{100}{100}}
\put(1350,1215){\blacken\ellipse{100}{100}}
\put(2025,990){\blacken\ellipse{100}{100}}
\put(2025,3240){\blacken\ellipse{100}{100}}
\put(3600,90){\blacken\ellipse{100}{100}}
\put(5400,990){\blacken\ellipse{100}{100}}
\put(5400,3240){\blacken\ellipse{100}{100}}
\put(3600,2340){\blacken\ellipse{100}{100}}
\put(4275,2115){\blacken\ellipse{100}{100}}
\put(4725,1215){\blacken\ellipse{100}{100}}
\put(900,3240){\shade\ellipse{100}{100}}
\put(1350,2340){\shade\ellipse{100}{100}}
\put(2025,2115){\shade\ellipse{100}{100}}
\put(225,1215){\shade\ellipse{100}{100}}
\put(900,990){\shade\ellipse{100}{100}}
\put(1350,90){\shade\ellipse{100}{100}}
\put(3600,1215){\shade\ellipse{100}{100}}
\put(4275,3240){\shade\ellipse{100}{100}}
\put(4725,2340){\shade\ellipse{100}{100}}
\put(5400,2115){\shade\ellipse{100}{100}}
\put(4725,90){\shade\ellipse{100}{100}}
\put(4275,990){\shade\ellipse{100}{100}}
\put(45,30){$\scriptstyle0$}
\put(1440,30){$\scriptstyle1$}
\put(1035,835){$\scriptstyle2$}
\put(45,1145){$\scriptstyle4$}
\put(-50,2280){$\scriptstyle44$}
\put(1440,1145){$\scriptstyle14$}
\put(2115,930){$\scriptstyle12$}
\put(2115,2055){$\scriptstyle124$}
\put(2115,3180){$\scriptstyle1244$}
\put(525,3180){$\scriptstyle244$}
\put(620,2055){$\scriptstyle24$}
\put(3420,30){$\scriptstyle0$}
\put(4840,30){$\scriptstyle1$}
\put(5490,930){$\scriptstyle12$}
\put(3420,1145){$\scriptstyle4$}
\put(3335,2280){$\scriptstyle44$}
\put(3910,3180){$\scriptstyle244$}
\put(5490,3180){$\scriptstyle1244$}
\put(5490,2055){$\scriptstyle124$}
\put(3985,2010){$\scriptstyle24$}
\put(4745,2185){$\scriptstyle144$}
\put(4095,940){$\scriptstyle2$}
\put(4840,1180){$\scriptstyle14$}
\end{picture}
\end{center}
\caption{The elementary cells of $G^{(1)}$ and $G^{(0)}$}
\label{cell2}
\end{figure}
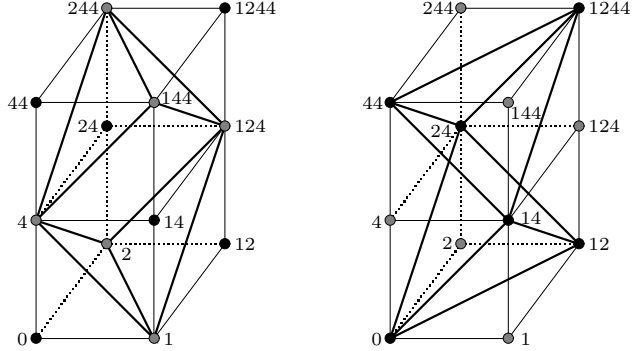

As in the case of reciprocal triangles, it is now possible to define lattices
which consist of reciprocal $(6,10)$ figures and configurations. Thus, we
consider two three-dimensional lattices on the complex plane of the same 
combinatorics and label them as follows. The sets $G^{(0)}$ and $G^{(1)}$
are defined as in Section 5, that~is
\bela{E48a} 
 \bear{rl}
   G^{(0)} & = \{(n_1,n_2,n_4)\in\mathbb{Z}^3: n_1+n_2+n_4\,\mbox{ even}\}\as
   G^{(1)} & = \{(n_1,n_2,n_4)\in\mathbb{Z}^3: n_1+n_2+n_4\,\mbox{ odd}\}.
 \ear
\ela
However, the edge structure of the lattices associated with $G^{(0)}$ and 
$G^{(1)}$ is slightly modified in order to take into account that the
reciprocal figures displayed in Figure \ref{sixten} consist of two pairs of
reciprocal triangles if the dotted lines are included. Accordingly, every
second `horizontal' edge which joins two tetrahedra placed on top of each 
other is removed from the lattices associated with $G^{(0)}$ and 
$G^{(1)}$. This is indicated in Figure \ref{cell2}.
Hence, the polyhedron inscribed in eight cubes of the
$\mathbb{Z}^3$ lattice constitutes a stella
octangula without the central horizontal edges (cf.~Figure \ref{stella2}).
Due to the complementary character of $G^{(0)}$ and~$G^{(1)}$, the associated
lattices $\Phi^{(0)}$ and $\Phi^{(1)}$ may now be combined to a lattice
\bela{E48b}
  \hat{\Phi}(\n) = \Phi^{(i)}(\n)\,\mbox{ if }\,\n\in G^{(i)},\quad 
  \n=(n_1,n_2,n_4).
\ela
In order to simplify the book-keeping which takes care of the missing 
horizontal edges, it is convenient to introduce two 
lattices $\Phi$ and $\Psi$ according to
\bela{E48c}
  \Phi(n_1,n_2,n_3) = \hat{\Phi}(n_1,n_2,2n_3),\quad
  \Psi(n_1,n_2,n_3) = \hat{\Phi}(n_1,n_2,2n_3+1).
\ela

As a consequence of Theorem \ref{T4}, the consistency conditions
(\ref{E45}) and (\ref{E46}) may be interpreted as the compatibility conditions
for the linear difference equations (\ref{E43}) 
which guarantee the existence of two three-dimensional lattices $\Phi$ and 
$\Psi$ on the complex plane which encapsulate an infinite number of
reciprocal $(6,10)$ figures or configurations via their constituent sublattices
$\Phi^{(0)}$ and $\Phi^{(1)}$. By virtue of the identities $cb_2=bc_1$, 
$ed_2=de_1$ and $cda_3 = ac_1d_2$, the dilations may be parametrized 
in terms of
two functions $\tau$ and $\sigma$. The relations (\ref{E47}), now regarded
as difference equations, then reduce to a system of two integrable 
equations. This is the content of the following theorem:

\begin{theorem} {\bf\boldmath ($(6,10)$ lattices).} 
Two three-dimensional lattices $\Phi$ and $\Psi$ or, equivalently, 
$\Phi^{(0)}$ and $\Phi^{(1)}$ on
the complex plane consist of reciprocal $(6,10)$ configurations (figures) 
with vertices $\Psi,\Psi_{12},\Phi_1,\Phi_2,\Phi_{13},
\Phi_{23}$ and $\Psi_1,\Psi_2,\Phi,\Phi_3$, $\Phi_{12},\Phi_{123}$ if
and only if the complex (real) dilations defined by
\bela{E49}
  \bear{rlrl}
   \Phi_{12} - \Phi = &\dis
   \frac{\tau_1\tau_2}{\tau\tau_{12}}(\Phi_1 - \Phi_2),\quad&
   \Psi_1 - \Phi = & \dis\frac{\tau_1\sigma}{\tau\sigma_1}(\Phi_1 - \Psi)\AS
   \Phi_{13} - \Psi = &\dis
   \frac{\sigma_1\tau_3}{\sigma\tau_{13}}(\Psi_1 - \Phi_3),\quad&
   \Psi_2 - \Phi = & \dis\frac{\tau_2\sigma}{\tau\sigma_2}(\Phi_2 - \Psi)\AS
   \Phi_{23} - \Psi = &\dis
   \frac{\sigma_2\tau_3}{\sigma\tau_{23}}(\Psi_2 - \Phi_3)
  \ear
\ela
may be parametrized in terms of solutions of the coupled BKP-type system
\bela{E50}
  \bear{rl}
  \sigma_{12}\tau + \sigma_1\tau_2 = & \sigma\tau_{12} + \sigma_2\tau_1\as
  \sigma\tau_{123} + \sigma_2\tau_{13} = &  \sigma_{12}\tau_3 
                                     +\sigma_1\tau_{23}.
  \ear
\ela
Lattices of this type are equivalently described by the cross-ratio
relations
\bela{E51}
  \bear{rl}
  Q(\Phi_1,\Phi_2,\Psi,\Psi_{12}) = & Q(\Psi_2,\Psi_1,\Phi_{12},\Phi)\as
  Q(\Psi_1,\Psi_2,\Phi_3,\Phi_{123}) = & Q(\Phi_{23},\Phi_{13},\Psi_{12},\Psi).
 \ear
\ela  
\end{theorem}

It turns out that the above results may be generalized. It is indeed
possible to define
integrable systems on three-dimensional hybrids of octahedral and hexahedral 
lattices which give rise to lattices on the complex plane composed of
Maxwell's complete octahedral figures and their hexahedral reciprocals. This 
is briefly discussed in the final section.

\section{Reciprocal octahedral and hexahedral figures}
\setcounter{equation}{0}

We begin with an observation associated with reciprocal triangles and
$(6,10)$ figures. Thus, the 8-point relation (\ref{E6}) may be
interpreted as the equality of the cross-ratios associated with any two
`parallel' quadrilaterals which reside in reciprocal triangles.
Indeed, for instance, the closed polygon
\bela{F0a}
\Phi_1\rightarrow\Phi_2\rightarrow\Phi_3\rightarrow\Phi_{123}\rightarrow
\Phi_1  
\ela
in Figure \ref{reciprocal} admits the counterpart
\bela{F0b}
\Phi_{12}\rightarrow\Phi\rightarrow\Phi_{23}\rightarrow\Phi_{13}
\rightarrow\Phi_{12}
\ela
with corresponding parallel edges and the 8-point
relation may be written as 
\bela{F0c}
Q(\Phi_1,\Phi_2,\Phi_3,\Phi_{123})=Q(\Phi_{12},\Phi,\Phi_{23},\Phi_{13}).
\ela
The cross-ratio relations~(\ref{E48})
associated with the reciprocal $(6,10)$ figures depicted in Figure 
\ref{sixten} admit the same interpretation. Moreover, these
imply that, for instance,
\bela{F1}
  M(\Phi_1,\Phi_2,\Psi,\Phi_{23},\Phi_{13},\Psi_{12}) =
  M(\Phi_{12},\Phi,\Psi_2,\Phi_3,\Phi_{123},\Psi_1),
\ela
where the multi-ratio of six points $P_1,\ldots,P_6$ on the complex plane is
defined~by
\bela{F2}
  M(P_1,P_2,P_3,P_4,P_5,P_6) = 
  \frac{(P_1-P_2)(P_3-P_4)(P_5-P_6)}{(P_2-P_3)(P_4-P_5)(P_6-P_1)}.
\ela
Inspection of Figure \ref{sixten} reveals that the points 
$\Phi_1,\Phi_2,\Psi,\Phi_{23},\Phi_{13},\Psi_{12}$
and $\Phi_{12},\Phi,\Psi_2,\Phi_3,\Phi_{123},\Psi_1$ constitute the vertices of
two `parallel' hexagons which are generated by moving along the edges 
$1, 2, 6, 10, 8$ and 4. We observe in passing that, up to complex conjugation, 
the multi-ratio is invariant under the group of inversive transformations.
This has been exploited in \cite{KonSch01} to relate the discrete Schwarzian
KP equation to the fundamental Theorem of Menelaus in the setting of
plane inversive geometry.

\begin{figure}
\begin{center}
\setlength{\unitlength}{0.00045489in}
\begin{picture}(8641,4146)(-500,-10)
\put(0203,3903){$P_3$}
\put(-208,2128){$P_2$}
\put(1178,1883){$P_4$}
\put(-1353,228){$P_1$}
\put(0478,878){$P_6$}
\put(2753,228){$P_5$}
\put(2978,1798){$P_6'$}
\put(6753,1968){$P_3'$}
\put(7183,658){$P_4'$}
\put(6928,3258){$P_2'$}
\put(7153,4248){$P_1'$}
\put(7543,-297){$P_5'$}
\dottedline{45}(6873,3483)(6238,1903)
\dottedline{45}(6243,1898)(7138,558)
\path(6878,3478)(6648,1873)
\path(6648,1873)(7133,563)
\path(6870,3471)(7253,4078)
\path(3428,1903)(7253,4078)
\path(7133,561)(7643,51)
\path(3428,1903)(7635,51)
\dottedline{45}(3428,1903)(6240,1903)
\dottedline{45}(7643,43)(8588,1408)
\dottedline{45}(7253,4071)(8588,1401)
\dottedline{45}(6653,1866)(8588,1401)
\dottedline{45}(-953,328)(2653,328)
\dottedline{45}(0403,3703)(2653,328)
\dottedline{45}(-953,328)(0403,3703)
\path(-953,328)(0628,1228)
\path(0628,1228)(2653,328)
\path(0403,3703)(1078,1903)
\path(0403,3703)(0178,2128)
\dottedline{45}(0178,2128)(0628,1228)
\dottedline{45}(0628,1228)(1078,1903)
\dottedline{45}(0178,2128)(1078,1903)
\path(0178,2128)(-953,328)
\path(1078,1903)(2653,328)
\put(0403,3703){\shade\ellipse{100}{100}}
\put(0178,2128){\shade\ellipse{100}{100}}
\put(1078,1903){\shade\ellipse{100}{100}}
\put(-943,328){\shade\ellipse{100}{100}}
\put(0628,1228){\shade\ellipse{100}{100}}
\put(2653,328){\shade\ellipse{100}{100}}
\put(3428,1898){\blacken\ellipse{100}{100}}
\put(6243,1898){\whiten\ellipse{100}{100}}
\put(6653,1868){\blacken\ellipse{100}{100}}
\put(7133,558){\blacken\ellipse{100}{100}}
\put(6878,3478){\blacken\ellipse{100}{100}}
\put(7253,4078){\blacken\ellipse{100}{100}}
\put(8588,1403){\whiten\ellipse{100}{100}}
\put(7643,53){\blacken\ellipse{100}{100}}
\put(0808,148){$\scriptscriptstyle1$}
\put(1663,1903){$\scriptscriptstyle2$}
\put(-500,1903){$\scriptscriptstyle3$}
\put(-1000,3000){\scriptsize (a)}
\put(3500,3000){\scriptsize (b)}
\put(-98,688){$\scriptscriptstyle4$}
\put(1348,688){$\scriptscriptstyle5$}
\put(0673,2488){$\scriptscriptstyle6$}
\put(0313,2488){$\scriptscriptstyle7$}
\put(0223,1588){$\scriptscriptstyle8$}
\put(0878,1408){$\scriptscriptstyle9$}
\put(0493,2083){$\scriptscriptstyle10$}
\put(-228,1228){$\scriptscriptstyle11$}
\put(1503,1093){$\scriptscriptstyle12$}
\put(6488,1138){$\scriptscriptstyle2$}
\put(6353,2623){$\scriptscriptstyle3$}
\put(5228,3028){$\scriptscriptstyle4$}
\put(5543,733){$\scriptscriptstyle5$}
\put(5138,1993){$\scriptscriptstyle1$}
\put(6918,1273){$\scriptscriptstyle6$}
\put(6828,2488){$\scriptscriptstyle7$}
\put(7973,2713){$\scriptscriptstyle8$}
\put(8198,643){$\scriptscriptstyle9$}
\put(7568,1723){$\scriptscriptstyle10$}
\put(7050,3568){$\scriptscriptstyle11$}
\put(7393,373){$\scriptscriptstyle12$}
\end{picture}
\end{center}
\caption{`Parallel' hexagons $H$}
\label{hexagon}
\end{figure}
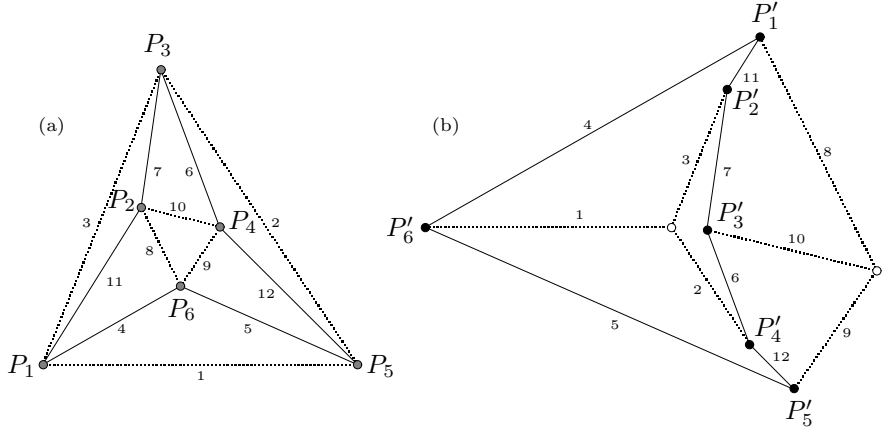

\begin{figure}
\begin{center}
\setlength{\unitlength}{0.00045489in}
\begin{picture}(8641,4146)(-500,-10)
\path(6873,3483)(6238,1903)
\blacken\path(6238,1903)(6310,2003)(6254,2025)(6238,1903)
\path(6243,1898)(7138,558)
\blacken\path(6243,1898)(6284,1781)(6334,1814)(6243,1898)
\path(6878,3478)(6648,1873)
\blacken\path(6648,1873)(6694,1987)(6635,1996)(6648,1873)
\path(6648,1873)(7133,563)
\blacken\path(6648,1873)(6661,1750)(6717,1770)(6648,1873)
\dottedline{45}(6870,3471)(7253,4078)
\dottedline{45}(3428,1903)(7253,4078)
\dottedline{45}(7133,561)(7643,51)
\dottedline{45}(3428,1903)(7635,51)
\path(3428,1903)(6240,1903)
\blacken\path(6240,1903)(6120,1933)(6120,1873)(6240,1903)
\dottedline{45}(7643,43)(8588,1408)
\dottedline{45}(7253,4071)(8588,1401)
\path(6653,1866)(8588,1401)
\blacken\path(6653,1866)(6762,1808)(6776,1867)(6653,1866)
\path(-953,328)(2653,328)
\blacken\path(2653,328)(2533,358)(2533,298)(2653,328)
\path(0403,3703)(2653,328)
\blacken\path(403,3703)(444,3586)(494,3619)(403,3703)
\path(-953,328)(0403,3703)
\blacken\path(-953,328)(-881,428)(-937,450)(-953,328)
\dottedline{45}(-953,328)(0628,1228)
\dottedline{45}(0628,1228)(2653,328)
\path(0403,3703)(1078,1903)
\blacken\path(1078,1903)(1063,2025)(1007,2004)(1078,1903)
\path(0403,3703)(0178,2128)
\blacken\path(403,3703)(356,3588)(415,3579)(403,3703)
\dottedline{45}(0178,2128)(0628,1228)
\dottedline{45}(0628,1228)(1078,1903)
\path(0178,2128)(1078,1903)
\blacken\path(178,2128)(287,2069)(301,2128)(178,2128)
\dottedline{45}(0178,2128)(-953,328)
\dottedline{45}(1078,1903)(2653,328)
\put(0403,3703){\shade\ellipse{100}{100}}
\put(0178,2128){\shade\ellipse{100}{100}}
\put(1078,1903){\shade\ellipse{100}{100}}
\put(-943,328){\shade\ellipse{100}{100}}
\put(0628,1228){\whiten\ellipse{100}{100}}
\put(2653,328){\shade\ellipse{100}{100}}
\put(3428,1898){\blacken\ellipse{100}{100}}
\put(6243,1898){\blacken\ellipse{100}{100}}
\put(6653,1868){\blacken\ellipse{100}{100}}
\put(7133,558){\blacken\ellipse{100}{100}}
\put(6878,3478){\blacken\ellipse{100}{100}}
\put(7253,4078){\whiten\ellipse{100}{100}}
\put(8588,1403){\blacken\ellipse{100}{100}}
\put(7643,53){\whiten\ellipse{100}{100}}
\put(0808,128){$\v_1$}
\put(1643,1903){$\v_3$}
\put(-658,1903){$\v_2$}
\put(-1000,3000){\scriptsize (a)}
\put(3500,3000){\scriptsize (b)}
\put(1000,2168){$\v_4$}
\put(0313,2488){$\v_5$}
\put(0493,1783){$\v_6$}
\put(6288,1138){$\v_3'$}
\put(6153,2623){$\v_2'$}
\put(5138,2013){$\v_1'$}
\put(6918,1273){$\v_4'$}
\put(6828,2488){$\v_5'$}
\put(7568,1723){$\v_6'$}
\end{picture}
\end{center}
\caption{`Parallel' figures $T$ and $T'$}
\label{triangle}
\end{figure}
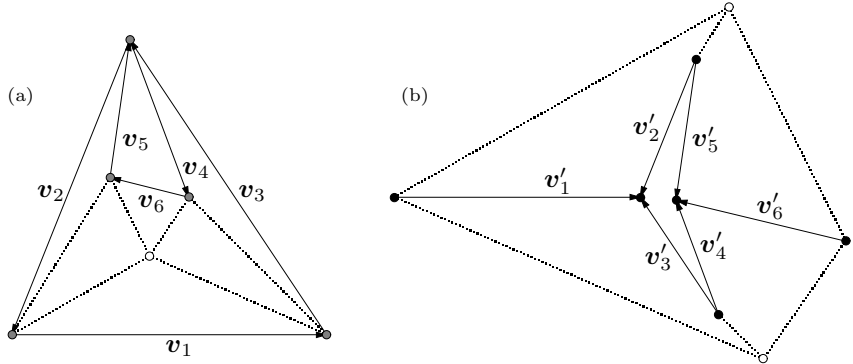

We now focus on Maxwell's octahedral figure and its
hexahedral reciprocals as displayed in Figure \ref{indeterminate}. It is 
readily seen that there exist four pairs of parallel hexagons. If the 
ordered collections of points
$P_1,\ldots,P_6$ and $P_1',\ldots,P_6'$ denote the vertices of a hexagon
\bela{F3}
  H(P_1,P_2,P_3,P_4,P_5,P_6)
\ela
and its parallel companion
\bela{F4}
  H(P_1',P_2',P_3',P_4',P_5',P_6')
\ela
respectively as shown in Figure \ref{hexagon}
then real dilations $\alpha_{ik}$ may be introduced according to
\bela{F5}
  P_i' - P_k' = \alpha_{ik}(P_i-P_k).
\ela
Another figure which turns out to be significant is that of two triangles
joined by a vertex. Maxwell's octahedral figure contains twelve objects of
this kind, an example of which together with the associated parallel 
figure in the hexahedral reciprocal is shown in Figure \ref{triangle}. 
The edges
of the pair of triangles
\bela{F6}
  T(\v_1,\v_2,\v_3,\v_4,\v_5,\v_6)
\ela
are denoted by $\v_1,\ldots,\v_6$ while the edges of the parallel figure
\bela{F7}
  T'(\v_1',\v_2',\v_3',\v_4',\v_5',\v_6')
\ela
are labelled by $\v_1',\ldots,\v_6'$. The orientation of the vectors
$\v_i$ and $\v_i'$ is prescribed as follows. The vectors $\v_i'$ point
towards the two vertices of $T'$ which are dual to the two triangles in
$T$, that is the vertices which are enclosed by $\v_1',\v_2',\v_3'$ and
$\v_4',\v_5',\v_6'$ respectively.
We assign a counterclockwise (or clockwise) orientation to the
quadrilateral in $T'$, that is we regard the quadrilateral as being 
generated by moving successively
along the edges associated with $\v_2',\v_3',\v_4',\v_5'$, say. 
The orientation of the degenerate 
hexagon~$T$ is defined to be that of the quadrilateral in the sense that we 
follow the edges in the same order, that is 
$\v_2,(\v_1),\v_3,\v_4,(\v_6),\v_5$.
This determines the direction of the vectors $\v_i$. Associated dilations 
$\alpha_i$ are then given by
\bela{F8}
  \v_i' = \alpha_i\v_i.
\ela

In Section 2, it has been demonstrated that any octahedral figure is
associated with a two-parameter family of hexahedral reciprocals. As usual,
we here regard figures which are similar as identical. This may now be
exploited to construct a one-parameter family of reciprocals which has
distinct properties. This is encapsulated in the following theorem 
which is a consequence of the analysis conducted below.

\begin{theorem} {\bf (A canonical class of reciprocal octahedral and 
hexahedral figures.)}
Any octahedral figure admits a one-parameter family of hexahedral reciprocals
such that any two parallel hexagons
$H(P_1,P_2,P_3,P_4,P_5,P_6)$ and $H(P_1',P_2',P_3',P_4',P_5',P_6')$ 
and any two parallel figures $T(\v_1,\v_2,\v_3,\v_4,\v_5,\v_6)$ 
and $T'(\v_1',\v_2',\v_3',\v_4',\v_5',\v_6')$ as defined above give rise to
the algebraic identities
\bela{F9}
  \bear{c}\dis
 \frac{\alpha_{12}\alpha_{34}\alpha_{56}}{\alpha_{23}\alpha_{45}\alpha_{61}}
  =1\AS\dis
   \alpha_2+\alpha_3+\frac{\alpha_2\alpha_3}{\alpha_1}
  +\alpha_4+\alpha_5+\frac{\alpha_4\alpha_5}{\alpha_6} = 0\AS
  \dis M(P_1,P_2,P_3,P_4,P_5,P_6) = M(P_1',P_2',P_3',P_4',P_5',P_6').
 \ear
\ela
\end{theorem}

\noindent
Even though there exist four pairs of parallel hexagons and twelve pairs of 
parallel figures $T$ and $T'$, only six of the identities (\ref{F9})$_{1,2}$ 
are independent. Moreover,
the relations~(\ref{F9})$_{1,3}$ are identical by virtue of the definitions
(\ref{F5}). In the context of octahedral and hexahedral integrable lattices
(cf.\ Section 9.2),
these may be satisfied identically by introducing `tau-functions'. The 
remaining (three) relations~(\ref{F9})$_2$ are interpreted once again 
as a discrete integrable system.

\subsection{An algebraic characterization}

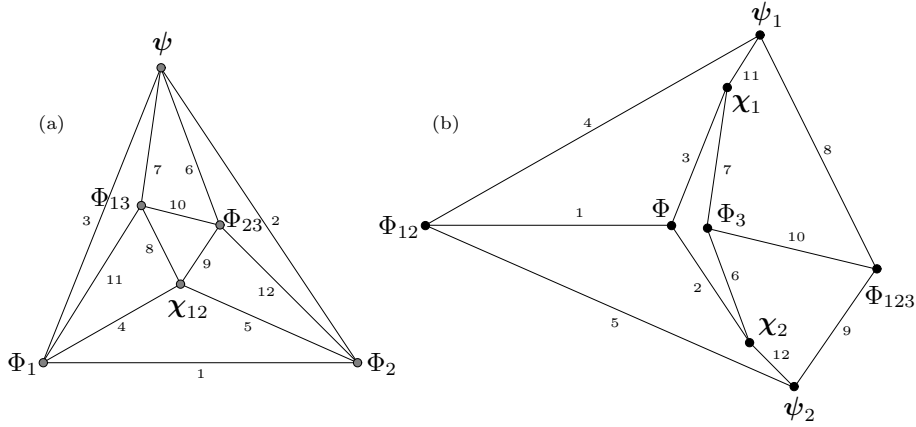
\begin{figure}
\begin{center}
\setlength{\unitlength}{0.00045489in}
\begin{picture}(8641,4146)(-500,-10)
\put(0303,3903){$\Psi$}
\put(-408,2128){$\Phi_{13}$}
\put(1108,1883){$\Phi_{23}$}
\put(-1353,228){$\Phi_1$}
\put(0478,908){$\Chi_{12}$}
\put(2753,228){$\Phi_2$}
\put(2888,1798){$\Phi_{12}$}
\put(6013,1998){$\Phi$}
\put(6753,1938){$\Phi_3$}
\put(7233,708){$\Chi_2$}
\put(6928,3258){$\Chi_1$}
\put(7153,4248){$\Psi_1$}
\put(8448,1003){$\Phi_{123}$}
\put(7523,-267){$\Psi_2$}
\path(6873,3483)(6238,1903)
\path(6243,1898)(7138,558)
\path(6878,3478)(6648,1873)
\path(6648,1873)(7133,563)
\path(6870,3471)(7253,4078)
\path(3428,1903)(7253,4078)
\path(7133,561)(7643,51)
\path(3428,1903)(7635,51)
\path(3428,1903)(6240,1903)
\path(7643,43)(8588,1408)
\path(7253,4071)(8588,1401)
\path(6653,1866)(8588,1401)
\path(-953,328)(2653,328)
\path(0403,3703)(2653,328)
\path(-953,328)(0403,3703)
\path(-953,328)(0628,1228)
\path(0628,1228)(2653,328)
\path(0403,3703)(1078,1903)
\path(0403,3703)(0178,2128)
\path(0178,2128)(0628,1228)
\path(0628,1228)(1078,1903)
\path(0178,2128)(1078,1903)
\path(0178,2128)(-953,328)
\path(1078,1903)(2653,328)
\put(0403,3703){\shade\ellipse{100}{100}}
\put(0178,2128){\shade\ellipse{100}{100}}
\put(1078,1903){\shade\ellipse{100}{100}}
\put(-943,328){\shade\ellipse{100}{100}}
\put(0628,1228){\shade\ellipse{100}{100}}
\put(2653,328){\shade\ellipse{100}{100}}
\put(3428,1898){\blacken\ellipse{100}{100}}
\put(6243,1898){\blacken\ellipse{100}{100}}
\put(6653,1868){\blacken\ellipse{100}{100}}
\put(7133,558){\blacken\ellipse{100}{100}}
\put(6878,3478){\blacken\ellipse{100}{100}}
\put(7253,4078){\blacken\ellipse{100}{100}}
\put(8588,1403){\blacken\ellipse{100}{100}}
\put(7643,53){\blacken\ellipse{100}{100}}
\put(0808,148){$\scriptscriptstyle1$}
\put(1663,1903){$\scriptscriptstyle2$}
\put(-500,1903){$\scriptscriptstyle3$}
\put(-1000,3000){\scriptsize (a)}
\put(3500,3000){\scriptsize (b)}
\put(-98,688){$\scriptscriptstyle4$}
\put(1348,688){$\scriptscriptstyle5$}
\put(0673,2488){$\scriptscriptstyle6$}
\put(0313,2488){$\scriptscriptstyle7$}
\put(0223,1588){$\scriptscriptstyle8$}
\put(0878,1408){$\scriptscriptstyle9$}
\put(0493,2083){$\scriptscriptstyle10$}
\put(-228,1228){$\scriptscriptstyle11$}
\put(1503,1093){$\scriptscriptstyle12$}
\put(6488,1138){$\scriptscriptstyle2$}
\put(6353,2623){$\scriptscriptstyle3$}
\put(5228,3028){$\scriptscriptstyle4$}
\put(5543,733){$\scriptscriptstyle5$}
\put(5138,1993){$\scriptscriptstyle1$}
\put(6918,1273){$\scriptscriptstyle6$}
\put(6828,2488){$\scriptscriptstyle7$}
\put(7973,2713){$\scriptscriptstyle8$}
\put(8198,643){$\scriptscriptstyle9$}
\put(7568,1723){$\scriptscriptstyle10$}
\put(7050,3568){$\scriptscriptstyle11$}
\put(7393,373){$\scriptscriptstyle12$}
\end{picture}
\end{center}
\caption{A parametrization of reciprocal octahedral and hexahedral figures}
\label{break}
\end{figure}

The octahedral figure \ref{indeterminate}(a) may be obtained from the
$(6,10)$ figure \ref{sixten}(a) by adding the edges 11 and 12. From a 
combinatorial point of view, this amounts to replacing each of  
the vertices $\Psi_1$ and $\Psi_2$ in the reciprocal figure \ref{sixten}(b)
by an edge and two vertices. Accordingly, it is natural to adopt the
parametrization shown in Figure \ref{break}. Indeed, if
the vertices $\Chi_1,\Chi_2$ and $\Psi_1,\Psi_2$ coincide then 
Figure~\ref{break}(b) reduces to Figure~\ref{sixten}(b).
The conditions of parallelism are now naturally split into the two groups of
linear equations
\bela{F10}
 \bear{rlrl}
 \Phi_{12}-\Phi = & a(\Phi_1-\Phi_2),\quad & \Chi_1-\Phi = & d(\Phi_1-\Psi)\as
 \Phi_{13}-\Psi = & b(\Chi_1-\Phi_3),\quad & \Chi_2-\Phi = & e(\Phi_2-\Psi)\as
 \Phi_{23}-\Psi = & c(\Chi_2-\Phi_3)&&
 \ear
\ela
and 
\bela{F11}
 \bear{rlrl}
 \Phi_{123}-\Phi_3 = & a_3(\Phi_{13}-\Phi_{23}),
  \quad & \Chi_{12}-\Phi_2 = & d_2(\Phi_{12}-\Psi_2)\as
 \Phi_{123}-\Psi_2 = & b_2(\Chi_{12}-\Phi_{23}),
  \quad & \Chi_{12}-\Phi_1 = & e_1(\Phi_{12}-\Psi_1)\as
 \Phi_{123}-\Psi_1 = & c_1(\Chi_{12}-\Phi_{13})&&
 \ear
\ela
and the pair
\bela{F12}
  \bear{rl}
    \Psi_1-\Chi_1 = &g(\Phi_{13}-\Phi_1)\as
    \Psi_2-\Chi_2 = &h(\Phi_{23}-\Phi_2)
  \ear
\ela
with real dilations $a,b,c,d,e,a_3,b_2,c_1,d_2,e_1$ and $g,h$.
Elimination of $\Chi_{12}$ and $\Phi_{123}$ from (\ref{F11}) leads to three
relations of the form
\bela{F13}
  E^{1k}(\Phi_1-\Psi)+E^{2k}(\Phi_2-\Psi)+E^{3k}(\Phi_3-\Phi) = 0,\quad
  k=1,2,3,
\ela
where the real coefficients $E^{ik}$ are given in terms of the dilations.
In contrast to the cases discussed in the preceding, these relations do
not imply that the coefficients $E^{ik}$ must vanish. However, in view of
a connection with discrete integrable systems, it is canonical to assume that
\bela{F14}
   E^{ik}=0.
\ela
It turns out that only six of the latter conditions on the 
dilations are independent. If we now regard the vertices 
$\Phi,\Phi_1,\Phi_2,\Phi_3$ and $\Psi$ as given then the system 
(\ref{F10})-(\ref{F12}) is compatible modulo (\ref{F13}). Since we have
five points and six dilations at our disposal, the class of reciprocal
octahedral and hexahedral figures constructed in this manner
contains sixteen arbitrary real parameters.
On the other hand, it is not difficult to show that any octahedral
figure is represented by this class. Accordingly, if we specify an octahedral
figure and the point $\Phi$, say, then there exist two degrees of freedom in
the construction of the hexahedral reciprocal. Thus, if we neglect the usual 
arbitrary scaling then the following theorem obtains:

\begin{theorem} {\bf (Construction of one-parameter families of
reciprocal 
\linebreak
octahedral and hexahedral figures.)} Any octahedral figure admits
a one-parameter family of hexahedral reciprocals with $E^{ik}=0$, that is
\bela{F15}
  \bear{rl}
     a_3bd = b_2d_2a,\quad a_3ce &= c_1e_1a,\quad b_2eg = c_1dh\as\dis
     \frac{1}{a_3bc}+\frac{1}{b}-\frac{1}{c}&\dis=\frac{de}{a}+d-e\AS\dis
     \frac{ae_1}{d_2}+\frac{1}{d_2}-a&\dis=ceh+e-h\AS\dis
     \frac{d}{bg}-\frac{1}{b}+d&\dis=\frac{e}{ch}-\frac{1}{c}+e,
  \ear
\ela
where the real dilations $a,\ldots,h$ are defined by (\ref{F10})-(\ref{F12}).
The vertices of these figures obey the multi-ratio relations
\bela{F16}
  \bear{rl}
    M(\Phi_{23},\Phi_{13},\Psi,\Phi_1,\Phi_2,\Chi_{12}) = &
    M(\Phi_{123},\Phi_3,\Chi_1,\Phi,\Phi_{12},\Psi_2)\as
    M(\Phi_{13},\Phi_{23},\Psi,\Phi_2,\Phi_1,\Chi_{12}) = &
    M(\Phi_{123},\Phi_3,\Chi_2,\Phi,\Phi_{12},\Psi_1)\as
    M(\Chi_{12},\Phi_{23},\Phi_2,\Psi,\Phi_1,\Phi_{13}) = &
    M(\Phi_{123},\Psi_2,\Chi_2,\Phi,\Chi_1,\Psi_1).
  \ear
\ela
Conversely, let $\Phi,\Phi_1,\Phi_2,\Phi_3,\Psi$ be five generic points on
the complex plane, $a,b,c,d,e,g$ be six arbitrary non-vanishing real
numbers and $a_3,b_2,c_1,d_2,e_1,h$ be given by (\ref{F15}). Then, the linear
system (\ref{F10})-(\ref{F12}) is compatible and the points $\Psi,\Chi_{12},
\Phi_1,\Phi_2,\Phi_{13},\Phi_{23}$ and $\Psi_1,\Psi_2,\Chi_1,\Chi_2,
\Phi,\Phi_3,\Phi_{12},\Phi_{123}$ constitute the vertices of a 
reciprocal pair of octahedral and hexahedral figures.
\end{theorem}

The relations (\ref{F15}) are nothing but a set of six independent relations 
of the form (\ref{F9})$_{1,2}$. Similarly, the multi-ratio relations
(\ref{F16}) constitute three independent relations of the type 
(\ref{F9})$_3$. In this connection, it is emphasized that for a given 
octahedral figure, the multi-ratio conditions (\ref{F9})$_3$
{\em define} the one-parameter family of hexahedral reciprocals.
  
\subsection{Integrable lattices}

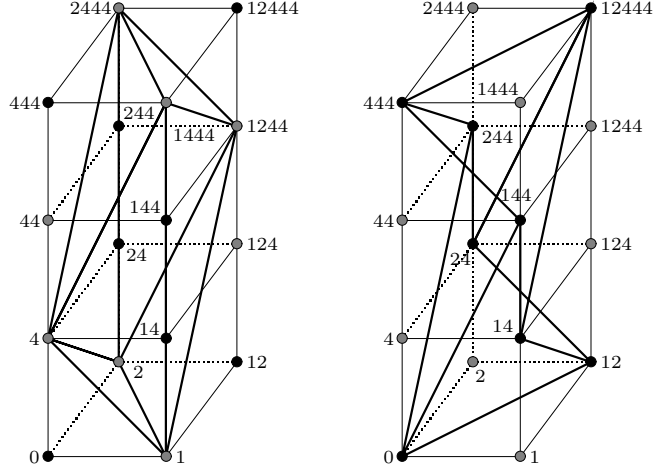
\begin{figure}
\begin{center}
\setlength{\unitlength}{0.00055in}
\begin{picture}(5859,4433)(0,-10)
\put(4055,1910){$\scriptstyle24$}
\put(1415,3100){$\scriptstyle1444$}
\put(1000,2410){$\scriptstyle144$}
\path(225,90)(225,3465)(1350,3465)(1350,90)(225,90)
\path(225,3465)(900,4365)(2025,4365)(2025,990)(1350,90)
\path(225,1215)(1350,1215)(2025,2115)
\path(225,2340)(1350,2340)(2025,3240)
\path(1350,3465)(2025,4355)
\path(3600,90)(3600,3465)(4725,3465)(4725,90)(3600,90)
\path(3600,3465)(4275,4365)(5400,4365)(5400,990)(4725,90)
\path(3600,1215)(4725,1215)(5400,2115)
\path(3600,2340)(4725,2340)(5400,3240)
\path(4725,3465)(5400,4355)
\dottedline{45}(225,90)(900,990)(900,3240)
\dottedline{45}(900,990)(2025,990)
\dottedline{45}(225,1215)(900,2115)(2025,2115)
\dottedline{45}(225,2340)(900,3240)(2025,3240)
\dottedline{45}(900,3240)(900,4365)
\dottedline{45}(3600,90)(4275,990)(4275,3240)
\dottedline{45}(4275,990)(5400,990)
\dottedline{45}(3600,1215)(4275,2115)(5400,2115)
\dottedline{45}(3600,2340)(4275,3240)(5400,3240)
\dottedline{45}(4275,3240)(4275,4365)
\thicklines
\path(1350,90)(225,1215)(900,4365)(1350,3465)(1350,90)(900,990)(225,1215)
     (1350,3465)(225,1215)(900,990)(900,4365)(2025,3240)(1350,90)
\path(900,990)(2025,3240)(1350,3465)
\path(3600,90)(4275,3250)(3600,3465)(4725,2340)(3600,90)(5400,990)(4275,2115)
     (5400,4365)(4725,1215)(5400,990)
\path(3600,3465)(5400,4365)(4275,2115)(4275,3240)
\path(4725,1215)(4725,2340)
\thinlines
\put(225,3465){\blacken\ellipse{100}{100}}
\put(900,4365){\shade\ellipse{100}{100}}
\put(2025,4365){\blacken\ellipse{100}{100}}
\put(1350,3465){\shade\ellipse{100}{100}}
\put(3600,3465){\blacken\ellipse{100}{100}}
\put(4275,4365){\shade\ellipse{100}{100}}
\put(5400,4365){\blacken\ellipse{100}{100}}
\put(4725,3465){\shade\ellipse{100}{100}}
\put(225,90){\blacken\ellipse{100}{100}}
\put(225,2340){\shade\ellipse{100}{100}}
\put(900,2115){\blacken\ellipse{100}{100}}
\put(1350,1215){\blacken\ellipse{100}{100}}
\put(2025,990){\blacken\ellipse{100}{100}}
\put(2025,3240){\shade\ellipse{100}{100}}
\put(3600,90){\blacken\ellipse{100}{100}}
\put(5400,990){\blacken\ellipse{100}{100}}
\put(5400,3240){\shade\ellipse{100}{100}}
\put(3600,2340){\shade\ellipse{100}{100}}
\put(4275,2115){\blacken\ellipse{100}{100}}
\put(4725,1215){\blacken\ellipse{100}{100}}
\put(900,3240){\blacken\ellipse{100}{100}}
\put(1350,2340){\blacken\ellipse{100}{100}}
\put(2025,2115){\shade\ellipse{100}{100}}
\put(225,1215){\shade\ellipse{100}{100}}
\put(900,990){\shade\ellipse{100}{100}}
\put(1350,90){\shade\ellipse{100}{100}}
\put(3600,1215){\shade\ellipse{100}{100}}
\put(4275,3240){\blacken\ellipse{100}{100}}
\put(4725,2340){\blacken\ellipse{100}{100}}
\put(5400,2115){\shade\ellipse{100}{100}}
\put(4725,90){\shade\ellipse{100}{100}}
\put(4275,990){\shade\ellipse{100}{100}}
\put(45,30){$\scriptstyle0$}
\put(1440,30){$\scriptstyle1$}
\put(1035,835){$\scriptstyle2$}
\put(45,1155){$\scriptstyle4$}
\put(-50,2280){$\scriptstyle44$}
\put(-145,3405){$\scriptstyle444$}
\put(1090,1245){$\scriptstyle14$}
\put(2115,930){$\scriptstyle12$}
\put(2115,2055){$\scriptstyle124$}
\put(2115,3180){$\scriptstyle1244$}
\put(2115,4305){$\scriptstyle12444$}
\put(430,4305){$\scriptstyle2444$}
\put(940,3310){$\scriptstyle244$}
\put(970,1950){$\scriptstyle24$}
\put(3420,30){$\scriptstyle0$}
\put(4815,30){$\scriptstyle1$}
\put(5490,930){$\scriptstyle12$}
\put(3420,1155){$\scriptstyle4$}
\put(3325,2280){$\scriptstyle44$}
\put(3230,3405){$\scriptstyle444$}
\put(3795,4305){$\scriptstyle2444$}
\put(4360,3080){$\scriptstyle244$}
\put(4310,3530){$\scriptstyle1444$}
\put(5490,3180){$\scriptstyle1244$}
\put(5490,4305){$\scriptstyle12444$}
\put(5490,2055){$\scriptstyle124$}
\put(4535,2520){$\scriptstyle144$}
\put(4295,800){$\scriptstyle2$}
\put(4460,1280){$\scriptstyle14$}
\end{picture}
\end{center}
\caption{The building blocks of $G^{(0)}$ and $G^{(1)}$}
\label{alter}
\end{figure}

In the case of reciprocal triangles and $(6,10)$ figures, the reciprocals
are of the same type as the original figures. Accordingly, the associated
pairs of integrable lattices consist of two lattices of the same
kind. In the present situation, the reciprocal figures exhibit different
combinatorics. It would therefore be natural to seek two lattices which
consist of octahedral and hexahedral figures respectively. Here, we adopt
an alternative approach and consider two correlated lattices of the same
kind, each of which contains both octahedral and hexahedral figures. Thus,
we are concerned with lattices of the form
\bela{F17}
   \Phi^{(0)} : G^{(0)} \rightarrow \mathbb{C},\quad
   \Phi^{(1)} : G^{(1)} \rightarrow \mathbb{C},
\ela
where the set $G^{(0)}$ is given by
\bela{F18}
  \bear{rl}
  G^{(0)} = \{(n_1,n_2,n_4)\in\mathbb{Z}^3: & n_4 = 0\bmod 3\mbox{ if }
    n_1+n_2 \mbox{ even}\\
    & n_4=1,2\bmod 3 \mbox{ if } n_1+n_2 \mbox{ odd}\}
  \ear
\ela
and $G^{(1)}$ is the complement of $G^{(0)}$. The edge structure of
$G^{(0)}$ and $G^{(1)}$ is obtained by alternating the octahedron and
hexahedron displayed in Figure \ref{alter} in the two horizontal directions. 
In the vertical direction, 
polyhedra of the same type are stacked on top of
each other. Thus, the lattices $\Phi^{(0)}$ and $\Phi^{(1)}$ possess the
combinatorics of three-dimensional hybrids of octahedral and hexahedral 
lattices. In order to show that the relations (\ref{F15}) may be interpreted 
as a discrete integrable system defined on these lattices of non-trivial
combinatorics, it is once again convenient to combine the (vertices of the)
lattices $\Phi^{(i)}$ to one lattice
\bela{F19}
  \hat{\Phi}(\n) = \Phi^{(i)}(\n)\,\mbox{ if }\, \n\in G^{(i)},\quad
  \n = (n_1,n_2,n_4)
\ela
and introduce the notation
\bela{F20}
  \bear{c}
  \Phi(n_1,n_2,n_3) = \hat{\Phi}(n_1,n_2,3n_3),\quad
  \Psi(n_1,n_2,n_3) = \hat{\Phi}(n_1,n_2,3n_3+1)\as
  \Chi(n_1,n_2,n_3) = \hat{\Phi}(n_1,n_2,3n_3+2).
  \ear
\ela

If we now regard indices on objects as shifts along corresponding edges
then the algebraic system (\ref{F10})-(\ref{F12}) may be regarded as a
system of linear difference equations for the lattices $\Phi^{(0)}$ and
$\Phi^{(1)}$ encapsulated in the fields $\Phi,\Psi$ and~$\Chi$. The 
compatibility conditions which guarantee the existence
of $\Phi$ and $\Chi$ are given by (\ref{F13}) and are therefore satisfied if 
we assume that $E^{ik}=0$. The compatibility
condition ${(\Psi_1)}_2={(\Psi_2)}_1$ applied to the pair (\ref{F12}) 
results in the additional constraint
\bela{F21}
   g_2 = h_1.
\ela
The latter and the relations (\ref{F15})$_{1,2,3}$, 
regarded as difference equations,
give rise to a parametrization of the dilations in terms of `tau-functions'
$\tau,\kappa$ and $\nu$ according~to
\bela{F22}
  \bear{c}\dis
   a = \frac{\tau_1\tau_2}{\tau\tau_{12}},\quad
   b = \frac{\kappa_1\tau_3}{\nu\tau_{13}},\quad
   c = \frac{\kappa_2\tau_3}{\nu\tau_{23}}\AS\dis
   d = \frac{\nu\tau_1}{\kappa_1\tau},\quad
   e = \frac{\nu\tau_2}{\kappa_2\tau},\quad
   g = \frac{\tau_1\tau_{13}}{\kappa_1\nu_1},\quad
   h = \frac{\tau_2\tau_{23}}{\kappa_2\nu_2}.
  \ear
\ela
The remaining relations (\ref{F15})$_{4,5,6}$ then reduce to a
`trilinear' integrable system with (\ref{F10}), (\ref{F12}) being its linear
representation. Thus, the following theorem holds:

\begin{theorem} {\bf (Integrable octahedral-hexahedral lattices.)} The lattice
equations
\bela{F23}
 \bear{rlrl}
 \Phi_{12}-\Phi = &\dis \frac{\tau_1\tau_2}{\tau\tau_{12}}(\Phi_1-\Phi_2),
  \quad & \Chi_1-\Phi = &\dis \frac{\nu\tau_1}{\kappa_1\tau}(\Phi_1-\Psi)\AS
 \Phi_{13}-\Psi = &\dis \frac{\kappa_1\tau_3}{\nu\tau_{13}}(\Chi_1-\Phi_3),
 \quad & \Chi_2-\Phi = &\dis \frac{\nu\tau_2}{\kappa_2\tau}(\Phi_2-\Psi)\AS
 \Phi_{23}-\Psi = &\dis \frac{\kappa_2\tau_3}{\nu\tau_{23}}(\Chi_2-\Phi_3),
   \quad & \Psi - \Chi = &\dis\frac{\tau\tau_3}{\kappa\nu}(\Phi_3-\Phi)
 \ear
\ela
are compatible modulo the difference equations
\bela{F24}
 \bear{rl}
   \nu(\tau\tau_{123}-\tau_3\tau_{12}) + \kappa_1(\tau_2\tau_3-\tau\tau_{23})
   +\kappa_2(\tau\tau_{13}-\tau_1\tau_3) = &0\as
   \kappa_2(\tau\kappa_{12}-\tau_1\nu_2) + 
   \tau_2(\nu_1\kappa_2-\tau_3\tau_{12}) + \tau_{12}(\tau\tau_{23}-\nu\nu_2)
   = & 0\as
   \nu(\kappa_1\nu_2-\kappa_2\nu_1)+\kappa_1(\tau_2\tau_3-\tau\tau_{23})
   +\kappa_2(\tau\tau_{13}-\tau_1\tau_3) = &0.
 \ear
\ela
The lattices $\Phi^{(0)}$ and $\Phi^{(1)}$ encapsulated in $\Phi,\Psi$ and
$\Chi$ are composed of reciprocal octahedral and hexahedral figures 
obeying the multi-ratio relations (\ref{F16}).
\end{theorem}

\noindent
We observe that the relations (\ref{F24})$_{1,3}$ imply that
\bela{F25}
  \tau\tau_{123} - \tau_3\tau_{12} = \kappa_1\nu_2 - \kappa_2\nu_1.
\ela

There exists a variety of other
reciprocal figures which may be associated with
cross-ratio and multi-ratio relations and are therefore naturally placed in the
setting of inversive geometry. Whether this property is generic to
reciprocal figures and to what extent reciprocal figures may be extended to
integrable lattices of diverse combinatorics is currently under investigation.

\end{document}